\documentclass[twocolumn,  twocolappendix]{aastex631}

\usepackage{graphicx, bm, amssymb, xcolor}
\usepackage{verbatim}
\usepackage[all]{hypcap}
\usepackage{xspace}
\usepackage{multirow,bigdelim}
\usepackage[fleqn]{amsmath}
\usepackage{subfigure}

\graphicspath{{./}{figures/}}

\begin{document}

\title{Time-scales of polycyclic aromatic hydrocarbon and dust continuum emission from gas clouds compared to molecular gas cloud lifetimes in PHANGS-JWST galaxies}

\shorttitle{Time-scales of mid-IR emission from gas clouds}
\shortauthors{Kim et al. }
\correspondingauthor{Jaeyeon Kim}
\email{jyeonkim@stanford.edu}
\suppressAffiliations
\newcommand{\Ox}{Sub-department of Astrophysics, Department of Physics, University of Oxford, Keble Road, Oxford OX1 3RH, UK}

\newcommand{\UGent}{Sterrenkundig Observatorium, Universiteit Gent, Krijgslaan 281 S9, B-9000 Gent, Belgium}

\newcommand{\STScI}{Space Telescope Science Institute, 3700 San Martin Drive, Baltimore, MD 21218, USA}

\newcommand{\MPIA}{Max-Planck-Institut f\"{u}r Astronomie, K\"{o}nigstuhl 17, D-69117, Heidelberg, Germany}

\newcommand{\AURA}{AURA for the European Space Agency (ESA), Space Telescope Science Institute, 3700 San Martin Drive, Baltimore, MD 21218, USA}

\newcommand{\UCSD}{Department of Astronomy \& Astrophysics, University of California, San Diego, 9500 Gilman Dr., La Jolla, CA 92093, USA}

\newcommand{\JHU}{Department of Physics and Astronomy, The Johns Hopkins University, Baltimore, MD 21218, USA}

\newcommand{\OSU}{Department of Astronomy, The Ohio State University, 140 West 18th Avenue, Columbus, OH 43210, USA}

\newcommand{\CCAPP}{Center for Cosmology and Astroparticle Physics (CCAPP), 191 West Woodruff Avenue, Columbus, OH 43210, USA}

\newcommand{\ARI}{Astronomisches Rechen-Institut, Zentrum f\"{u}r Astronomie der Universit\"{a}t Heidelberg, M\"{o}nchhofstr. 12-14, D-69120 Heidelberg, Germany}

\newcommand{\UConn}{Department of Physics, University of Connecticut, 196A Auditorium Road, Storrs, CT 06269, USA}

\newcommand{\UHawaii}{Institute for Astronomy, University of Hawaii, 2680 Woodlawn Drive, Honolulu, HI 96822, USA}

\newcommand{\UniCA}{Universit\'{e} C\^{o}te d'Azur, Observatoire de la C\^{o}te d'Azur, CNRS, Laboratoire Lagrange, 06000, Nice, France}

\newcommand{\UAlberta}{Dept. of Physics, University of Alberta, 4-183 CCIS, Edmonton, Alberta, T6G 2E1, Canada}

\newcommand{\Arcetri}{INAF — Osservatorio Astrofisico di Arcetri, Largo E. Fermi 5, I-50125, Florence, Italy}

\newcommand{\UWyoming}{Department of Physics and Astronomy, University of Wyoming, Laramie, WY 82071, USA}

\newcommand{\LJMU}{Astrophysics Research Institute, Liverpool John Moores University, 146 Brownlow Hill, Liverpool L3 5RF, UK}

\newcommand{\ITA}{Universit\"{a}t Heidelberg, Zentrum f\"{u}r Astronomie, Institut f\"{u}r Theoretische Astrophysik, Albert-Ueberle-Str 2, D-69120 Heidelberg, Germany}

\newcommand{\CfA}{Center for Astrophysics $\mid$ Harvard \& Smithsonian, 60 Garden St., 02138 Cambridge, MA, USA}

\newcommand{\MPE}{Max-Planck-Institut f\"{u}r Extraterrestrische Physik (MPE), Giessenbachstr. 1, D-85748 Garching, Germany}

\newcommand{\Surrey}{Department of Physics, University of Surrey, Guildford GU2 7XH, UK}

\newcommand{\ESO}{European Southern Observatory, Karl-Schwarzschild Stra{\ss}e 2, D-85748 Garching bei M\"{u}nchen, Germany}

\newcommand{\IWR}{Universit\"{a}t Heidelberg, Interdisziplin\"{a}res Zentrum f\"{u}r Wissenschaftliches Rechnen, Im Neuenheimer Feld 205, D-69120 Heidelberg, Germany}

\newcommand{\ulyon}{Univ Lyon, Univ Lyon1, ENS de Lyon, CNRS, Centre de Recherche Astrophysique de Lyon UMR5574, F-69230 Saint-Genis-Laval France}

\newcommand{\COOL}{Cosmic Origins Of Life (COOL) Research DAO, \href{https://coolresearch.io}{https://coolresearch.io}}

\newcommand{\OAN}{Observatorio Astron{\'o}mico Nacional (IGN), C/ Alfonso XII 3, E-28014 Madrid, Spain}

\newcommand{\UBonn}{Argelander-Institut f\"{u}r Astronomie, Universit\"{a}t Bonn, Auf dem H\"{u}gel 71, 53121 Bonn, Germany}

\newcommand{\kipac}{Kavli Institute for Particle Astrophysics \& Cosmology (KIPAC), Stanford University, CA 94305, USA}

\newcommand{\Umanc}{Jodrell Bank Centre for Astrophysics, Department of Physics and Astronomy, University of Manchester, Oxford Road, Manchester M13 9PL, UK}

\newcommand{\NRAO}{National Radio Astronomy Observatory, 520 Edgemont Road, Charlottesville, VA 22903, USA}

\newcommand{\ANU}{Research School of Astronomy and Astrophysics, Australian National University, Canberra, ACT 2611, Australia}

\newcommand{\AThreeD}{ARC Centre of Excellence for All Sky Astrophysics in 3 Dimensions (ASTRO 3D), Australia}

\newcommand{\IAC}{Instituto de Astrof\'isica de Canarias, C/ V\'ia L\'actea s/n, E-38205, La Laguna, Spain}

\newcommand{\ULL}{Departamento de Astrof\'isica, Universidad de La Laguna, Av. del Astrof\'isico Francisco S\'anchez s/n, E-38206, La Laguna, Spain}

\newcommand{\Princeton}{Department of Astrophysical Sciences, Princeton University, 4 Ivy Lane, Princeton, NJ 08544, USA}

\newcommand{\IRAM}{IRAM, 300 rue de la Piscine, 38400 Saint Martin d'H\'{e}res, France}

\newcommand{\LERMA}{LERMA, Observatoire de Paris, PSL Research University, CNRS, Sorbonne Universit\'{e}s, 75014 Paris}

\newcommand{\YB}{Centro de Desarrollos Tecnol\'ogicos, Observatorio de Yebes (IGN), 19141 Yebes, Guadalajara, Spain}

\newcommand{\Whitman}{Whitman College, 345 Boyer Avenue, Walla Walla, WA 99362, USA}

\newcommand{\IRAP}{IRAP, OMP, UPS, Université de Toulouse, 9 Av. du Colonel Roche, BP 44346, F-31028 Toulouse cedex 4, France}

\newcommand{\ESOChile}{European Southern Observatory (ESO), Alonso de Córdova 3107, Casilla 19, Santiago 19001, Chile}

\author[0000-0002-0432-6847]{Jaeyeon Kim}\thanks{Kavli Postdoctoral Fellow}
\affiliation{\kipac}

\author[0000-0002-5635-5180]{M\'elanie Chevance}
\affiliation{\ITA}
\affiliation{\COOL}

\author[0000-0002-9190-9986]{Lise~Ramambason}
\affiliation{\ITA}

\author[0000-0001-6551-3091]{Kathryn Kreckel}
\affiliation{Astronomisches Rechen-Institut, Zentrum f\"ur Astronomie der Universit\"at Heidelberg, M\"onchhofstr.\ 12-14, D-69120 Heidelberg, Germany}

 \author[0000-0002-0560-3172]{Ralf S.\ Klessen}
\affiliation{Universit\"at Heidelberg, Zentrum f\"ur Astronomie, Institut f\"ur Theoretische Astrophysik, Albert-Ueberle-Str. 2, 69120 Heidelberg, Germany \label{ITA}}
\affiliation{Universit\"{a}t Heidelberg, Interdisziplin\"{a}res Zentrum f\"{u}r Wissenschaftliches Rechnen, Im Neuenheimer Feld 205, 69120 Heidelberg, Germany \label{IWR}}
\affiliation{Harvard-Smithsonian Center for Astrophysics, 60 Garden Street, Cambridge, MA 02138, U.S.A. \label{CfA}}
\affiliation{Elizabeth S. and Richard M. Cashin Fellow at the Radcliffe Institute for Advanced Studies at Harvard University, 10 Garden Street, Cambridge, MA 02138, U.S.A. \label{Radcliffe}}

\author[0000-0002-5782-9093]{Daniel~A.~Dale}
\affiliation{Department of Physics and Astronomy, University of Wyoming, Laramie, WY 82071, USA}

\author[0000-0002-2545-1700]{Adam~K.~Leroy}
\affiliation{\OSU}
\affiliation{\CCAPP}

\author[0000-0002-4378-8534]{Karin~Sandstrom}
\affiliation{\UCSD}

\author[0000-0001-8241-7704]{Ryan~Chown}
\affiliation{\OSU}

\author[0000-0002-0012-2142]{Thomas G. Williams}
\affiliation{\Ox}

\author[0000-0002-4781-7291]{Sumit K. Sarbadhicary}
\affiliation{\OSU, \CCAPP, \JHU}

\author[0000-0002-2545-5752]{Francesco~Belfiore}
\affiliation{INAF — Osservatorio Astrofisico di Arcetri, Largo E. Fermi 5, I-50125, Florence, Italy}

\author[0000-0003-0166-9745]{Frank Bigiel}
\affiliation{Argelander-Institut f\"ur Astronomie, Universit\"at Bonn, Auf dem H\"ugel 71, 53121 Bonn, Germany}

\author[0000-0002-8549-4083]{Enrico Congiu}
\affiliation{\ESOChile}

\author[0000-0002-4755-118X]{Oleg V. Egorov}\affiliation{\ARI}

\author[0000-0002-6155-7166]{Eric Emsellem}
\affiliation{\ESO}

\author[0000-0001-6708-1317]{Simon C.~O.\ Glover}
\affiliation{\ITA}

\author[0000-0002-3247-5321]{Kathryn Grasha}\thanks{ARC DECRA Fellow}
\affiliation{\ANU}

\author[0000-0002-9181-1161]{Annie Hughes}
\affiliation{\IRAP}

\author[0000-0002-8804-0212]{J.~M.~Diederik~Kruijssen}
\affiliation{\COOL}

\author[0000-0002-2278-9407]{Janice C. Lee}
\affiliation{\STScI}

\author[0000-0003-2721-487X]{Debosmita Pathak}
\affiliation{Department of Astronomy, Ohio State University, 180 W. 18th Ave, Columbus, Ohio 43210}
\affiliation{Center for Cosmology and Astroparticle Physics, 191 West Woodruff Avenue, Columbus, OH 43210, USA}

\author[0000-0002-0873-5744]{Ismael Pessa}
\affiliation{Leibniz-Institut for Astrophysik Potsdam (AIP), An der Sternwarte 16, 14482 Potsdam, Germany}

\author[0000-0002-5204-2259]{Erik Rosolowsky}
\affiliation{\UAlberta}

\author[0000-0003-0378-4667]{Jiayi~Sun}
\altaffiliation{NASA Hubble Fellow}
\affiliation{\Princeton}

\author[0000-0002-9183-8102]{Jessica Sutter}
\affiliation{\Whitman}

\author[0000-0002-8528-7340]{David A. Thilker}
\affiliation{\JHU}



\begin{abstract} 
Recent JWST mid-infrared (mid-IR) images, tracing polycyclic aromatic hydrocarbons (PAHs) and dust continuum emission, provide detailed views of the interstellar medium (ISM) in nearby galaxies. Leveraging PHANGS-JWST Cycle 1 and PHANGS-MUSE data, we measure the PAH and dust continuum emission lifetimes of gas clouds across 17 nearby star-forming galaxies by analyzing the relative spatial distributions of mid-IR (7.7-11.3\,$\mu$m) and H$\alpha$ emission at various scales. We find that the mid-IR emitting time-scale of gas clouds in galaxy disks (excluding centers) ranges from 10 to 30\,Myr. After star formation is detected in H$\alpha$, mid-IR emission persists for 3-7\,Myr during the stellar feedback phase, covering 70-80\% of the H$\alpha$ emission. This significant overlap is due to intense radiation from star-forming regions, illuminating the surrounding PAHs and dust grains. In most galaxies, the mid-IR time-scale closely matches the molecular cloud lifetime measured with CO. Although mid-IR emission is complex as influenced by ISM distribution, radiation, and abundances of dust and PAHs, the similarity between the two time-scales suggests that once gas clouds form with compact mid-IR emission, they quickly provide sufficient shielding for stable CO formation. This is likely due to our focus on molecular gas-rich regions of galaxies with near-solar metallicity. Finally, we find that the mid-IR emitting time-scale is longer in galaxies with well-defined \textsc{Hii} regions and less structured backgrounds, allowing photons to more efficiently heat the ambient ISM surrounding the \textsc{Hii} regions, rather than contributing to diffuse emission. This suggests that the shape of the ISM also influences mid-IR emission.
\end{abstract}



\section{Introduction} \label{sec:intro}

Star formation begins in the diffuse interstellar medium (ISM), where giant molecular clouds (GMCs) assemble. When these clouds collapse and form stars, young stars feed back energy and matter to their surroundings, creating \textsc{Hii} regions and exploding as supernovae, revealing young stars within their parental clouds \citep{klessen16, chevance22_rev}. Together, these processes drive the baryon cycle \citep{tumlinson17, schinnerer24}. On a galactic scale, observations have shown a tight correlation between molecular gas and star formation rate (SFR) surface densities, well-known as the ``star formation relation'' \citep{silk97, kennicutt98_rev, bigiel08, leroy13}. However, high-resolution observations at scales of $\lesssim$500\,pc illustrate a spatial decorrelation between molecular clouds and young stellar regions, often traced using CO and H$\alpha$  \citep{elmegreen00, engargiola03,  bigiel08,blitz07, kawamura09, onodera10, schruba10, miura12, meidt15, corbelli17, leroy17, hirota18, kreckel18, kruijssen19, schinnerer19, barnes20, pan22}. 

The spatial decorrelation observed between CO and H$\alpha$ emission causes the global star formation relation to break down at scales of $\lesssim 500$\,pc and is a direct consequence of the gas and young stellar phases  representing different stages in the lifespan of GMCs \citep{schruba10, feldmann11, kruijssen14}. Various studies have utilized the high resolution ($\sim 10-200$\,pc) spatial distributions of CO and H$\alpha$ flux to infer time-scales associated with the evolution of GMCs \citep{corbelli17,grasha18, grasha19, kruijssen19, chevance20, chevance20_rev, zabel20, chevance22, kim22, turner22}. These studies conclude that GMCs typically survive for about 10-30\,Myr before being rapidly dispersed by stellar feedback within 1-5\,Myr after massive star formation is detected in H$\alpha$. In simulations, the observed small-scale spatial offset is used as a diagnostic tool to test \citep{fujimoto19, jeffreson21b, semenov21} and inform \citep{keller22} sub-grid physics for stellar feedback, as well as the assumptions regarding gas and SFR tracers commonly adopted in observations \citep{hu24}. 

\citet{kruijssen14} and \citet{kruijssen18} developed a rigorous method that translates the relative distributions of cold gas and SFR tracer emission into their underlying time-scales. Initial applications to small samples of galaxies indicated that GMCs undergo vigorous, feedback-driven lifecycles \citep{kruijssen19,chevance20,chevance22,ward20_HI,ward22,zabel20,kim21,kim23,lu22} and these findings have contributed to a comprehensive understanding of the necessary conditions for extending these analyses to larger, more systematic galaxy surveys. By applying this method to cloud-scale CO and narrow-band H$\alpha$ observations from PHANGS \citep[][A. Razza et al, in prep]{leroy21_survey}, \citet{kim22} systematically measured the evolutionary sequence of GMCs, from inert molecular gas phase to \textsc{Hii} regions, across 54 main-sequence galaxies. The GMC lifetime varies from 5 to 30\,Myr with an average and 1$\sigma$ range of $16 \pm 6$\,Myr, which is 1{-}3 times the GMC turbulence crossing time-scale. The CO and H$\alpha$ emission overlaps for $1-5$\,Myr, with an average and 1$\sigma$ range of $3.2\pm1.1$\,Myr. These intervals greatly exceed the uncertainties and thus represent physical variation. The CO emission is dispersed before the first supernova (SN) explodes, which typically takes place 4-20\,Myr after massive star formation \citep{chevance22}. Even when the deeply embedded star-forming phase (missed in H$\alpha$) is considered with emission from dust in thermal equilibrium with very intense radiation fields using \textit{Spitzer} 24\,$\mu$m, the feedback time-scale is still short as this phase typically lasts less than 4\,Myr \citep[][L.~Ramambason et al. in preparation with JWST 21\,$\mu$m]{kim21, kim23}. The short overlapping phase indicates that the pre-SN feedback, such as the photoionization and stellar winds of massive stars, is important for molecular cloud dispersal. Supernovae then explode in a pre-processed environment and deposit their energy into the diffuse interstellar medium, potentially contributing to large-scale turbulence \citep{lucas20}, as evidenced by the lack of supernova remnants associated with dense molecular gas \citep{chen23, chen24}.  

The large sample of 54 main-sequence galaxies, together with Local Group galaxies from \citet{kim21}, allowed \citet{kim22} to identify quantitative links between small-scale GMCs and large-scale host galaxy properties. In particular, GMC lifetime increases with increasing stellar mass, which can be explained by two physical arguments, where 1) a large fraction of molecular gas in low-mass and low-metallicity environment is not detected in CO emission, being CO-dark, resulting in apparently short-lived clouds, or 2) high-mass galaxies have a higher mid-plane pressure, creating conditions for longer-lived GMCs. To distinguish how much this trend is affected by the presence of CO-dark gas, high-resolution tracers of cold gas that are independent of their ability to emit in CO are crucially needed.

The main limitation in extending this GMC timeline to represent the multi-phase picture of star formation, including the gas cloud phase dark in CO, has been the lack of cloud-scale observations tracing atomic and molecular gas dark in CO emission. While \textsc{Hi} is the ideal tracer for the neutral ISM, existing \textsc{Hi} observations by the Karl G. Jansky Very Large Array (VLA) and MeerKAT have a physical resolution of $\sim 1$\,kpc \citep{chiang24, eibensteiner24}, which is not sufficient to resolve the spatial offset between clouds and star-forming regions required for robust time-scale measurements \citep{kruijssen18}. So far, such high-resolution observations have been limited to the Local Group galaxies. \citet{ward20_HI, ward22} have shown that in the Large Magellanic Cloud (LMC), the \textsc{Hi} cloud lifetime is $48^{+13}_{-8}$\,Myr and is driven by the gravitational collapse of the mid-plane ISM. In contrast, the measured GMC lifetime in the LMC is much shorter ($11.8^{+2.7}_{-2.2}$\,Myr) and is similar to the GMC free-fall time-scale. The significant difference between the atomic gas and molecular gas cloud lifetimes indicates that GMCs are decoupled from the effects of galactic dynamics and have lifetimes set by internal processes. However, this was based on one galaxy, the LMC, and may be specific to its unique environment such as its low mass, low metallicity and lack of strong spiral arms. 

JWST imaging in the mid-infrared (mid-IR; $\lambda=7.7{-}21\,\mu$m) is transforming our understanding of the ISM, uncovering a detailed network of filamentary structures that are widespread within galaxy disks \citep{lee23, sandstrom23, thilker23, williams24}. The mid-IR emission, in particular at 7.7, 10, and 11.3\,$\mu$m, originates from small dust grains and/or polycyclic aromatic hydrocarbons (PAHs) that are excited by ultraviolet and optical radiation fields \citep{draine07, galliano18, lai20}. \citet{chown24} have characterized the relation between CO and PAH emission across the full PHANGS-JWST Cycle~1 and 2 surveys. CO exhibits a strong and approximately linear relation with PAH emission at $\sim 100$\,pc scales. The scaling relationships between CO and PAH emission are similar in both \textsc{Hii} and non-\textsc{Hii} regions, but they exhibit a dependence on specific star formation rates. \citet{sandstrom23} have used PAH emission to probe ISM in the \textsc{Hi}-dominated regime with a gas surface density below $\sim 7\,M_{\odot}\rm pc^{-2}$.

Using the first four PHANGS-JWST galaxies \citep{lee23}, \citet{leroy23} have investigated correlations among mid-IR, CO and H$\alpha$ emission at $\sim$100\,pc scale. Empirically, the mid-IR emission correlates strongly with both molecular gas and SFR, traced by CO and H$\alpha$ emission, respectively. The CO and H$\alpha$ relations with mid-IR are tighter with steeper slopes compared to the correlation between CO and H$\alpha$ emission. \citet{leroy23} also demonstrated that mid-IR emission can be described as a linear combination of scaled CO and H$\alpha$ fluxes. Across PHANGS-JWST Cycle~1 galaxies \citep{williams24}, \citet{pathak24} have measured that on average 30\% of the flux at 7.7{-}11.3\,$\mu$m can be attributed to mid-IR emission powered by young massive stars and tend to be associated with a high mid-IR intensity. At lower intensities, mid-IR emission originates from relatively diffuse ISM (mostly molecular), which contributes 60-70\% of the total mid-IR flux.

These studies suggest that mid-IR (or PAH) emission traces multiple gas phases. Away from \textsc{Hii} regions, where the diffuse interstellar radiation field dominates heating, mid-IR emission primarily traces the distribution of the neutral ISM, assuming that dust grains and PAHs are well mixed with the neutral gas phases \citep{chown21, gao22, hensley23, sandstrom23}. In contrast, near \textsc{Hii} regions, mid-IR emission is powered by radiation from young stars. The PAH emission is also sensitive to PAH abundances. For instance, PAH emission at 7.7 and 11.3$\,\mu$m is typically suppressed in \textsc{Hii} regions relative to the hot dust continuum at 21$\mu$m, due to a lower PAH abundance in the ionized gas compared to the neutral ISM \citep{gordon08, chastenet19, chastenet23, egorov23, sutter24}.

The mid-IR emission from JWST offers a new, exciting, high-resolution, and high-sensitivity view of the gas distribution for galaxies outside the Local Group. In this work, we leverage PHANGS-JWST Cycle~1 observations to measure mid-IR emitting time-scales, which is an extension of previous studies on GMC lifetimes. We capitalize on mid-IR observations in F770W, F1000W and F1130W (7.7, 10, and 11.3\,$\mu$m, respectively). While F770W and F1130W are regarded as more direct tracers of PAHs, we also include F1000W in our analysis because previous studies have indicated similarities between F1000W and both F770W and F1130W \citep{leroy23, pathak24}. The JWST observations allow us to characterize mid-IR (or PAH) emitting time-scales during the star formation lifecycle and relate this to the successive phases of gas that participates in the star formation process, from neutral gas reservoir (from JWST) to compact molecular gas (from ALMA) and finally to ionized \textsc{Hii} regions (from MUSE H$\alpha$) that are free of cold gas. 

The outline of the paper is as follows. In Section~\ref{sec:obs}, we provide an overview of the observational data used in our analysis. In Section~\ref{sec:method}, we summarize the method and describe relevant input parameters. In Section~\ref{sec:result}, we present our measurements of the evolutionary lifecycle of neutral gas clouds and other derived physical quantities. In Section~\ref{sec:disc}, we examine correlations between our measurements and galactic-scale properties. Finally, we summarize and conclude in Section~\ref{sec:summ}. 

\begin{deluxetable*}{lcccccccccccc}
\tabletypesize{\scriptsize}
\tablewidth{0pt} 
\tablecaption{Summary of physical and observed properties \label{tab:prop}}
\tablehead{
\colhead{} & \colhead{(a)} & \colhead{(b)} & \colhead{(c)} & \colhead{(d)} & \colhead{(e)} & \colhead{(f)} & \colhead{(g)} & \colhead{(h)} & \colhead{(i)} & \colhead{(j)} & \colhead{(k)} & \colhead{(l)}  \\
\colhead{Galaxy} & \colhead{$M_{\rm *}^{\rm global}$} & \colhead{$Z$}& \colhead{$\rm SFR^{\rm global}$} & \colhead{$M_{\rm HI}^{\rm  global}$} & \colhead{$M_{\rm H_{2}}^{\rm global}$} & \colhead{$L_{\rm CO}^{\rm global}$}& \colhead{$\rm\Delta$MS}& \colhead{$R_{\rm eff}$} & \colhead{Dist.} & \colhead{Incl.} & \colhead{P.A.} & \colhead{Hubble}\\
\colhead{} & \colhead{[$\rm log\,M_{\odot}$]} & \colhead{} & \colhead{[$\rm log$\,$M_{\rm\odot}\rm yr^{-1}$]} & \colhead{[$\rm log\,M_{\odot}$]} & \colhead{[$\rm log\,M_{\odot}$]} & \colhead{[$\rm log\,K\,km\,pc^{2}\,s^{-1}$]} & \colhead{[dex]}  & \colhead{[kpc]} & \colhead{[Mpc]} & \colhead{[deg]} & \colhead{[deg]} & \colhead{T-type} 
} 
\startdata 
NGC0628&10.3&8.5&0.2&9.7&9.4&8.4&0.18&3.90&9.84&8.9&20.7&5.2\\
NGC1087&9.9&8.4&0.1&9.1&9.2&8.3&0.33&3.23&15.85&42.9&359.1&5.2\\
NGC1300&10.6&8.5&0.1&9.4&9.4&8.5&-0.18&6.53&18.99&31.8&278.0&4.0\\
NGC1365&11.0&8.5&1.2&9.9&10.3&9.5&0.72&2.78&19.57&55.4&201.1&3.2\\
NGC1385&10.0&8.4&0.3&9.2&9.2&8.4&0.50&3.37&17.22&44.0&181.3&5.9\\
NGC1433&10.9&8.6&0.1&9.4&9.3&8.5&-0.36&4.30&18.63&28.6&199.7&1.5\\
NGC1512&10.7&8.6&0.1&9.9&9.1&8.3&-0.21&4.76&18.83&42.5&261.9&1.2\\
NGC1566&10.8&8.6&0.7&9.8&9.7&8.9&0.29&3.17&17.69&29.5&214.7&4.0\\
NGC1672&10.7&8.6&0.9&10.2&9.9&9.1&0.56&3.39&19.4&42.6&134.3&3.3\\
NGC2835&10.0&8.4&0.1&9.5&8.8&7.7&0.26&3.30&12.22&41.3&1.0&5.0\\
NGC3351&10.4&8.6&0.1&8.9&9.1&8.2&0.05&3.04&9.96&45.1&193.2&3.1\\
NGC3627&10.8&8.6&0.6&9.1&9.8&9.0&0.19&3.64&11.32&57.3&173.1&3.1\\
NGC4254&10.4&8.6&0.5&9.5&9.9&8.9&0.37&2.41&13.1&34.4&68.1&5.2\\
NGC4303&10.5&8.6&0.7&9.7&9.9&9.0&0.54&3.43&16.99&23.5&312.4&4.0\\
NGC4321&10.7&8.6&0.6&9.4&9.9&9.0&0.21&5.50&15.21&38.5&156.2&4.0\\
NGC4535&10.5&8.6&0.3&9.6&9.6&8.6&0.14&6.26&15.77&44.7&179.7&5.0\\
NGC5068&9.4&8.3&-0.6&8.8&8.4&7.3&0.02&1.97&5.2&35.7&342.4&6.0\\
\enddata
\tablecomments{(a) (c), (e), and (f) -- Global stellar mass, SFR, molecular gas mass, and integrated CO luminosity from PHANGS--ALMA \citep{leroy19, leroy21_survey}. (b) -- Molecular gas mass weighted average of gas phase metallicity [12+log(O/H)] from \citet{kreckel19} and \citet{williams22}. (d) -- Literature \textsc{Hi} mass adopted from Hyper-LEDA \citep{makarov14}. (g) and (h) -- Offset from the star-forming main sequence and effective radius from \citet{leroy21_survey}. (i), (j), and (k) -- Distance from \citet{anand21} and inclination and position angle from \citet{lang20}. (l) -- Hubble type from LEDA \citep{paturel98}.}
\end{deluxetable*}

\section{Observational Data}\label{sec:obs}
The PHANGS-JWST Cycle~1 Treasury program (GO 2107, PI Lee; \citealp{lee23, williams24}) has constructed near- and mid-IR observations of 19 nearby star-forming galaxies. In this work, we focus on PHANGS-JWST Cycle 1 galaxies with GMC evolutionary time-scale measurements from \citet{kim22}. This excludes IC\,5332 as the number of CO emission peaks was not enough to statistically sample clouds in various stages of the evolutionary sequence (see Appendix~\ref{app:robust}). We further remove NGC\,7496 as the resolution was not sufficient to resolve the spatial offset between mid-IR emission and star-forming regions for a robust time-scale measurement. As a result, our sample is composed of 17 galaxies in total. Table~\ref{tab:prop} lists the physical and observational properties of these galaxies. As an example, Figure~\ref{fig:obs} shows three-color composite images of NGC\,0628 made using JWST bands, CO, and H$\alpha$ observations. We also show locations of identified mid-IR, CO, and H$\alpha$ emission peaks used in our analysis (see Section~\ref{sec:method}). Three color images and associated emission peaks of the full sample can be found in Appendix~\ref{app:images}. 

\subsection{Mid-IR bands}\label{ssec:gastracer}
We use mid-IR emission at 7.7\,$\mu$m, 10\,$\mu$m, and 11.3\,$\mu$m observed with F770W, F1000W, and F1130W filters on MIRI on board JWST\footnote{F770W band includes contributions from stellar continuum, especially the galaxy centers. Using the stellar continuum subtracted F770W map from \citet{chown24} and \citet{sutter24}, we confirm that the change in our results is negligible, mostly due to galaxy centers being excluded in our analysis.}. The observations are obtained from the Mikulski Archive for Space Telescope at the Space Telescope Science Institute\footnote{Data can be accessed via \url{https://archive.stsci.edu/hlsp/phangs} or \citet{doi}}. For comprehensive explanations of the survey and the data reduction process, we refer readers to \citet{lee23} and \citet{williams24}. The F770W and F1130W filters capture strong emission from PAHs while F1000W is expected to be dominated by dust continuum \citep{whitcomb23} with its behavior similar to the PAH bands \citep{leroy23}. 

These mid-IR bands are known to best correlate with the cold gas \citep{chown21, gao22, leroy23, whitcomb23, chown24} and currently provide the highest resolution and highest sensitivity view of the neutral gas distribution of the PHANGS-JWST galaxies. This makes them a valuable tool for tracing gas distribution that is missed in PHANGS-ALMA CO observations \citep{leroy21_survey}, which includes atomic gas, CO-dark molecular gas or CO-bright molecular gas below the detection threshold. 

However, the physical origin of these emission is complex, as they are influenced by various factors like gas column density, heating, dust and PAH abundances. We do not include the dust continuum at 21\,$\mu$m (F2100W) in our analysis, as it includes emission from dust in thermal equilibrium with very high radiation fields and therefore is more associated with star-forming regions than cold gas \citep{draine07b,kennicutt12, leroy13, belfiore23, egorov23, hassani23, leroy23, whitcomb23, pathak24}. Indeed, this longer wavelength has been successfully used to trace the embedded star-forming period with \textit{Spitzer} 24\,$\mu$m observations \citep{kim21} and with JWST F2100W \citep[][L. Ramambason et al., in prep.]{kim23}. The resolution of F770W, F1000W and F1130W is 0\farcs24, 0\farcs31, and 0\farcs36, respectively, which roughly translates into a physical resolution of $5-35$\,pc across the range of distances of our galaxy sample ($5-20$\,Mpc). 

\subsection{MUSE H$\alpha$}\label{ssec:sfrtracer}
As a SFR tracer, we use H$\alpha$ observations from the PHANGS-MUSE survey \citep{emsellem22}. The full spectral fitting to MUSE observations allows us to properly measure H$\alpha$ emission without contamination from the [NII] line. This is different from the SFR tracer maps adopted in \citet{kim22}, which were narrow-band H$\alpha$ observations taken with the du Pont 2.5-m telescope at the Las Campanas Observatory and the Wide Field Imager instrument at the MPG-ESO 2.2-m telescope at the La Silla Observatory. We confirm that the results are consistent within error bars when the narrow-band H$\alpha$ observations are used. Detailed explanations of the PHANGS-MUSE survey and data reduction can be found in \citet{emsellem22}. In brief, the observations have mapped spectra of 19 nearby star-forming galaxies with the Multi Unit Spectroscopic Explorer (MUSE) integral field spectrograph on the ESO/Very Large Telescope (VLT), where the sample overlaps with PHANGS-JWST Cycle~1. From data cubes with homogenized point spread functions (PSF; `copt' version), we use H$\alpha$ line maps obtained by a Gaussian fit to the H$\alpha$ line, accounting for the stellar continuum and Balmer line absorption. The resolution ranges 0\farcs6{-}1\farcs3, which corresponds to $50-150$\,pc in physical scale for our galaxy sample. The typical 3$\sigma$ H$\alpha$ flux sensitivity is $\rm 4{-}7 \times 10^{37} erg\,s^{-1}kpc^{-2}$ \citep{emsellem22}. 

We note that to be consistent with our analysis in \citet{kim22}, we do not use H$\alpha$ maps that have been corrected for the dust extinction using the Balmer decrement \citep{belfiore22}. In order to assess the impact of extinction correction, we used Balmer decrement-corrected H$\alpha$ observations for one of the nearest galaxy NGC\,0628 from \citet{belfiore23} and find that the changes in our derived time-scale are within the $1\sigma$ uncertainties. However, we observe a slightly smaller average separation between independent star-forming regions after applying the extinction correction ($52_{-8}^{+8}$\,pc) compared to the value of $67_{-8}^{+7}$\,pc reported in Table~\ref{tab:result}. The extinction correction would only have a significant effect if the decrement varied considerably within individual \textsc{Hii} regions, which is unlikely since these regions are typically not well-resolved.

\subsection{Post-processing images}\label{ssec:post}
To proceed with our analysis (see Section~\ref{sec:method}), we apply additional procedures to the data. We first convolve finer-resolution JWST observations to match the coarser resolution of the H$\alpha$ observations, where Table~\ref{tab:input} lists the matched resolution ($l_{\rm ap, min}$). We then regrid these convolved maps to align with the pixel grid of the H$\alpha$ maps. During the convolution, we use a kernel that transforms the JWST PSF to a circular Gaussian PSF, generated following \citet{aniano11}. 

As described in Section~\ref{sec:method}, our time-scale measurements are based on gas-to-SFR tracer flux ratios. Therefore, exceedingly bright regions that are not likely to represent the overall cloud (or \textsc{Hii} region) population but constitute a significant fraction in the total emission have the potential to bias our measurements (e.g., 30~Doradus in the LMC by \citealp{ward22}). Accordingly, in our previous analysis of the same galaxies using CO as a tracer for the molecular gas \citep{kim22}, we masked such bright regions (defined below) detected in CO and H$\alpha$ emission. Most of the galactic centers that appeared crowded were also masked because they complicate the identification of star-forming regions and gas clouds within the environment. 

\begin{figure*}
\includegraphics[scale=0.67]{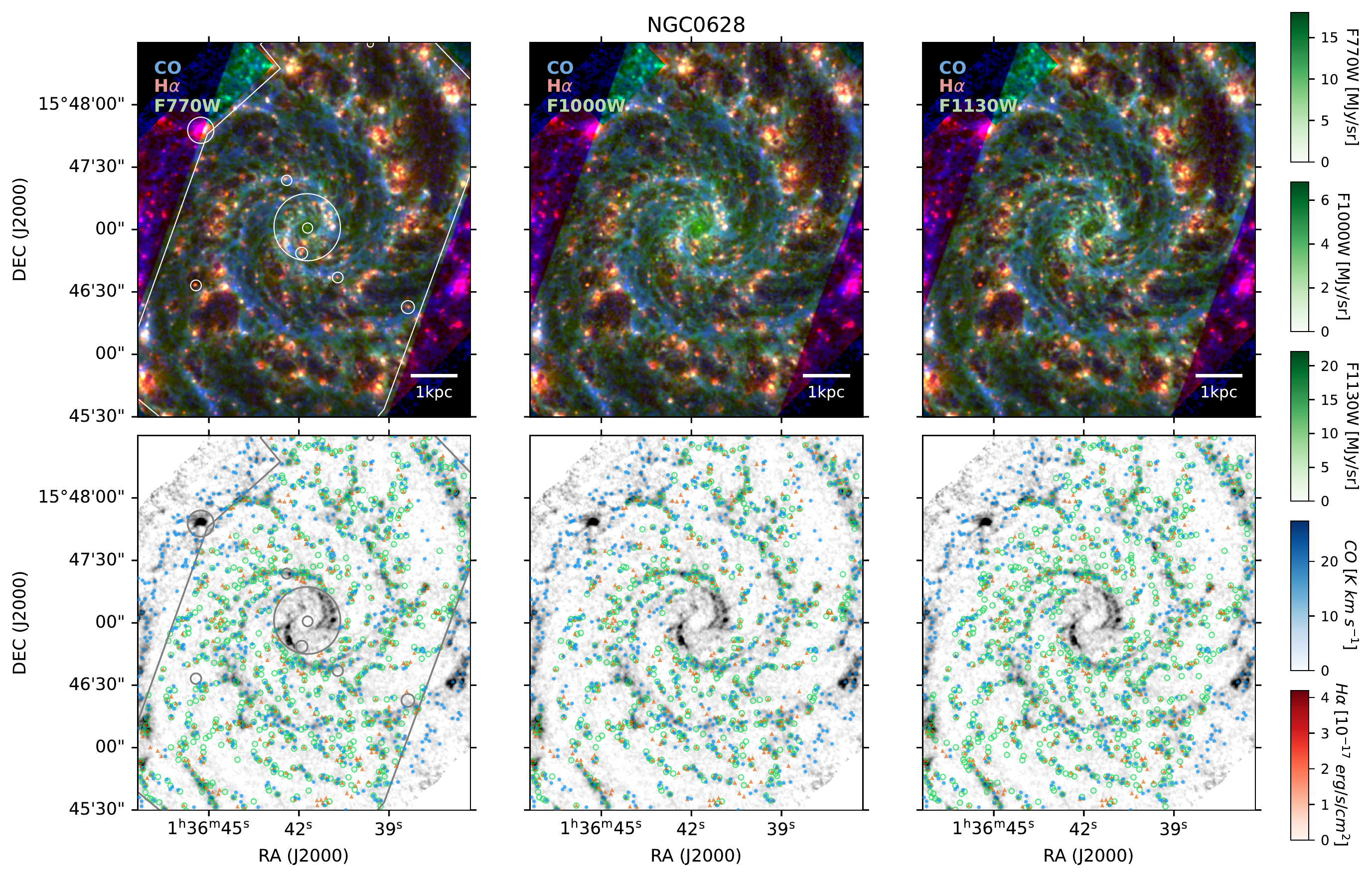}
\caption{\textit{Top:} Composite three color images created using CO (blue), H$\alpha$ (red), and mid-IR (green) observations, where each panel from left to right uses mid-IR emission of F770W, F1000W, and F1130W, respectively. The mid-IR observations have been convolved and regridded to match the coarser resolution and pixel grid of of H$\alpha$ observations (see Section~\ref{ssec:post}). For visualization purposes, a power-law brightness scale with gamma correction ($\gamma=2$) has been applied in the top panels. The color bars on the right reflect the true flux ranges in each observation. The left panel shows the area included in our analysis as a polygon. Crowded galaxy center (ellipse in the center), as well as artifacts and extremely bright peaks (circles) are excluded. \textit{Bottom:} Locations of identified H$\alpha$ (orange triangles), CO (blue filled circles), and mid-IR (green open circles) emission peaks (see Section~\ref{sec:method}) are overlaid on the CO map, which is shown in grayscale with a linear brightness scale. Again, from left to right, the mid-IR emission peaks correspond to peaks identified in F770W, F1000W, and F1130W maps respectively. The CO grayscale image uses the same intensity range as the CO emission in the top panels shown in blue, with range of flux indicated in the color bar on the right.\label{fig:obs}}
\end{figure*}

In this work, to be consistent with previous studies, we use the same masks as \citet{kim22} and apply additional masks if extremely bright regions are present in mid-IR observations. We adopt the same criteria for detecting exceedingly bright regions as in \citet{kim22}, which uses luminosity functions of the identified emission peaks (see Section~\ref{sec:method}). Specifically, we first sort the identified peaks in descending order of luminosity. Then, starting from the top, we search for gaps in the distribution, defined as a peak being at least twice as luminous as the next brightest peak. Whenever such a gap is identified, we mask all peaks that surpass the brightness of the next brightest peak. We check for extremely bright peaks in all three mid-IR bands and mask them in all our analyses, even if the criterion is met in only one of the bands. As a result, we apply additional masking to bright peaks in several galaxies (NGC\,1087, 1300, 1365, 1672, 2835), with up to five extra peaks masked per galaxy. Appendix~\ref{app:masking} shows that masking bright regions has a negligible impact on the derived properties, with results consistent within the 1$\sigma$ uncertainty. Lastly, we limit our analysis to areas where both JWST and MUSE observations overlap (polygon in Figure~\ref{fig:obs}). The analyzed region is similar to \citet{kim22}, which was defined using ALMA and narrowband H$\alpha$ observations.

\begin{deluxetable*}{lcccccccccccccccccccc}
\tabletypesize{\scriptsize}
\tablewidth{0pt} 
\tablecaption{Main input parameters adopted in our analysis. For other parameters not listed here, related to error propagation and model fitting, we use the default values from 
\citet{kruijssen18}. \label{tab:input}}
\tablehead{
\colhead{Galaxy} & \colhead{$r_{\rm min}$} & \colhead{$l_{\rm ap, min}$ } & \colhead{$l_{\rm ap, max}$} & \colhead{$N_{\rm pix, min}$}& \colhead{$t_{\rm ref}$}& \multicolumn{14}{c}{$\Delta\rm log_{10}\mathcal{F}$, $\delta\rm log_{10}\mathcal{F}$, $n_{\lambda}^{a}$} \\
\cline{7-8}\cline{10-12}\cline{14-16}\cline{18-20}
\colhead{} & \colhead{[kpc]} & \colhead{[pc]} & \colhead{[pc]} & \colhead{}& \colhead{[Myr]}& \multicolumn{2}{c}{H$\alpha$}&&\multicolumn{3}{c}{F770W}&&\multicolumn{3}{c}{F1000W}&&\multicolumn{3}{c}{F1130W}
}
\startdata NGC0628&0.1&44&3000&50&$7.45_{-0.64}^{+0.62}$&2.8&0.05&&2.4&0.05&13&&2.4&0.05&13&&2.4&0.05&13\\
NGC1087&0.1&83&3000&40&$8.20_{-1.37}^{+2.27}$&3.2&0.05&&2.5&0.05&14&&2.5&0.05&14&&2.5&0.05&14\\
NGC1300&0.1&89&3000&15&$7.80_{-0.88}^{+0.70}$&2.8&0.05&&2.5&0.05&13&&2.5&0.05&13&&2.5&0.05&13\\
NGC1365&0.2&145&3000&10&$8.15_{-1.09}^{+1.20}$&3.5&0.05&&1.7&0.05&15&&1.7&0.05&15&&1.7&0.05&15\\
NGC1385&0.1&76&3000&20&$6.92_{-0.92}^{+1.56}$&3.0&0.05&&2.2&0.05&14&&2.2&0.05&14&&2.2&0.05&14\\
NGC1433&0.1&88&3000&30&$6.35_{-0.50}^{+0.51}$&2.0&0.05&&1.9&0.10&13&&1.9&0.10&13&&1.9&0.10&13\\
NGC1512&0.1&133&3000&20&$6.02_{-0.51}^{+0.31}$&2.0&0.03&&1.8&0.03&11&&1.8&0.03&14&&1.8&0.03&14\\
NGC1566&0.1&74&2000&20&$8.96_{-1.07}^{+1.21}$&2.6&0.07&&2.8&0.07&14&&2.8&0.07&14&&2.8&0.07&14\\
NGC1672&0.2&105&3000&50&$8.74_{-1.59}^{+1.49}$&2.0&0.05&&2.0&0.05&13&&2.0&0.05&13&&2.0&0.05&13\\
NGC2835&0.1&79&3000&20&$5.59_{-0.47}^{+0.57}$&3.0&0.05&&2.2&0.05&14&&2.2&0.05&11&&2.2&0.05&13\\
NGC3351&0.1&60&3000&20&$6.78_{-0.66}^{+0.98}$&2.0&0.05&&2.8&0.05&13&&2.8&0.05&13&&2.8&0.05&13\\
NGC3627&0.1&78&3000&30&$<7.00^{b}$&3.1&0.05&&3.2&0.05&13&&3.2&0.05&13&&3.2&0.05&13\\
NGC4254&0.1&62&1500&15&$9.01_{-1.03}^{+1.31}$&3.5&0.05&&2.5&0.05&13&&2.5&0.05&13&&2.5&0.05&13\\
NGC4303&0.2&67&3000&40&$8.28_{-1.23}^{+1.79}$&2.5&0.05&&2.6&0.05&13&&2.6&0.05&14&&2.6&0.05&13\\
NGC4321&0.1&97&3000&10&$7.39_{-0.68}^{+0.75}$&2.2&0.05&&2.2&0.05&15&&2.2&0.05&15&&2.2&0.05&15\\
NGC4535&0.1&51&2500&20&$8.87_{-1.08}^{+2.26}$&3.0&0.05&&3.0&0.05&12&&3.0&0.05&11&&3.0&0.05&12\\
NGC5068&0.0&29&3000&70&$5.49_{-0.33}^{+0.40}$&3.4&0.05&&2.6&0.05&13&&2.6&0.05&11&&2.6&0.05&13\\
\enddata
\tablecomments{$^{a}$ For a given galaxy, the same input parameters are used for all the analysis adopting different mid-IR bands except parameters related to peak identification ($\Delta\rm log_{10}\mathcal{F}$) and diffuse emission filtering ($n_{\lambda}$), which are listed separately for each mid-IR band.\\
$^{b}$ Only an upper limit was constrained in \citet{kim22}.}
\end{deluxetable*}

\section{Method}\label{sec:method}
We characterize evolutionary time-scales from mid-IR emitting gas phase to \textsc{Hii} regions by applying a robust statistical method \citep{kruijssen14, kruijssen18} to mid-IR and H$\alpha$ observations. In this section, we briefly describe the method (formalized in the \textsc{Heisenberg} code\footnote{https://github.com/mustang-project/Heisenberg}) and list the main input parameters used. We direct readers to \citet{kruijssen14} for an in-depth explanation of the methodological framework and \citet{kruijssen18} for a demonstration and validation of the code using simulated galaxies, along with a comprehensive description of its input parameters. Including the galaxies in this paper, this method has been applied to $\sim$ 60 observed galaxies to measure GMC evolutionary time-scales in diverse environments (stellar mass $M_{*}=10^{9}-10^{11}~M_\odot$; \citealp{kruijssen19, chevance20, chevance22, haydon20,  hygatePhD, zabel20, ward20_HI, ward22, kim21, kim22, kim23, lu22}). These studies mostly used CO and H$\alpha$ as tracers for cold gas and young star-forming regions, respectively. 

As first demonstrated by \citet{schruba10}, we quantify the small-scale differences in the flux distributions of cold gas and SFR tracers by measuring the gas-to-SFR tracer flux ratios across various spatial scales, ranging from cloud scales ($<$100\,pc) to global, kpc-scale.  The small-scale decorrelation between gas and young stars is naturally explained by the transient nature of the cloud life cycle, during which clouds assemble, form stars and are subsequently dispersed by stellar feedback, only leaving young stars to be detected without associated cold gas. Therefore, the level of (de-)correlation on small scales compared to the galactic-scale is linked to the evolutionary time-scales of clouds. For example, when measuring the gas-to-SFR tracer flux ratio of cold gas peaks on a cloud scale, the deviation from the galactic average value will depend on the flux of SFR tracer peaks included in these small apertures. This information allows us to constrain how long clouds remain inert and the duration required for stellar feedback to disperse them \citep{kruijssen14}.

To determine mid-IR-to-H$\alpha$ flux ratios of emission peaks at various spatial scales, we first identify emission peaks in both mid-IR and H$\alpha$ maps (at the highest PSF-matched resolution; $l_{\rm ap, min}$) using \textsc{Clumpfind} \citep{williams94}. This algorithm detects peaks by contouring the data at multiple flux levels. The contours are spaced apart by a step size of $\rm{\delta}log_{10}\mathcal{F}$, covering the entire range of $\rm{\Delta}log_{10}\mathcal{F}$ starting from the maximum flux. Peaks containing fewer than $N_{\rm pix, min}$ pixels are disregarded. Table~\ref{tab:input} lists adopted values of $\rm{\delta}log_{10}\mathcal{F}$ and $\rm{\Delta}log_{10}\mathcal{F}$ for different mid-IR wavelengths. A single value for $\rm N_{pix, min}$ is used for each galaxy as the spatial resolutions are matched across wavelengths. As shown in Figure~\ref{fig:obs} and Appendix~\ref{app:images}, we have visually verified that all the obvious emission peaks are identified. 

Next, we center apertures of varying sizes, ranging from $\rm l_{ap,min}\approx 100\,pc$ to $\rm l_{ap,max}\approx 3\,kpc$, on each peak of a given type. For each aperture size and peak type, we sum the total enclosed mid-IR and H$\alpha$ flux across all apertures and compute their ratio. This gives the aperture-averaged mid-IR-to-H$\alpha$ flux ratio as a function of aperture size. Because we integrate all the flux within the apertures, our results are not highly sensitive to the specific peak identification parameters.

\begin{figure*}
\centering
\includegraphics[scale=0.45]{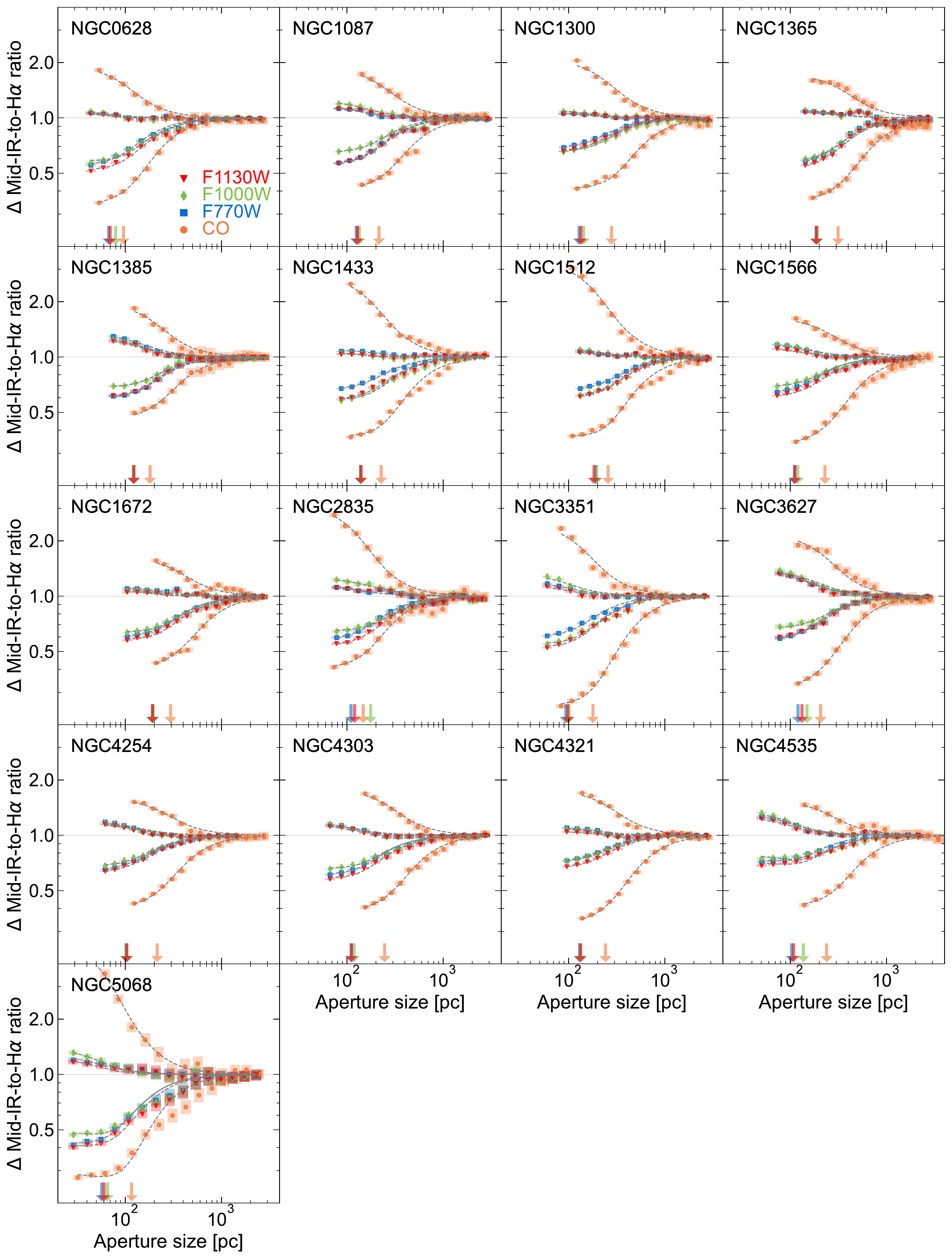}
\caption{The measured deviations of mid-IR-to-H$\alpha$ flux ratio compared to the galactic average value are shown as a function of size-scale. The top branch is when the apertures are focused on mid-IR peaks and the bottom is when the apertures are focused on H$\alpha$ peaks. Shaded region indicates the effective 1$\sigma$ error, after the covariance between data points is taken into account. The same measurements but using CO as the gas tracer from \citet{kim22} is also shown for comparison. We also show the galactic average value, which equals 1 (solid line), as well as our best-fitting model (dashed line) to the measured flux ratios. The arrow indicates our best-fitting $\lambda$, with its value listed in Table~\ref{tab:result} including other constrained parameters ($t_{\rm g}$ and $t_{\rm fb}$). } \label{fig:tuningforks}
\end{figure*}

There could be a case where some apertures will overlap, especially when aperture sizes are big. To avoid double-counting pixels and ensure statistical independence, we generate Monte Carlo realizations of non-overlapping aperture samples for each peak type and aperture size. For each realization, we randomly draw a subset of peaks such that no apertures overlap. We then compute the total fluxes enclosed within these independent samples and repeat this process 1000 times, following the description in \citet{kruijssen18}. The final mid-IR-to-H$\alpha$ flux ratio at a given aperture size is obtained by averaging over all realizations. This ensures that large-scale aperture statistics are not skewed by overlapping regions. Figure~\ref{fig:tuningforks} shows mid-IR-to-H$\alpha$ flux ratios normalized by the ratio on large-scales as a function of aperture size.

To these observed flux ratios, we fit an analytical function which depends on the relative durations of successive phases in the timeline of cloud evolution (mid-IR bright, H$\alpha$ bright and overlap phase), as well as the typical distance between regions undergoing independent evolution ($\lambda$). The relative durations are converted into absolute values by using the duration of the H$\alpha$-bright phase ($t_{\rm s}$) as a reference time-scale ($t_{\rm ref}$) which is measured in \citet{kim22}. The analytical function is then described by three independent quantities: the mid-IR bright phase ($t_{\rm g}^{X}$), the mid-IR and H$\alpha$ overlap phase ($t_{\rm fb}^{X}$), and the region separation length ($\lambda^{X}$), where we use $X$ to denote the mid-IR wavelength. We refer the overlap phase, during which mid-IR and H$\alpha$ emission is found coincident, as the feedback time-scale since it indicates the time for intense radiation from \textsc{Hii} regions to illuminate the surrounding PAH and dust and the time for PAHs and/or dust dispersal at the scales of $\sim$100\,pc.

Our analytical model assumes that the emission enclosed in an aperture originates solely from regions inside it. Diffuse emission on large spatial-scales affects the measured cloud evolutionary time-scales by contributing flux that is not physically associated with the compact emission peaks identified in the maps. For instance, substantial diffuse emission in the gas tracer map could lead to an elevated global gas-to-SFR tracer flux ratio compared to that measured locally centered on gas emission peaks \citep{hygatePhD}. More importantly, the diffuse components in mid-IR and H$\alpha$ maps often have a different physical origin from the compact star-forming structures we aim to trace. For example, large-scale mid-IR emission traces diffuse gas that is not likely to be dense enough to form massive stars \citep{calzetti07, sandstrom23}. One of the major contributions to the diffuse ionized gas is ionizing photons leaking from \textsc{Hii} regions that are no longer co-spatial with their parent molecular clouds. We therefore filter out large-scale emission to avoid biasing the inferred time-scales by emission that does not participate in the evolutionary cycle of compact star-forming regions and clouds we seek to characterize \citep[see also][]{kruijssen18, hygate19, chevance20}.  

We remove emission on scales larger than $n_{\lambda}$ times the region separation length $\lambda$ using a Gaussian high-pass filter in Fourier space, where $n_{\lambda}$ is the scaling factor. We adopt the lowest possible integer value of $n_{\lambda}$ for which the flux loss from the compact emission remains $<$10\%, as recommended by \citet{hygate19}. This approach has been shown to achieve a balance between removing diffuse background emission and preserving the structure of compact regions, and has been adopted in all of our previous analyses \citep{hygatePhD, kruijssen19, chevance20, kim21, kim22, kim23}\footnote{We note that in the case of NGC\,1512, where we measure the highest diffuse emission fraction in 7.7\,$\mu$m, a filtering scale smaller than $n_{\lambda}\times \lambda=4$\,kpc is required to obtain a higher (lower) mid-IR-to-H$\alpha$ flux ratio compared to the galactic average value on small scale, when focusing on mid-IR (H$\alpha$) peaks. The time-scale measurements using $n_{\lambda}\times \lambda=4$\,kpc and our best-fitting model ($n_{\lambda}\times \lambda\approx2$\,kpc) are consistent within 1$\sigma$ error, although the measured diffuse emission fraction at 7.7\,$\mu$m decreases by 10\% due to less emission being filtered out when $n_{\lambda}\times \lambda=4$\,kpc.}. This procedure is carried out iteratively until the convergence criterion is met, defined as when the change in $\lambda$ is less than 5\% for three consecutive runs. 

As a result, we filter out 30{-}80\% of the flux from mid-IR observations with an average of 60\%, which agrees well with the diffuse emission fraction measured in \textit{Spitzer} 8\,$\mu$m observations of M33 obtained by separating discrete sources \citep{verley09}. In H$\alpha$ emission maps, we remove about 40{-}60\% of the total emission, with an average of 50\%. This is in line with \citet{belfiore22}, who find that the fraction of ionized gas emission located outside the compact \textsc{Hii} regions ranges from 20{-}60\%, with an average of 40\% using the same PHANGS-MUSE observations \citep{emsellem22}. We discuss this further in Section~\ref{ssec:fdiff}.

\begin{deluxetable*}{lccccccccccc}
\tabletypesize{\scriptsize}
\tablewidth{0pt} 
\tablecaption{Physical quantities describing the cloud evolution from mid-IR emitting gas phase to young stellar \textsc{Hii} regions. Columns are mid-IR wavelengths (Band), mid-IR emitting time-scale ($t_{\rm g}$), feedback time-scale ($t_{\rm fb}$), H$\alpha$ emitting time-scale ($t_{\rm s}$), average region separation length ($\lambda$), diffuse emission fractions of mid-Ir and H$\alpha$ maps ($f_{\rm diffuse}^{\rm mid-IR}$ and $f_{\rm diffuse}^{\rm H\alpha}$), fraction of the mid-IR emitting phase associated with H$\alpha$ emission ($t_{\rm fb}/ t_{\rm g}$), fraction of the H$\alpha$ emitting phase associated with mid-IR emission ($t_{\rm fb}/ t_{\rm s}$), difference in feedback time-scales when using mid-IR and CO as the gas tracer ($\Delta t_{\rm fb, CO}$), and the duration of gaseous mid-IR emitting phase dark in CO ($t_{\rm CO-dark}$). For galaxies affected by blending of sources (see Section~\ref{app:robust}), we show lower and/or upper limits when possible.  \label{tab:result}}
\tablehead{
\colhead{Galaxy} & \colhead{Band} & \colhead{$t_{\rm g}$} & \colhead{$t_{\rm fb}$} & \colhead{$t_{\rm s}$}& \colhead{$\lambda$}& \colhead{$f_{\rm diffuse}^{\rm mid-IR}$}& \colhead{$f_{\rm diffuse}^{\rm H\alpha}$} & \colhead{$t_{\rm fb}/ t_{\rm g}$}& \colhead{$t_{\rm fb}/ t_{\rm s}$} &\colhead{$\Delta t_{\rm fb, CO}$} & \colhead{$t_{\rm CO-dark}$}\\
\colhead{} & \colhead{$\mu m$} & \colhead{[Myr]} & \colhead{[Myr]} & \colhead{[Myr]}& \colhead{[pc]}& \colhead{[-]}& \colhead{[-]} & \colhead{[-]}& \colhead{[-]} &\colhead{[Myr]} & \colhead{[Myr]}\\
}
\startdata 
NGC0628&$7.7$&$25.7_{-1.6}^{+3.4}$&$6.0_{-0.7}^{+1.0}$&$7.4_{-0.5}^{+0.5}$&$67_{-8}^{+7}$&$0.60_{-0.03}^{+0.03}$&$0.50_{-0.03}^{+0.04}$&$0.23_{-0.05}^{+0.05}$&$0.80_{-0.14}^{+0.16}$&$2.8_{-0.9}^{+1.1}$&$-1.1_{-4.2}^{+2.9}$\\
&$10$&$25.6_{-3.3}^{+2.1}$&$6.8_{-1.1}^{+0.5}$&$7.4_{-0.6}^{+0.4}$&$79_{-9}^{+9}$&$0.59_{-0.02}^{+0.03}$&$0.45_{-0.03}^{+0.04}$&$0.27_{-0.06}^{+0.05}$&$0.92_{-0.19}^{+0.14}$&$3.7_{-1.3}^{+0.7}$&$-2.1_{-3.2}^{+4.1}$\\
&$11.3$&$34.6_{-4.2}^{+2.5}$&$7.2_{-1.3}^{+0.3}$&$7.4_{-0.5}^{+0.4}$&$70_{-7}^{+6}$&$0.57_{-0.03}^{+0.03}$&$0.47_{-0.03}^{+0.04}$&$0.21_{-0.05}^{+0.04}$&$0.97_{-0.21}^{+0.13}$&$4.1_{-1.4}^{+0.6}$&$6.5_{-3.5}^{+4.8}$\\
\hline
NGC1087&$7.7$&$21.5_{-3.4}^{+2.3}$&$6.6_{-1.1}^{+1.1}$&$8.2_{-1.3}^{+0.7}$&$125_{-17}^{+23}$&$0.66_{-0.02}^{+0.03}$&$0.61_{-0.03}^{+0.03}$&$0.31_{-0.08}^{+0.09}$&$0.80_{-0.19}^{+0.22}$&$2.7_{-2.5}^{+1.7}$&$-0.9_{-4.3}^{+7.1}$\\
&$10$&$16.4_{-3.2}^{+1.7}$&$5.9_{-1.0}^{+1.1}$&$8.2_{-1.4}^{+0.9}$&$135_{-19}^{+36}$&$0.67_{-0.03}^{+0.02}$&$0.59_{-0.04}^{+0.03}$&$0.36_{-0.10}^{+0.12}$&$0.73_{-0.21}^{+0.23}$&$2.1_{-2.5}^{+1.7}$&$-5.4_{-4.0}^{+7.0}$\\
&$11.3$&$22.4_{-3.8}^{+1.9}$&$6.5_{-1.1}^{+1.0}$&$8.2_{-1.3}^{+0.6}$&$129_{-14}^{+21}$&$0.67_{-0.03}^{+0.03}$&$0.61_{-0.03}^{+0.03}$&$0.29_{-0.07}^{+0.08}$&$0.79_{-0.19}^{+0.21}$&$2.6_{-2.5}^{+1.7}$&$0.2_{-4.1}^{+7.3}$\\
\hline
NGC1300&$7.7$&$19.7_{-1.7}^{+1.8}$&$<7.4$&$<8.3$&$<150$&$0.66_{-0.02}^{+0.04}$&$0.54_{-0.03}^{+0.04}$&$<0.41$&$-$&$<4.2$&$-2.5$ - $8.4$\\
&$10$&$22.3_{-1.7}^{+2.2}$&$6.5_{-0.8}^{+0.9}$&$7.8_{-0.6}^{+0.5}$&$144_{-16}^{+16}$&$0.63_{-0.02}^{+0.03}$&$0.51_{-0.02}^{+0.04}$&$0.29_{-0.06}^{+0.06}$&$0.83_{-0.15}^{+0.16}$&$2.9_{-1.1}^{+1.2}$&$2.8_{-3.3}^{+2.5}$\\
&$11.3$&$21.3_{-1.6}^{+2.2}$&$6.6_{-0.7}^{+0.9}$&$7.8_{-0.6}^{+0.5}$&$135_{-19}^{+17}$&$0.66_{-0.02}^{+0.03}$&$0.53_{-0.02}^{+0.04}$&$0.31_{-0.06}^{+0.06}$&$0.85_{-0.15}^{+0.16}$&$3.1_{-1.0}^{+1.2}$&$1.6_{-3.3}^{+2.5}$\\
\hline
NGC1365&$7.7$&$18.2_{-2.7}^{+4.1}$&$<6.9$&$<9.0$&$<237$&$0.79_{-0.03}^{+0.03}$&$0.65_{-0.04}^{+0.04}$&$<0.45$&$-$&$<3.4$&$-8.8$ - $4.9$\\
&$10$&$19.3_{-2.5}^{+4.0}$&$<6.9$&$<9.0$&$<234$&$0.81_{-0.02}^{+0.02}$&$0.64_{-0.03}^{+0.03}$&$<0.41$&$-$&$<3.4$&$-7.4$ - $5.9$\\
&$11.3$&$21.4_{-2.5}^{+4.6}$&$<7.1$&$<8.9$&$<225$&$0.80_{-0.02}^{+0.02}$&$0.64_{-0.03}^{+0.03}$&$<0.38$&$-$&$<3.6$&$-5.6$ - $8.6$\\
\hline
NGC1385&$7.7$&$11.3_{-1.5}^{+1.9}$&$4.2_{-0.8}^{+0.7}$&$6.9_{-0.9}^{+1.1}$&$122_{-27}^{+34}$&$0.61_{-0.05}^{+0.05}$&$0.58_{-0.06}^{+0.07}$&$0.37_{-0.12}^{+0.10}$&$0.61_{-0.19}^{+0.17}$&$1.6_{-1.7}^{+1.1}$&$-3.7_{-3.2}^{+5.3}$\\
&$10$&$11.0_{-1.7}^{+1.5}$&$4.4_{-1.1}^{+1.2}$&$6.9_{-1.0}^{+0.9}$&$124_{-23}^{+37}$&$0.58_{-0.09}^{+0.07}$&$0.57_{-0.09}^{+0.09}$&$0.40_{-0.14}^{+0.16}$&$0.63_{-0.23}^{+0.25}$&$1.7_{-1.8}^{+1.5}$&$-4.2_{-3.0}^{+5.3}$\\
&$11.3$&$13.6_{-1.7}^{+1.9}$&$4.5_{-0.9}^{+1.2}$&$6.9_{-0.9}^{+0.8}$&$123_{-16}^{+22}$&$0.62_{-0.05}^{+0.05}$&$0.58_{-0.06}^{+0.06}$&$0.33_{-0.11}^{+0.12}$&$0.65_{-0.20}^{+0.23}$&$1.9_{-1.8}^{+1.5}$&$-1.8_{-3.2}^{+5.3}$\\
\hline
NGC1433&$7.7$&$12.0_{-1.3}^{+1.4}$&$4.8_{-0.5}^{+0.6}$&$6.3_{-0.5}^{+0.4}$&$139_{-26}^{+35}$&$0.76_{-0.02}^{+0.02}$&$0.52_{-0.02}^{+0.03}$&$0.40_{-0.08}^{+0.08}$&$0.76_{-0.12}^{+0.14}$&$2.7_{-0.6}^{+0.7}$&$-5.3_{-2.2}^{+2.1}$\\
&$10$&$21.7_{-2.7}^{+1.2}$&$6.2_{-1.0}^{+0.2}$&$6.3_{-0.4}^{+0.3}$&$141_{-16}^{+19}$&$0.61_{-0.02}^{+0.03}$&$0.52_{-0.02}^{+0.03}$&$0.28_{-0.06}^{+0.05}$&$0.97_{-0.19}^{+0.11}$&$4.0_{-1.1}^{+0.5}$&$3.1_{-2.1}^{+3.1}$\\
&$11.3$&$19.1_{-1.6}^{+1.5}$&$5.8_{-0.8}^{+0.4}$&$6.3_{-0.4}^{+0.3}$&$140_{-18}^{+13}$&$0.71_{-0.01}^{+0.02}$&$0.51_{-0.01}^{+0.02}$&$0.31_{-0.06}^{+0.05}$&$0.92_{-0.16}^{+0.12}$&$3.7_{-0.9}^{+0.6}$&$0.8_{-2.3}^{+2.3}$\\
\hline
NGC1512&$7.7$&$11.7_{-1.2}^{+1.8}$&$<5.5$&$<6.4$&$<238$&$0.79_{-0.02}^{+0.02}$&$0.56_{-0.03}^{+0.03}$&$<0.52$&$-$&$<3.8$&$-5.1$ - $3.3$\\
&$10$&$14.8_{-1.3}^{+1.7}$&$<5.4$&$<6.3$&$<225$&$0.71_{-0.02}^{+0.02}$&$0.53_{-0.02}^{+0.03}$&$<0.40$&$-$&$<3.8$&$-2.2$ - $6.3$\\
&$11.3$&$16.9_{-2.2}^{+1.2}$&$<5.8$&$<6.3$&$<209$&$0.74_{-0.02}^{+0.02}$&$0.54_{-0.02}^{+0.03}$&$<0.40$&$-$&$<4.3$&$-1.3$ - $8.0$\\
\hline
NGC1566&$7.7$&$17.5_{-1.7}^{+2.3}$&$6.0_{-0.7}^{+0.9}$&$9.0_{-0.9}^{+0.9}$&$111_{-16}^{+22}$&$0.62_{-0.03}^{+0.03}$&$0.56_{-0.04}^{+0.03}$&$0.34_{-0.08}^{+0.08}$&$0.67_{-0.14}^{+0.15}$&$1.3_{-1.4}^{+1.4}$&$-7.6_{-4.0}^{+3.9}$\\
&$10$&$16.8_{-1.8}^{+1.6}$&$6.0_{-0.9}^{+0.9}$&$9.0_{-0.8}^{+0.9}$&$120_{-19}^{+24}$&$0.69_{-0.03}^{+0.03}$&$0.55_{-0.06}^{+0.04}$&$0.36_{-0.08}^{+0.09}$&$0.68_{-0.16}^{+0.16}$&$1.3_{-1.5}^{+1.4}$&$-8.4_{-3.6}^{+3.9}$\\
&$11.3$&$25.0_{-2.2}^{+2.3}$&$7.0_{-1.1}^{+1.2}$&$9.0_{-0.7}^{+0.7}$&$112_{-15}^{+21}$&$0.63_{-0.04}^{+0.03}$&$0.56_{-0.05}^{+0.04}$&$0.28_{-0.07}^{+0.07}$&$0.78_{-0.17}^{+0.18}$&$2.3_{-1.6}^{+1.6}$&$-1.1_{-4.0}^{+4.1}$\\
\hline
NGC1672&$7.7$&$19.0_{-3.3}^{+2.2}$&$7.0_{-0.9}^{+1.1}$&$8.7_{-1.0}^{+0.9}$&$194_{-30}^{+104}$&$0.53_{-0.10}^{+0.06}$&$0.50_{-0.08}^{+0.05}$&$0.37_{-0.08}^{+0.10}$&$0.80_{-0.17}^{+0.19}$&$2.5_{-1.7}^{+1.9}$&$-6.8_{-5.3}^{+5.3}$\\
&$10$&$21.0_{-3.3}^{+3.0}$&$7.6_{-1.2}^{+1.1}$&$8.7_{-1.1}^{+0.9}$&$194_{-26}^{+70}$&$0.59_{-0.06}^{+0.05}$&$0.50_{-0.06}^{+0.05}$&$0.36_{-0.10}^{+0.10}$&$0.87_{-0.23}^{+0.23}$&$3.1_{-1.9}^{+1.9}$&$-5.5_{-5.7}^{+5.3}$\\
&$11.3$&$25.6_{-4.0}^{+3.0}$&$8.0_{-1.3}^{+0.7}$&$8.7_{-0.9}^{+0.8}$&$192_{-26}^{+61}$&$0.57_{-0.07}^{+0.05}$&$0.50_{-0.06}^{+0.05}$&$0.31_{-0.08}^{+0.08}$&$0.91_{-0.22}^{+0.19}$&$3.5_{-1.9}^{+1.7}$&$-1.3_{-5.7}^{+5.8}$\\
\hline
NGC2835&$7.7$&$15.7_{-2.1}^{+2.1}$&$<4.9$&$<6.0$&$<148$&$0.64_{-0.04}^{+0.03}$&$0.57_{-0.04}^{+0.04}$&$<0.36$&$-$&$<3.8$&$1.6$ - $10.8$\\
&$10$&$10.3_{-1.5}^{+1.2}$&$3.9_{-0.5}^{+0.6}$&$5.6_{-0.5}^{+0.5}$&$177_{-40}^{+74}$&$0.61_{-0.04}^{+0.03}$&$0.51_{-0.05}^{+0.04}$&$0.38_{-0.09}^{+0.10}$&$0.69_{-0.14}^{+0.16}$&$2.6_{-0.7}^{+0.7}$&$-0.7_{-1.8}^{+2.1}$\\
&$11.3$&$19.1_{-2.7}^{+1.8}$&$4.4_{-0.6}^{+0.6}$&$5.6_{-0.5}^{+0.4}$&$120_{-20}^{+39}$&$0.58_{-0.04}^{+0.03}$&$0.57_{-0.04}^{+0.04}$&$0.23_{-0.05}^{+0.06}$&$0.79_{-0.15}^{+0.17}$&$3.1_{-0.8}^{+0.8}$&$7.6_{-2.2}^{+3.1}$\\
\hline
NGC3351&$7.7$&$17.7_{-1.8}^{+1.7}$&$5.0_{-0.6}^{+0.8}$&$6.8_{-0.6}^{+0.6}$&$94_{-12}^{+16}$&$0.74_{-0.02}^{+0.02}$&$0.57_{-0.02}^{+0.02}$&$0.28_{-0.06}^{+0.07}$&$0.74_{-0.14}^{+0.17}$&$2.5_{-1.1}^{+1.0}$&$-7.5_{-3.2}^{+5.3}$\\
&$10$&$20.6_{-1.7}^{+2.3}$&$4.7_{-0.6}^{+1.0}$&$6.8_{-0.6}^{+0.6}$&$101_{-8}^{+10}$&$0.72_{-0.01}^{+0.01}$&$0.55_{-0.02}^{+0.02}$&$0.23_{-0.05}^{+0.06}$&$0.69_{-0.15}^{+0.19}$&$2.2_{-1.1}^{+1.1}$&$-4.3_{-3.5}^{+5.2}$\\
&$11.3$&$24.3_{-1.6}^{+2.7}$&$5.5_{-0.6}^{+0.8}$&$6.8_{-0.5}^{+0.4}$&$98_{-10}^{+9}$&$0.70_{-0.01}^{+0.02}$&$0.56_{-0.01}^{+0.02}$&$0.23_{-0.05}^{+0.05}$&$0.81_{-0.14}^{+0.16}$&$3.0_{-1.1}^{+1.0}$&$-1.4_{-3.8}^{+5.2}$\\
\hline
NGC3627&$7.7$&$<14.7$&$<4.5$&$<7.0$&$121_{-23}^{+32}$&$0.63_{-0.05}^{+0.04}$&$0.60_{-0.04}^{+0.04}$&$-$&$-$&$-$&$-$\\
&$10$&$<11.8$&$<4.6$&$<7.0$&$150_{-46}^{+66}$&$0.64_{-0.05}^{+0.06}$&$0.55_{-0.06}^{+0.07}$&$-$&$-$&$-$&$-$\\
&$11.3$&$<15.3$&$<4.7$&$<7.0$&$133_{-31}^{+44}$&$0.68_{-0.05}^{+0.04}$&$0.59_{-0.05}^{+0.05}$&$-$&$-$&$-$&$-$\\
\hline
NGC4254&$7.7$&$17.1_{-1.3}^{+1.8}$&$6.5_{-0.6}^{+0.6}$&$9.0_{-0.7}^{+0.9}$&$102_{-18}^{+16}$&$0.62_{-0.02}^{+0.03}$&$0.58_{-0.02}^{+0.04}$&$0.38_{-0.07}^{+0.06}$&$0.72_{-0.12}^{+0.11}$&$1.8_{-1.4}^{+1.1}$&$-2.4_{-2.6}^{+3.2}$\\
&$10$&$17.4_{-1.2}^{+1.4}$&$6.4_{-0.6}^{+0.7}$&$9.0_{-0.6}^{+0.7}$&$102_{-16}^{+16}$&$0.68_{-0.02}^{+0.03}$&$0.57_{-0.03}^{+0.04}$&$0.37_{-0.06}^{+0.06}$&$0.72_{-0.11}^{+0.12}$&$1.7_{-1.4}^{+1.2}$&$-2.0_{-2.4}^{+3.2}$\\
&$11.3$&$19.5_{-1.2}^{+1.9}$&$6.8_{-0.5}^{+0.7}$&$9.0_{-0.6}^{+0.7}$&$104_{-16}^{+19}$&$0.66_{-0.02}^{+0.02}$&$0.57_{-0.03}^{+0.03}$&$0.35_{-0.05}^{+0.05}$&$0.75_{-0.11}^{+0.12}$&$2.0_{-1.3}^{+1.2}$&$-0.2_{-2.7}^{+3.2}$\\
\hline
NGC4303&$7.7$&$21.3_{-1.8}^{+1.8}$&$6.8_{-0.8}^{+0.9}$&$8.3_{-0.8}^{+0.6}$&$112_{-14}^{+18}$&$0.60_{-0.03}^{+0.04}$&$0.55_{-0.04}^{+0.04}$&$0.32_{-0.06}^{+0.06}$&$0.82_{-0.16}^{+0.17}$&$2.8_{-1.9}^{+1.5}$&$-2.2_{-3.7}^{+5.0}$\\
&$10$&$21.0_{-1.7}^{+1.8}$&$6.6_{-0.9}^{+1.1}$&$8.3_{-0.7}^{+0.6}$&$119_{-13}^{+16}$&$0.61_{-0.03}^{+0.04}$&$0.52_{-0.04}^{+0.05}$&$0.31_{-0.07}^{+0.07}$&$0.79_{-0.17}^{+0.18}$&$2.5_{-2.0}^{+1.6}$&$-2.3_{-3.7}^{+5.0}$\\
&$11.3$&$28.3_{-2.1}^{+1.8}$&$7.1_{-0.9}^{+0.8}$&$8.3_{-0.6}^{+0.5}$&$111_{-12}^{+15}$&$0.61_{-0.03}^{+0.03}$&$0.56_{-0.03}^{+0.04}$&$0.25_{-0.05}^{+0.04}$&$0.86_{-0.15}^{+0.15}$&$3.1_{-2.0}^{+1.4}$&$4.6_{-3.6}^{+5.1}$\\
\hline
NGC4321&$7.7$&$12.2_{-1.1}^{+1.2}$&$<6.2$&$<8.0$&$<156$&$0.67_{-0.03}^{+0.03}$&$0.55_{-0.03}^{+0.04}$&$<0.56$&$-$&$<3.2$&$-11.9$ - $-3.4$\\
&$10$&$15.4_{-1.1}^{+1.5}$&$<6.8$&$<7.9$&$<151$&$0.72_{-0.02}^{+0.02}$&$0.53_{-0.03}^{+0.04}$&$<0.47$&$-$&$<3.8$&$-9.3$ - $0.1$\\
&$11.3$&$18.4_{-1.5}^{+1.3}$&$<7.1$&$<7.7$&$<150$&$0.67_{-0.02}^{+0.02}$&$0.53_{-0.02}^{+0.03}$&$<0.42$&$-$&$<4.2$&$-7.0$ - $2.9$\\
\hline
NGC4535&$7.7$&$18.2_{-2.3}^{+2.5}$&$6.2_{-1.2}^{+1.4}$&$8.9_{-1.1}^{+1.1}$&$104_{-12}^{+12}$&$0.54_{-0.05}^{+0.07}$&$0.49_{-0.04}^{+0.09}$&$0.34_{-0.10}^{+0.11}$&$0.70_{-0.21}^{+0.23}$&$1.5_{-2.5}^{+1.8}$&$-7.8_{-4.4}^{+8.8}$\\
&$10$&$17.6_{-2.1}^{+3.0}$&$6.6_{-1.4}^{+1.4}$&$8.9_{-1.2}^{+0.9}$&$137_{-19}^{+17}$&$0.49_{-0.11}^{+0.08}$&$0.43_{-0.12}^{+0.11}$&$0.38_{-0.13}^{+0.12}$&$0.75_{-0.22}^{+0.24}$&$2.0_{-2.6}^{+1.8}$&$-8.9_{-4.7}^{+8.7}$\\
&$11.3$&$27.0_{-3.1}^{+2.5}$&$6.7_{-1.0}^{+1.2}$&$8.9_{-0.9}^{+0.8}$&$108_{-10}^{+12}$&$0.52_{-0.05}^{+0.06}$&$0.48_{-0.04}^{+0.07}$&$0.25_{-0.06}^{+0.06}$&$0.75_{-0.17}^{+0.19}$&$2.1_{-2.4}^{+1.6}$&$0.5_{-4.4}^{+9.0}$\\
\hline
NGC5068&$7.7$&$27.9_{-3.7}^{+3.8}$&$3.8_{-0.4}^{+0.7}$&$5.5_{-0.4}^{+0.3}$&$57_{-7}^{+25}$&$0.32_{-0.14}^{+0.08}$&$0.44_{-0.09}^{+0.05}$&$0.14_{-0.03}^{+0.04}$&$0.69_{-0.11}^{+0.16}$&$2.7_{-0.5}^{+0.8}$&$13.7_{-4.2}^{+4.4}$\\
&$10$&$21.8_{-2.4}^{+4.2}$&$3.5_{-0.5}^{+0.8}$&$5.5_{-0.4}^{+0.3}$&$66_{-10}^{+16}$&$0.34_{-0.12}^{+0.08}$&$0.42_{-0.08}^{+0.05}$&$0.16_{-0.05}^{+0.05}$&$0.64_{-0.13}^{+0.17}$&$2.5_{-0.6}^{+0.9}$&$7.9_{-4.6}^{+3.3}$\\
&$11.3$&$32.2_{-3.4}^{+5.4}$&$4.2_{-0.5}^{+0.9}$&$5.5_{-0.4}^{+0.3}$&$60_{-8}^{+10}$&$0.28_{-0.11}^{+0.09}$&$0.36_{-0.07}^{+0.06}$&$0.13_{-0.04}^{+0.04}$&$0.77_{-0.14}^{+0.19}$&$3.2_{-0.6}^{+0.9}$&$17.7_{-5.7}^{+4.1}$\\
\hline
\enddata
\end{deluxetable*}

\begin{figure*}
\includegraphics[scale=0.73]{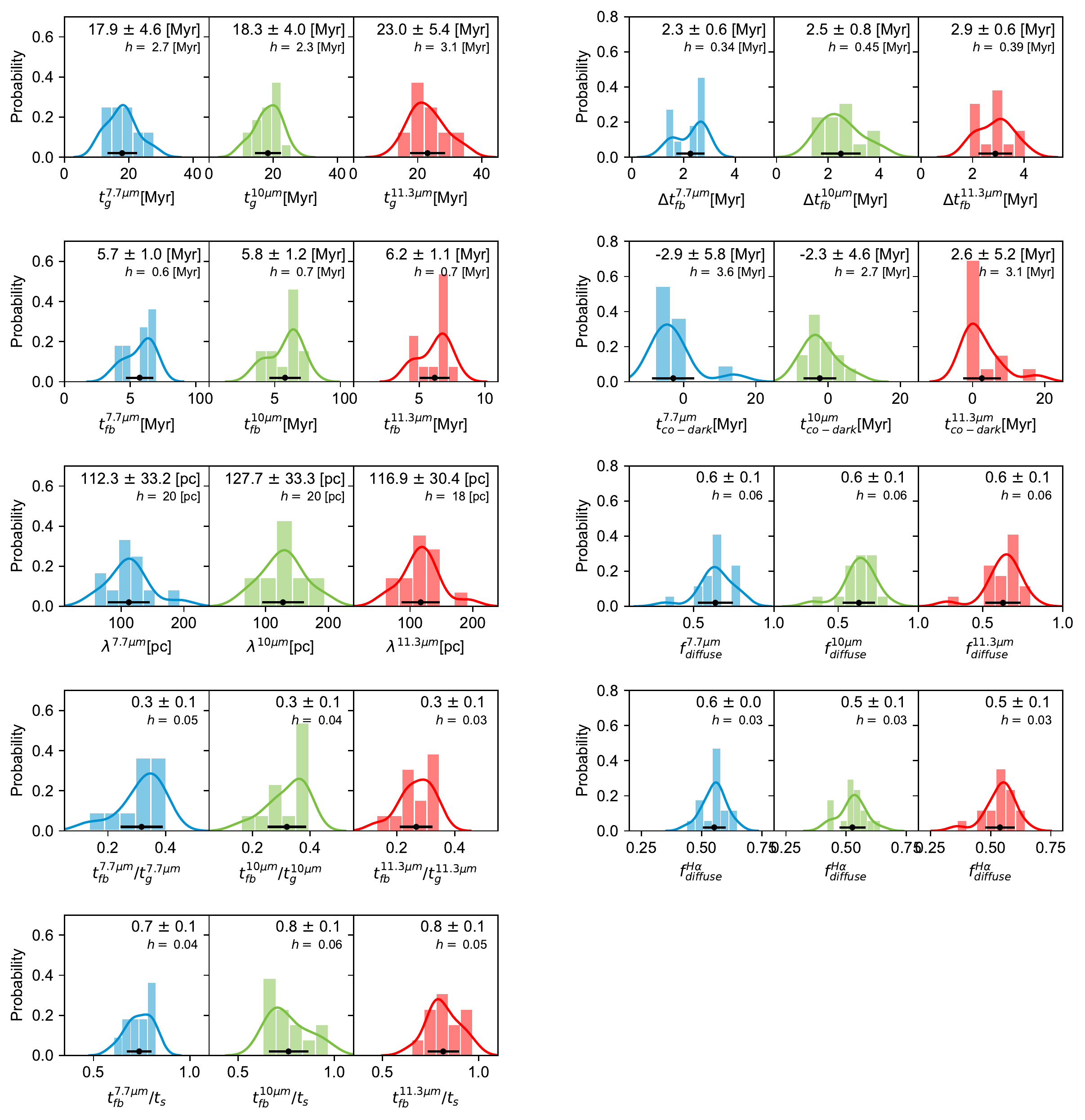}
\caption{Distributions of physical quantities listed in Table~\ref{tab:result}, which are from the application of the method described in Section~\ref{sec:method} to mid-IR and H$\alpha$ observations of 17 galaxies. For each physical quantity, results with 7.7, 10, and 11.3\,$\mu$mare shown in blue, green, and red, respectively. The solid line shows the smoothed distribution based on a Gaussian kernel density estimate, with the bandwidth ($h$; shown in each panel) selected according to Scott’s rule \citep{scott}. Table~\ref{tab:result} lists actual values and galaxies with only upper or lower limit constraints are excluded (see Section~\ref{app:robust}). In each histogram, the mean (black dot) and 16{-}84\% range (horizontal line) are shown, as well as in the upper right corner. } \label{fig:histo}
\end{figure*}

\begin{figure*}
\centering
\includegraphics[scale=0.75]{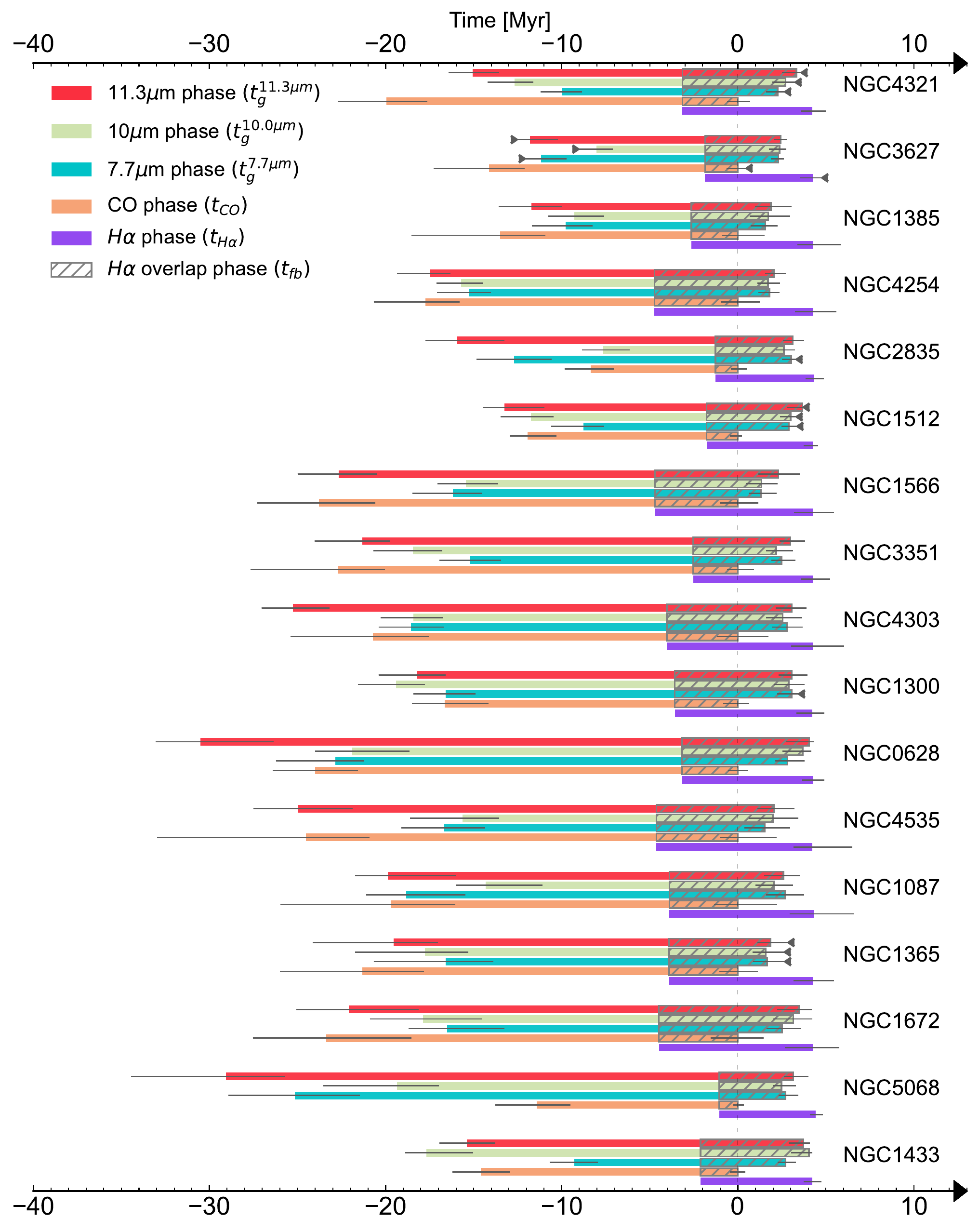}
\caption{The multi-tracer timeline of cloud evolution in PHANGS-JWST galaxies. From left to right, clouds are initially inert, being detected only in CO and mid-IR, tracing the cold gaseous phase. Star formation then takes place, causing gas tracers to become coincident with H$\alpha$ emission (dashed region). In most galaxies, mid-IR bands (7.7, 10, and 11.3~$\mu$m) fully cover the molecular gas phase traced with CO. The dashed line at Time $= 0$\,Myr indicates the moment when the CO emission disperses. The mid-IR emission persists for a longer period after the CO emission has been dispersed by stellar feedback (dashed line) and remains detectable for 2-3\,Myr, on average, covering a significant fraction of the H$\alpha$ emitting phase. Galaxies are ordered by the increasing H$\alpha$ flux density contrast measured between the H$\alpha$ peak and galactic average ( $\epsilon_{\rm H\alpha}$), which  shows a strong correlation with 10\,$\mu$m emitting time-scales (see Figure~\ref{fig:hm} and Section~\ref{sssec:corr_meaning}). Galaxies with only upper limit constraints are indicated with left or right-pointing triangles.} \label{fig:timeline}
\end{figure*}

\section{Timeline of cloud evolution from mid-IR emitting phase to \textsc{Hii} regions}\label{sec:result}
In this section, we present results from applying the method described in Section~\ref{sec:method} to mid-IR (F770W, F1000W, and F1130W) and H$\alpha$ observations across 17 galaxies in our sample. 

\subsection{Relative spatial distribution into evolutionary time-scales}\label{ssec:tuningfork}

Figure~\ref{fig:tuningforks} shows the measured deviation of mid-IR-to-H$\alpha$ tracer flux ratios compared to the galactic average value at different spatial scales ranging from $\sim 100$\,pc to 3\,kpc. We also show results using CO as the gas tracer for comparison from \citet{kim22}. Our best-fit model is shown as a dashed line, allowing us to derive time-scales associated with different phases of the star formation process. In all of our measurements using different gas tracers (CO and three mid-IR bands), we find that the measured flux ratio deviates from the galactic average value on small scales. A higher (lower) mid-IR-to-H$\alpha$ flux ratio than the galactic average is measured when apertures are placed on mid-IR (H$\alpha$) peaks on small scales, resulting in a relative deviation at small-scale higher (lower) than 1, demonstrating a spatial offset between mid-IR and H$\alpha$ emission. However, we find that the degree of divergence is much more significant when CO is used as the gas tracer compared to when mid-IR bands are used. This indicates that H$\alpha$ spatially correlates better with compact mid-IR than CO, which is consistent with other empirical studies \citep{chown21, gao22, leroy23, whitcomb23}. 

In Table~\ref{tab:result}, we list our best-fit parameters: the mid-IR emitting phase ($t_{\rm g}$; see Section~\ref{ssec:tg}), the feedback phase ($t_{\rm fb}$; see Section~\ref{ssec:tfb}), and the average separation length between independent star-forming regions ($\lambda$; see Section~\ref{ssec:lambda}). We also present other derived quantities, such as the duration of the mid-IR emitting phase without associated CO and H$\alpha$ emission (CO-dark mid-IR emitting phase, $t_{\rm CO-dark}^{\rm mid-IR}$; see Section~\ref{ssec:tg}), the fraction of the mid-IR emitting phase associated with the SFR tracer, the fraction of the SFR tracer emitting phase associated with the mid-IR ($t_{\rm fb}/t_{\rm g}$ and $t_{\rm fb}/t_{\rm s}$; see Section~\ref{ssec:tfb}), the difference in feedback time-scales when using mid-IR versus CO ($\Delta t_{\rm fb, CO}$; see Section~\ref{ssec:tfb}), and the diffuse emission fractions measured in mid-IR and SFR tracer maps ($f_{\rm diffuse}^{\rm mid-IR}$ and $f_{\rm diffuse}^{\rm H\alpha}$; see Section~\ref{ssec:fdiff}). Figure~\ref{fig:histo} shows the distributions of these physical quantities, highlighting significant galaxy-to-galaxy variation. A more detailed discussion is provided in the subsequent sections.

Figure~\ref{fig:timeline} shows the multi-tracer evolutionary timeline of regions transitioning from gas to stars, obtained by combining the mid-IR emitting phase measured in this work with GMC evolutionary timeline from \cite{kim22}. Regions are initially inert and detected only in CO and mid-IR tracing the gaseous cloud phase. Later H$\alpha$ becomes visible as the newly formed stars partially emerge from the parental cloud. Finally, the cloud disperses, allowing H$\alpha$ to be detected alone without associated CO and mid-IR emission. For most galaxies, the mid-IR emitting phase encompasses the CO-emitting phase and continues to be detected after the CO has disappeared, covering a significant fraction of the H$\alpha$-emitting phase.

\subsection{Mid-IR emitting time-scale}\label{ssec:tg}
\begin{figure*}
\includegraphics[scale=0.48]{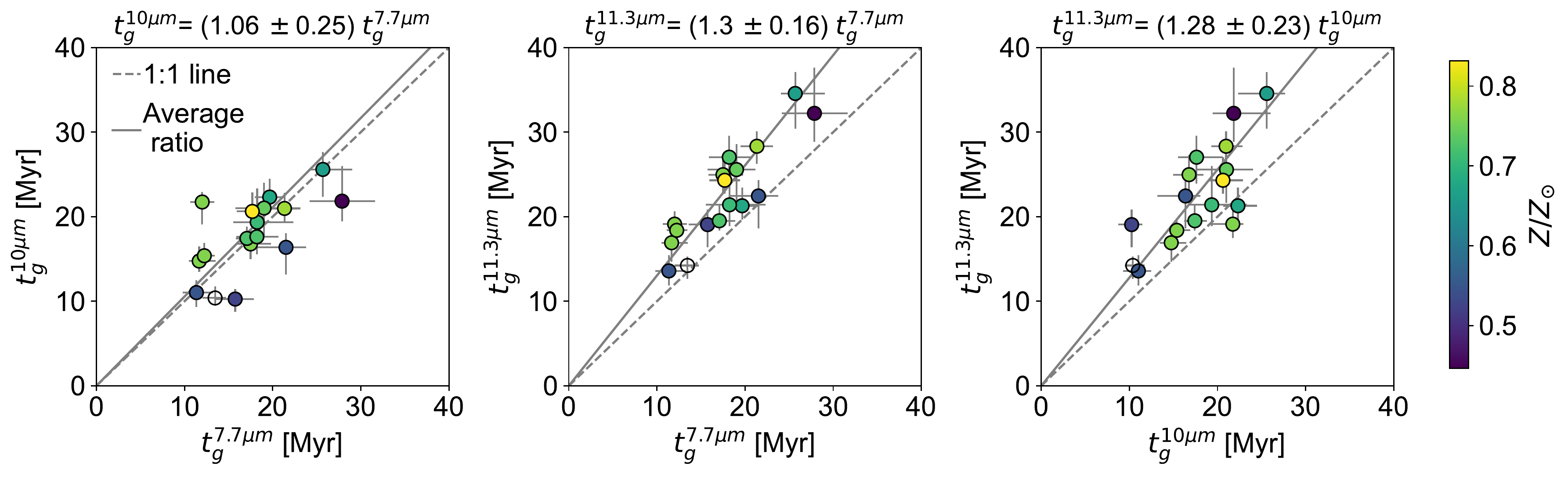}
\caption{Comparisons of mid-IR emitting time-scales obtained using different bands ($t_{\rm g}^{\rm 7.7,\mu m}$, $t_{\rm g}^{\rm 10,\mu m}$, and $t_{\rm g}^{\rm 11.3,\mu m}$). The average ratio between two time-scales and the 1$\sigma$ distributions are indicated in each panel. The solid line indicates the relation using the average ratio, whereas the dashed line shows a one-to-one correlation. Data points are color-coded by metallicity relative to the solar value ($Z/Z_{\odot}$), whereas the empty circle indicates NGC\,3627 where our measurements are upper limits. While the three mid-IR emitting time-scale generally agree regardless of the wavelength, the left panel shows that $t_{\rm g}^{\rm 7.7,\mu m}$ is longer than $t_{\rm g}^{\rm 10,\mu m}$ for metal-poor galaxies. Additionally, $t_{\rm g}^{\rm 11.3,\mu m}$ is slightly longer than both $t_{\rm g}^{\rm 7.7,\mu m}$ and $t_{\rm g}^{\rm 10,\mu m}$.} \label{fig:tgs}
\end{figure*}

Across our sample of galaxies, the mid-IR emitting time-scale of gas clouds in the galactic disk is measured to range $t_{\rm g}^{\rm 7.7\,\mu m}=11-28$\,Myr, $t_{\rm g}^{\rm 10\,\mu m}=10-26$\,Myr, and $t_{\rm g}^{\rm 11.3\,\mu m}=14-35$\,Myr when 7.7, 10, and 11.3\,$\mu m$ are used, respectively. The average and $16{-}84$\% range for each mid-IR band are $18\pm5$\,Myr, $18\pm4$\,Myr, and $23\pm5$\,Myr, for 7.7\,$\mu$m, 10\,$\mu$m, and 11.3\,$\mu$m, respectively. Figure~\ref{fig:tgs} compares mid-IR emitting time-scales across different mid-IR bands and shows that $t_{\rm g}^{7.7\,\mu m}$ are longer than $t_{\rm g}^{10\,\mu m}$ for relatively metal poor galaxies. The $t_{\rm g}^{\rm 11.3\,\mu m}$ is somewhat longer than the mid-IR emitting time-scale measured with the other two bands. The F1130W band, while dominated by the $11.3\,\mu m$ PAH feature, likely contains a higher level of dust continuum emission in thermal equilibrium with the radiation field in and around \textsc{Hii} regions, compared to the dust continuum in the two shorter-wavelength bands \citep{smith07, lai22}. This will increase the overlap between mid-IR and H$\alpha$, as shown in Figure~\ref{fig:histo} and Figure~\ref{fig:timeline}, and therefore the total 11.3\,$\mu$m emitting time-scale. 

As described in Section~\ref{ssec:gastracer}, the mid-IR emission traces multiple gas phases and is influenced by a combination of gas column density, radiation, and abundances of dust and PAHs. Therefore, the physical interpretation of the mid-IR emitting time-scale is not trivial. The strong correlation between mid-IR (or PAH) bands and cold gas \citep{chown21, gao22, leroy23, whitcomb23, chown24} suggests that the measured mid-IR emitting time-scale likely reflects the neutral gas phase, particularly the compact neutral gas structures \citep{sandstrom23} with our diffuse emission filtering process. However, this assumes that dust and PAHs are mainly illuminated by the diffuse interstellar radiation field, which may not hold true, especially for bright mid-IR peaks often associated with star forming regions \citep{leroy23, pathak24}. In addition, we also note that our method is flux weighted, which naturally biases our measurements toward bright mid-IR emission peaks, which could make the mid-IR emitting time-scale more related to gas near \textsc{Hii} regions and not fully capture the entire neutral gas cloud phase.

\begin{figure*}
\includegraphics[scale=0.48]{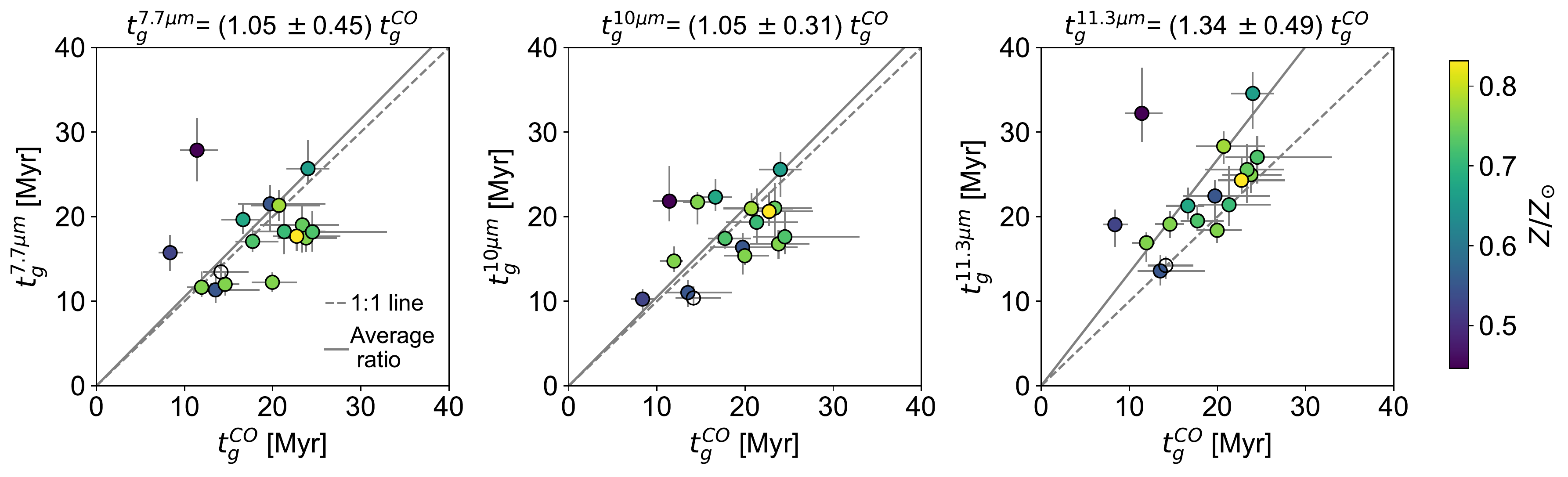}
\caption{Similar to Figure~\ref{fig:tgs}, GMC lifetimes compared to mid-IR emitting time-scales obtained using different bands ($t_{\rm g}^{\rm 7.7,\mu m}$, $t_{\rm g}^{\rm 10,\mu m}$, and $t_{\rm g}^{\rm 11.3,\mu m}$) are shown. we find that the two  time-scales agree well, suggesting that once gas clouds form being associated with compact mid-IR emission, they quickly provide enough shielding for stable CO formation. This is likely due to our focus on molecular gas-rich, central regions of galaxies with near-solar metallicity. }\label{fig:tgs_co}
\end{figure*}

With these caveats in mind, in Figure~\ref{fig:tgs_co} we have compared our measurements of the mid-IR emitting time-scale and GMC lifetimes from \citet{kim22}. We find that the mid-IR emitting time-scale is similar to the CO emitting time-scale (molecular cloud lifetime $t_{\rm g}^{\rm CO}$ from \citealp{kim22}) with averages and 1$\sigma$ distributions of their ratio ($t_{\rm g}^{\rm mid-IR}/t_{\rm g}^{\rm CO}$) being  $1.1\pm0.4$, $1.1\pm0.3$, and $1.3\pm0.5$, for 7.7\,$\mu$m, 10\,$\mu$m, and 11.3\,$\mu$m, respectively. Compared to the characteristic times-cales of GMCs \citep[][A. Hughes in prep.]{rosolowsky21, sun22}, the mid-IR emitting time-scale is, on average, 2-3 times longer than the average GMC free-fall time-scale and GMC turbulence crossing time-scales (see Section~\ref{ssec:tcomp}). 

The similarity between the mid-IR and CO emitting time-scales might seem surprising, as Figure~\ref{fig:tuningforks} shows significant differences in the flux ratio deviations, where branches diverge more significantly when CO is used compared to when mid-IR bands are used. The top branch with mid-IR, where flux ratios are measured by focusing apertures on mid-IR peaks, is close to the galactic average value being almost flat in some galaxies even on small-scales. The main reason for the much narrower branches with mid-IR compared to CO is because of the longer overlapping time-scale between mid-IR and H$\alpha$ (see Section~\ref{ssec:tfb}) than that measured between CO and H$\alpha$, which makes the deviation of flux ratios smaller. We also note that, while the small deviation of the top branch with mid-IR might indicate that all the mid-IR peaks have associated H$\alpha$ emission, we still measure a long, isolated mid-IR emitting phase (see Figure~\ref{fig:timeline}). This is related to the flux evolution of independent star-forming regions considered within our model. We make use of the ratios of time-scales ($t_{\rm fb}/t_{\rm g}$ and $t_{\rm fb}/t_{\rm s}$) to constrain the flux ratios of mid-IR and H$\alpha$ peaks in their overlap phase relative to their isolated phases ($\beta_{\rm mid-IR}$ and $\beta_{\rm H\alpha}$, respectively; \citealp{kruijssen18}). We find that, for most of our galaxies, $\beta_{\rm H\alpha}$ is quite high (3 on average), implying that H$\alpha$ peaks are brighter in their initial phase when overlapping with mid-IR compared to when they are in an isolated H$\alpha$ emitting phase later on. This makes sense given that the overlap phase lasts 6 Myr, covering the period where H$\alpha$ emission is the brightest during the \textsc{Hii} region evolution \citep{leitherer99}. As a result, even though there are many mid-IR peaks not overlapping with H$\alpha$ (and therefore a long isolated mid-IR emitting phase; see also Figure~\ref{fig:obs}), H$\alpha$ peaks in the overlapping phase with mid-IR are much brighter and this makes the top branch appear flatter on small-scales and deviate less from the galactic average ratio. In contrast, when CO is adopted, we find $\beta_{\rm H\alpha}$ is 1 on average and a shorter $t_{\rm fb}\approx 3$\,Myr.

Before star formation is detected in H$\alpha$, we define the mid-IR emitting phase without associated CO (and H$\alpha$) emission as the CO-dark mid-IR emitting cloud phase ($t_{\rm CO-dark}^{\rm mid-IR}$ or $t_{\rm CO-dark}^{\rm X}$ for a specific mid-IR band $X$), which can be obtained as $t_{\rm CO-dark}^{X}=t_{\rm g}^{X}-t_{\rm fb}^{X}-(t_{\rm g}^{\rm CO}-t_{\rm fb}^{\rm CO})$, where $t_{\rm fb}^{\rm CO}$ is the feedback time-scale measured with CO. We find that the average and 1$\sigma$ range of $t_{\rm CO-dark}^{\rm mid-IR}$ are $-3\pm6$\,Myr, $-2\pm5$\,Myr, and $3\pm5$\,Myr, respectively, for 7.7, 10, and 11.3\,$\mu$m. In most galaxies in our sample, regardless of the mid-IR bands, the measured $t_{\rm CO-dark}^{\rm mid-IR}$ is consistent with being zero, indicating that the distributions of CO and compact mid-IR emission are similar before star formation commences. This implies that once gas clouds assemble, being associated with compact mid-IR peaks, they immediately provide sufficient shielding for CO to become stable. The conversion from neutral gas clouds (traced by compact mid-IR emission) to CO-emitting GMCs occurs very rapidly, which is likely given that our observations are focused on the molecular gas-rich central regions of galaxies with high gas surface density and near solar metallicity. In the case of lower density and diffuse ISM regime, simulations suggest that there may be a few Myr when compact gas clouds are present, but the conditions are insufficient to provide the shielding required for CO formation during the molecular gas cloud assembly \citep{clark12, smith14}. We also found exceptions (NGC\,1566 and NGC\,4321), where we measure strongly negative $t_{\rm CO-dark}^{\rm mid-IR}$. This is likely due to the breakdown of the mid-IR and CO flux correlation and is discussed in more detail in Section~\ref{ssec:copahdecorr}.

We have also compared our measurements to the neutral gas cloud lifetime measured with \textsc{Hi} in the LMC \citep{ward20_HI, ward22}. We find that our measurements of a very short $t_{\rm CO-dark}^{\rm mid-IR}$ contrast the results of the LMC, where the neutral gas cloud lifetime is $\sim$5 times longer than the molecular gas cloud lifetime. The neutral gas cloud phase traced with \textsc{Hi} but dark in CO ($t_{\rm CO-dark}^{\rm HI}$) is found to be $\approx 35$\,Myr. Using [CII] and [CI] lines in the far-infrared, \citet{pineda17} have found that $70{–}80$\% of the molecular gas in the LMC and Small Magellanic Cloud (SMC) is dark in CO.

While it is difficult to make a direct comparison as our measurements are biased towards bright mid-IR peaks which are associated with star-forming regions rather than the neutral gas distribution, we suspect that the significant difference in $t_{\rm CO-dark}^{\rm HI}$ and $t_{\rm CO-dark}^{\rm mid-IR}$, between the LMC  \citep{ward20_HI, ward22} and our galaxy sample is mainly due to the fact that our JWST observations are focused on the central part of the galaxy, where the majority of the ISM mass is in the form of molecular gas and has a high gas surface density. The average molecular gas fraction ($f_{\rm H_{2}}$) of our galaxy sample is $0.5\pm0.2$, whereas in the LMC, the fraction is quite low ($f_{\rm H_{2}}=0.09$; \citealp{schruba19}). The high molecular gas fraction enhances the stability of CO, making it a more reliable tracer of cold molecular gas \citep{clark12}. In an atomic gas dominated environment, CO is only emitted from the central region of the HI clumps, causing the time-scales between the two gas phases to be significantly different. Furthermore, our sample of galaxies has almost solar metallicity ($Z/Z_{\odot}=0.7\pm0.1)$, unlike the LMC, which has a lower metallicity of ($Z/Z_{\odot}=0.5)$. This allows CO to be photodissociated more easily in the LMC due to the lack of shielding, resulting in a large reservoir of neutral gas clouds that are dark in CO emission in a metal-poor environment \citep{bolatto13, pineda17}. Indeed, $t_{\rm CO-dark}^{\rm mid-IR}$ shows negative trends with metallicity (see Figure~\ref{fig:hm} and Figure~\ref{fig:corr_codark}). For the lowest metallicity and the lowest metallicity galaxy in our sample, NGC\,5068, we measure $t_{\rm CO-dark}^{\rm mid-IR}$ to be the longest, ranging from 10 to 20\,Myr depending on the mid-IR bands. This is discussed in more detail in Section~\ref{sssec:theory}.

\subsection{Mid-IR and H$\alpha$ overlapping time-scale} \label{ssec:tfb}
The duration for which the mid-IR emission overlaps with H$\alpha$ ranges from $3.8{-}7.0$\,Myr, $3.5{-}7.6$\,Myr, $4.2{-}8.0$\,Myr, with an average and 16{-}84\% range of $5.7\pm1.0$\,Myr, $5.8\pm1.2$\,Myr, and $6.2\pm1.1$\,Myr, for 7.7, 10, and 11.3\,$\mu$m, denoted as $t_{\rm fb}^{7.7\mu m}$, $t_{\rm fb}^{10\mu m}$, and $t_{\rm fb}^{11.3\mu m}$, respectively. The majority of the H$\alpha$ emitting phase has associated mid-IR emission, with an average ratio $t_{\rm fb}^{\rm mid-IR}/t_{\rm s}$ of 70{-}80\%. 

While there is evidence of PAH destruction within \textsc{Hii} regions \citep{chastenet19, chastenet23, egorov23, sutter24}, we are not likely resolving the ISM shells and bubbles created by young stars on scales of 100\,pc. Therefore, the significant overlap between mid-IR and H$\alpha$ emission is because \textsc{Hii} regions are major contributors to ultraviolet and optical radiations, which illuminate the PAHs and dust grains near \textsc{Hii} regions and thus allow mid-IR emission to continue during the H$\alpha$ emitting phase. Our measurements also show that mid-IR emitting phase ends before H$\alpha$, implying that the overlapping time-scale can be referred to as the feedback time-scales at scales of $\sim$100\,pc for PAHs (and dust grains) dispersal. If we were to resolve the PAH suppression at a smaller scale, the feedback time-scale would likely be shorter. The range of the overlap time-scale is similar to the ages of clusters associated with compact PAH and \textsc{Hii} region morphologies, which have an age span of 1 to 6\,Myr, with an average of 4\,Myr \citep{pedrini24}.

As shown in Figure~\ref{fig:timeline}, we find that the overlapping or feedback time-scale obtained with mid-IR is always longer than 3\,Myr and is also longer than that obtained with CO as a cold gas tracer ($t_{\rm fb}^{\rm CO}$). We measure the difference in the feedback time-scales between mid-IR bands and CO as $\Delta t_{\rm fb, CO}^{X}=t_{\rm fb}^{X}- t_{\rm fb}^{\rm CO}$, where $X$ denotes mid-IR bands. We find the difference to range $1.3{-}2.8$\,Myr, $1.3{-}4.0$\,Myr, and $1.9{-}4.1$\,Myr, for $\Delta t_{\rm fb, CO}^{7.7\mu m}$, $\Delta t_{\rm fb, CO}^{10\mu m}$, and $\Delta t_{\rm fb, CO}^{11.3\mu m}$, respectively. The longer feedback time-scale with mid-IR compared to CO is likely because PAH or dust emission is enhanced by the presence of nearby intense radiation from young stars. This leads to enhanced mid-IR emission near (but not inside) \textsc{Hii} regions, whereas CO is expected to be dispersed quickly. Assuming that our detection of emission peaks is symmetric in terms of mass traced across different tracers, our measurements imply that in and near \textsc{Hii} regions, $30{-}70\%$ of the compact mid-IR emission is dark in CO, with an average of $\sim 45\%$. This is somewhat smaller than the mass fraction of CO-dark molecular gas ($\sim75\%$) measured in a star bursting region in the LMC \citep[30~Doradus][]{chevance20lmc}, possibly due to the lower metallicity and sparser ISM density of the LMC. However, we also note that limited sensitivity of CO emission could also contribute to the differences. It is possible that when the cloud is dispersed CO emission falls below the detection limit earlier on compared to the deeper JWST observations.

\subsection{Average separation length between regions}\label{ssec:lambda}
Figure~\ref{fig:obs} demonstrates that gas and SFR tracers show distinct distributions, reflecting the fact that they trace different phases of region evolution from gas to stars. Our method constrains the average separation length between these regions ($\lambda^{\rm mid-IR}$), which is measured to range 60{-}190\,pc across our analysis with different mid-IR bands, with their average and 16{-}84\% range of $112\pm33$, $128\pm33$, and $117\pm30$\,pc for $\lambda^{\rm 7.7\mu m}$, $\lambda^{\rm 10\mu m}$, and $\lambda^{\rm 11.3\mu m}$, respectively.

\begin{figure*}
\centering
\includegraphics[scale=0.3]{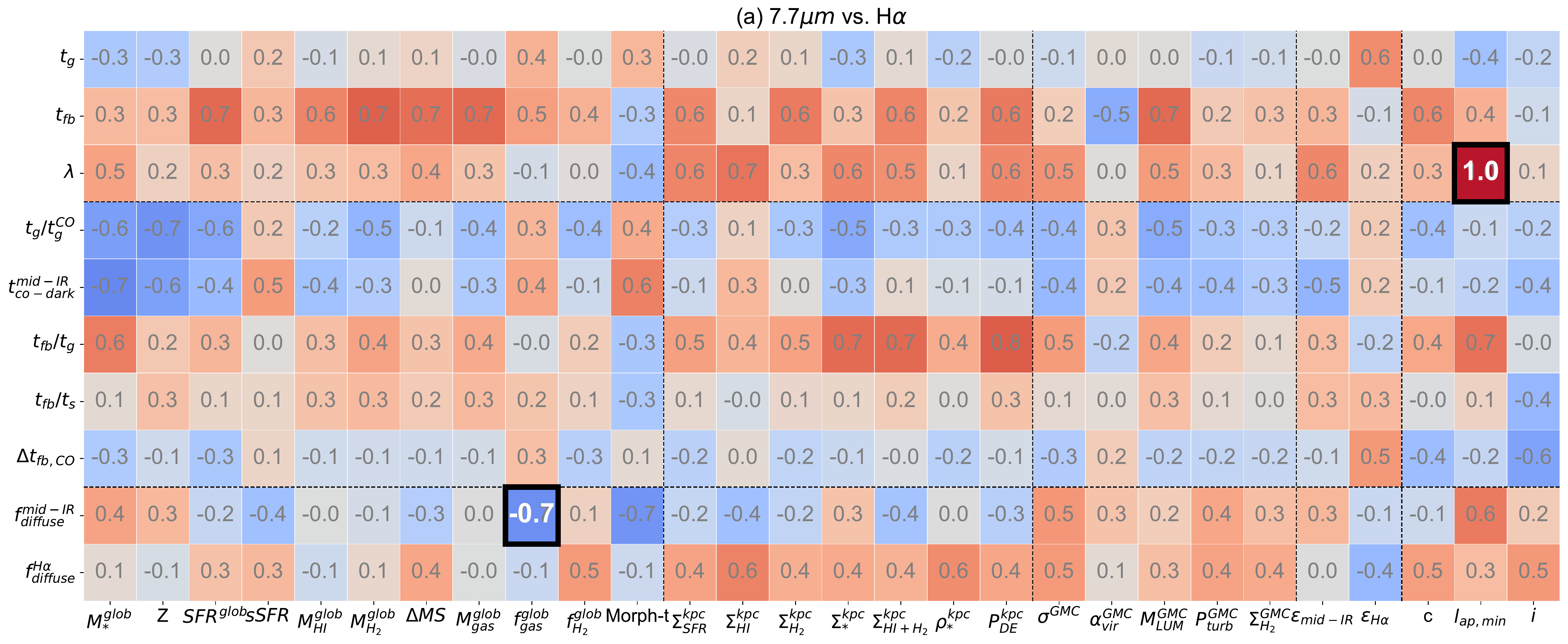}
\includegraphics[scale=0.3]{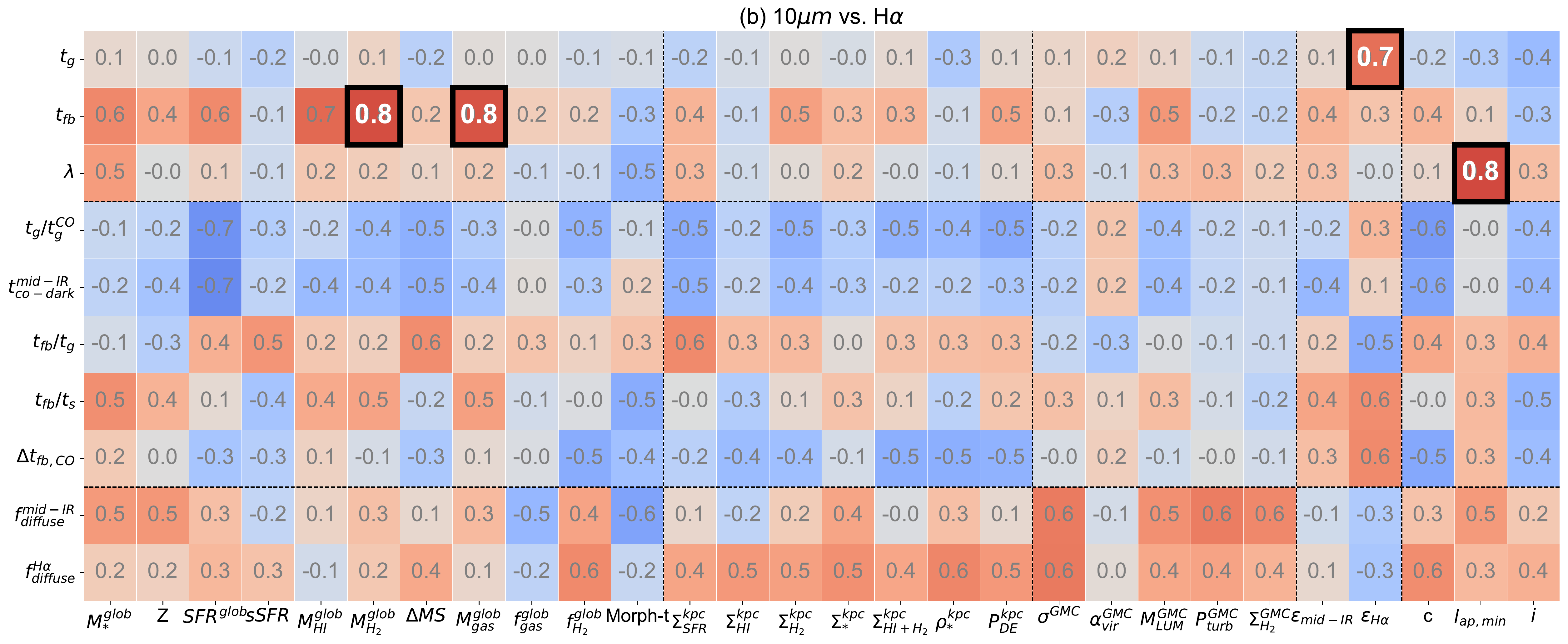}
\includegraphics[scale=0.3]{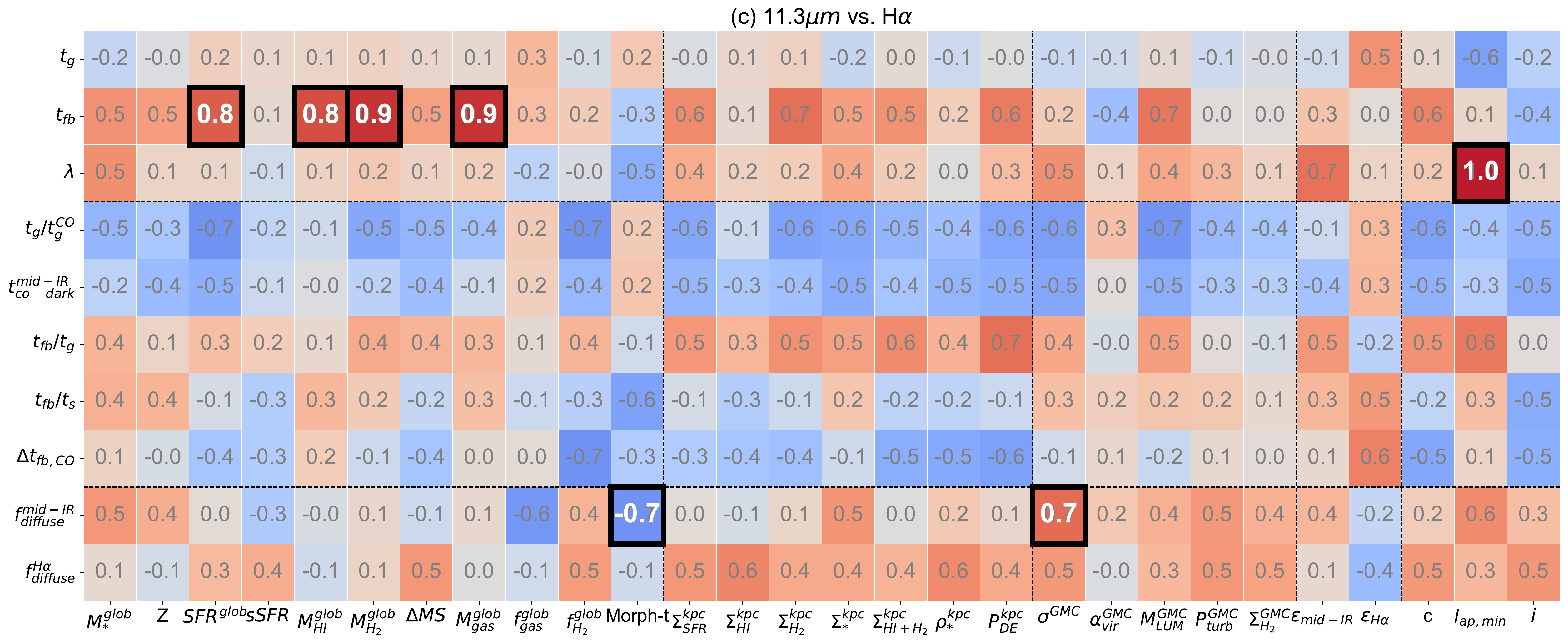}
\caption{Spearman's rank correlation coefficients measured between galaxy properties (columns; Section~\ref{ssec:env}) and our measurements (rows; Section~\ref{sec:result}). From (a) to (c), correlations obtained using different mid-IR bands (7.7, 10, and 11.3\,$\mu$m) are shown, respectively. Statistically strong correlations are highlighted as black squares, with red indicating a positive correlation and blue indicating a negative one. Our measurements are the mid-IR emitting time-scale ($t_{\rm g}$), feedback time-scale ($t_{\rm fb}$), region separation length ($\lambda$), ratios between time-scales ($t_{\rm g}/t_{\rm g}^{\rm CO}$, $t_{\rm fb}/t_{\rm g}$, $t_{\rm fb}/t_{\rm s}$), duration of the mid-IR emitting phase dark in CO and H$\alpha$ ($t_{\rm CO-dark}$), difference between the overlap time-scales measured with mid-IR versus CO ($\Delta t_{\rm fb, CO}$), and diffuse emission fractions ($f_{\rm diffuse}^{\rm mid-IR}$ and $f_{\rm diffuse}^{\rm H\alpha}$). We correlate these measurements with global galaxy properties, average kpc-scale galaxy properties, average GMC properties, average surface flux density contrasts between emission peaks and galactic average measured in mid-IR and H$\alpha$ maps, and systematic biases described in Section~\ref{ssec:env}.
} \label{fig:hm}
\end{figure*}

As indicated by the arrow in each panel of Figure~\ref{fig:tuningforks}, the measured $\lambda$ with mid-IR as the neutral gas tracer is smaller than that measured with CO as the molecular gas tracer ($\lambda^{\rm CO}=100-310$\,pc, with an average of $230\pm 60$\,pc). In \citet{kim22}, we pointed out that the $\lambda$ is the most sensitive to the resolution while other constrained parameters such as the cloud lifetime and the feedback time-scale do not show strong correlations with the resolution. This is also illustrated in Figure~\ref{fig:hm}, where $\lambda$ strongly correlates with resolution $l_{\rm ap, min}$, while the correlations with mid-IR emitting and overlapping time-scales are weak. The difference in $\lambda$ between mid-IR and CO is most likely due to the difference in resolution and sensitivity between the two observations. The minimum aperture sizes in Figure~\ref{fig:tuningforks} indicate the resolutions adopted in runs with different gas tracers. It shows that CO observations have the coarsest resolution, possibly leading to a larger region separation length.

\subsection{Diffuse emission fractions}\label{ssec:fdiff}
As our method is based on the measurements of the deviation of the local flux ratio compared to the global flux ratio, it is important that flux ratios are not biased by the existence of large-scale diffuse emission that is not associated with identified emission peaks. We therefore filter out large-scale diffuse emission with a scale larger than $n_{\lambda}\lambda\approx 1.5$\,kpc using a Gaussian high-pass filter in Fourier space (see Section~\ref{sec:method} for the justification of $n_{\lambda}$). The adopted filtering-scale is similar to the size of the most extended coherent, primarily atomic \textsc{HI} filament identified in the Milky Way (the ``Maggie'' Filament with a total length of 1.2\,kpc; \citealp{syed22}), confirming that our chosen scale effectively removes large-scale diffuse emission while preserving emission from compact structures. This allows us to measure diffuse emission fractions in mid-IR and H$\alpha$ tracer maps, which are derived solely from the morphological features. We find diffuse emission fractions in mid-IR emission maps ($f_{\rm diffuse}^{\rm mid-IR}$ or $f_{\rm diffuse}^{X}$ when referring to a specific band) to range $32-79$\%, $34-81$\%, and, $28-80$\% for 7.7, 10, and 11.3\,$\mu$m, respectively. The average and 1$\sigma$ range of diffuse mid-IR emission fraction are very similar among different wavelengths ($60\pm10\%$). As for the diffuse emission fraction in the H$\alpha$ map, $f_{\rm diffuse}^{\rm H\alpha}$, we find it to range from $40\%$ to $60$\%, with an average of $50-60\%$.

Using the \textit{Spitzer} 8$\mu$m emission map of M33, \citet{verley09} have measured a similar diffuse mid-IR emission fraction, which ranges $40-60$\% with the fraction increasing at larger radii (see also \citealp{li13}). Our measurements of $f_{\rm diff}^{\rm H\alpha}$ also agree well with estimates from \citet{belfiore22}, who find the diffuse ionized gas fraction to range 20-55\%, using the same PHANGS-MUSE observations. \citet{pan22} employed an unsharp masking technique to estimate the H$\alpha$ diffuse emission fraction in PHANGS galaxies using narrowband H$\alpha$ observations (Razza et al. in prep.), which ranges from 40 to 90\%. 

Using the locations and sizes of \textsc{Hii} regions from \citet{GROVES_HIICAT} for PHANGS–JWST Cycle 1 galaxies, \citet{pathak24} found that mid-IR emission is brighter and more compact when associated with \textsc{Hii} regions. In contrast, mid-IR emission in non-\textsc{Hii} regions is generally fainter and more extended. Excluding galaxy centers, \citet{pathak24} reports that around 50-80\% of the mid-IR emission in galaxy disks originates from outside of \textsc{Hii} regions and is therefore likely to be diffuse. We find good agreement between the fraction of mid-IR emission in non-\textsc{Hii} regions and the diffuse emission fraction measured using Fourier filtering. This suggests that the time-scales derived from this method trace the evolutionary cycle of gas clouds involved in star formation, while excluding the more diffuse gas that likely does not participate in star formation.

\section{Discussion}\label{sec:disc}

\subsection{Correlations between global galaxy properties and cloud evolutionary time-scales}\label{ssec:env}
In order to understand the effects of environment on the cloud evolution, we have correlated our measurements listed in Table~\ref{tab:result} with galaxy and average GMC properties. Galaxy properties include globally observed properties listed in Table~\ref{tab:prop} ($M_{*}^{\rm glob}$, $\rm SFR^{\rm glob}$, $M_{\rm HI}^{\rm glob}$, $M_{\rm H_{2}}^{\rm glob}$, $\rm \Delta MS$, $Z$, and Hubble T-type), as well as other derived properties such as specific SFR ($\rm{sSFR}= SFR^{\rm glob}/M_{*}^{\rm glob}$), total gas mass ($M_{\rm gas}^{\rm glob}=M_{\rm HI}^{\rm glob}+M_{\rm H_{2}}^{\rm glob}$), gas fraction ($f_{\rm gas}^{\rm glob}=M_{\rm gas}^{\rm glob}/(M_{\rm gas}^{\rm glob}+M_{*}^{\rm glob})$), molecular gas fraction ($f_{\rm H_{2}}^{\rm glob}=M_{\rm H_{2}}^{\rm glob}/M_{\rm gas}^{\rm glob}$). 

We include mass weighted averages of kpc-scale galaxy properties measured by \citet{sun22}. These are surface densities of the SFR, atomic gas, molecular gas, and neutral gas ($\Sigma_{\rm SFR}^{\rm kpc}$, $\Sigma_{\rm HI}^{\rm kpc}$, $\Sigma_{\rm H_{2}}^{\rm kpc}$, and $\Sigma_{\rm HI+H_{2}}^{\rm kpc}$), the density of the stellar mass volume near the mid-plane of the disk ($\rho_{*}^{\rm kpc}$), and the dynamical equilibrium pressure ($P_{\rm DE}^{\rm kpc}$), which is the pressure required to support ISM within the gravitational potential of a galaxy \citep[e.g.][]{sun20b}. 

For the average GMC properties, we obtain mass-weighted averages of the velocity dispersion ($\sigma^{\rm GMC}$), virial parameter ($\alpha_{\rm vir}^{\rm GMC}$), GMC mass ($M^{\rm GMC}$), internal turbulent pressure ($P_{\rm turb}^{\rm GMC}$), and molecular gas surface density ($\Sigma_{\rm H_{2}}^{\rm GMC}$), using the GMC catalog generated using the CPROPS algorithm \citep[][A. Hughes et al. in prep.]{rosolowsky21} for PHANGS-ALMA \citep{leroy21_survey} galaxies. When calculating the averages, we exclude GMCs located in the central regions of the galaxy, which have been masked from our analysis (Table~\ref{tab:input}). 

We also explore correlations with environmental metrics constrained within the method, $\epsilon_{\rm mid-IR}$ ($\epsilon_{\rm H\alpha}$), which represents the surface flux density contrast measured between the average emission peak of mid-IR (H$\alpha$) and the galactic average value, on a diffuse emission filtered map. The higher density contrast implies that the emission peak is sharper with less crowding and less background. Finally, we test the effects of systematic biases by examining correlations with the minimum aperture size ($l_{\rm ap, min}$; see Table~\ref{tab:input}), the inclination ($i$; see Table~\ref{tab:prop}), and the mass-weighted average of CO flux completeness on the kpc-scale ($c$; from \citealp{leroy21_pipe} and \citealp{sun22}). The latter $c$ is measured as the ratio between the total flux in moment-0 map when strict signal identification criteria are adopted (``strict'' moment-0 map) compared to that when a more generous criteria is used to include all regions of the cube that are likely to contain signal (``broad'' moment-0 map),  $c=\Sigma {\rm strict~mom0}/\Sigma {\rm broad~mom0}$\footnote{While we refer readers to Section 7.3 of \citet{leroy21_pipe} for full details on how these masks are generated, in brief, the strict mask includes regions with $S/N>2$ over two consecutive velocity channels, but only retains those that are spatially and spectrally connected to a core mask defined by $S/N>4$ over two channels. The broad mask is defined as the union of all strict masks across resolutions, capturing extended and faint emission and encompassing all emission from the galaxy.}. Therefore the completeness ($c$) compares the total CO flux recovered using high-confidence emission regions to that using the full, inclusive mask. Deeper observations, with higher signal-to-noise ratio, would allow more of the true CO emission to satisfy the strict mask criteria, making $c$ closer to 1. In shallower data, much of the real emission is only captured by the broad mask, resulting in lower $c$. Therefore, $c$ serves as a proxy for the depth and flux recovery efficiency of the CO data \citep{leroy21_survey, leroy21_pipe, sun22}. In Figure~\ref{fig:hm}, we illustrate Spearman's rank correlation coefficients computed between our measurements and the galaxy environmental properties described above. In the following subsections, we discuss the significance and physical implications of the observed correlations.

\subsubsection{Theoretical expectations}\label{sssec:theory}
In theory, the \textsc{HI} and $\rm H_{2}$ equilibrium time-scale ($t_{\rm eq}$), which is the time required to reach a balance between formation and destruction rates of $\rm H_{2}$, is described as
\begin{equation}
t_{\rm eq}=\frac{1}{2Rn+D+\zeta},
\end{equation}
where $R$ is the $\rm H_{2}$ formation rate, $n$ is the mean H nucleon number density or the root mean square density in the presence of turbulence, $D$ is the local photodissociation rate including $\rm H_{2}$ self-shielding, and $\zeta$ is the cosmic-ray ionization rate. Assuming the case of a well shielded region ($D\ll Rn$) with a negligible $\rm H_{2}$ dissociation by cosmic rays \citep{sternberg}, 
\begin{equation}
t_{\rm eq}=\frac{1}{2Rn}\approx \frac{10^{9}}{Z'_{d}n}~\rm yr,
\end{equation}
where $R$ is assumed to be $3\times10^{-17}Z'_{d}\,\rm cm^{3}s^{-1}$ and $Z'_{d}$ is the dust-to-gas ratio relative to the solar neighborhood value \citep{chevance22_rev, bialy24}. This implies that when the dust-to-gas ratio and gas density are higher, $t_{\rm eq}$ becomes shorter and the conversion from atomic gas to molecular gas is rapid, allowing the equilibrium between \textsc{HI} and $\rm H_{2}$ to be reached faster. In the case of photodissociation dominated regions, related to the mid-IR and H$\alpha$ overlap phases, theories suggest that the fraction of CO-dark gas should increase at lower metallicities due to reduced shielding of CO \citep{wolfire10, madden20, chevance20lmc, wolfire22}. 

We therefore expect the mid-IR and CO emitting time-scales to be more similar with their ratio ($t_{\rm g}^{\rm mid-IR}/t_{\rm g}^{\rm CO}$) close to one at higher metallicity and gas surface density environments. However, we note that our analysis here with mid-IR and \citet{kim22} with CO have differences in the spatial resolution and sensitivity. The difference in resolution is not expected to significantly bias the trends observed with $t_{\rm g}^{\rm mid-IR}/t_{\rm g}^{\rm CO}$ as shown in Figure~\ref{fig:hm}, with Spearman's rank correlation coefficients between $t_{\rm g}^{\rm mid-IR}/t_{\rm g}^{\rm CO}$ and resolution ($l_{\rm ap, min}$) being almost zero for 7.7 and 10\,$\mu$m and -0.4 with 11.3\,$\mu$m. On the other hand, $t_{\rm g}^{\rm mid-IR}/t_{\rm g}^{\rm CO}$ and CO flux completeness ($c$), indicating how deep the CO observation is, shows stronger trends with coefficients ranging from -0.4 to -0.6. This suggests that trends observed with $t_{\rm g}^{\rm mid-IR}/t_{\rm g}^{\rm CO}$ could be influenced by the differences in CO sensitivity. This is discussed in more detail below.

\begin{figure*}
\includegraphics[scale=0.67]{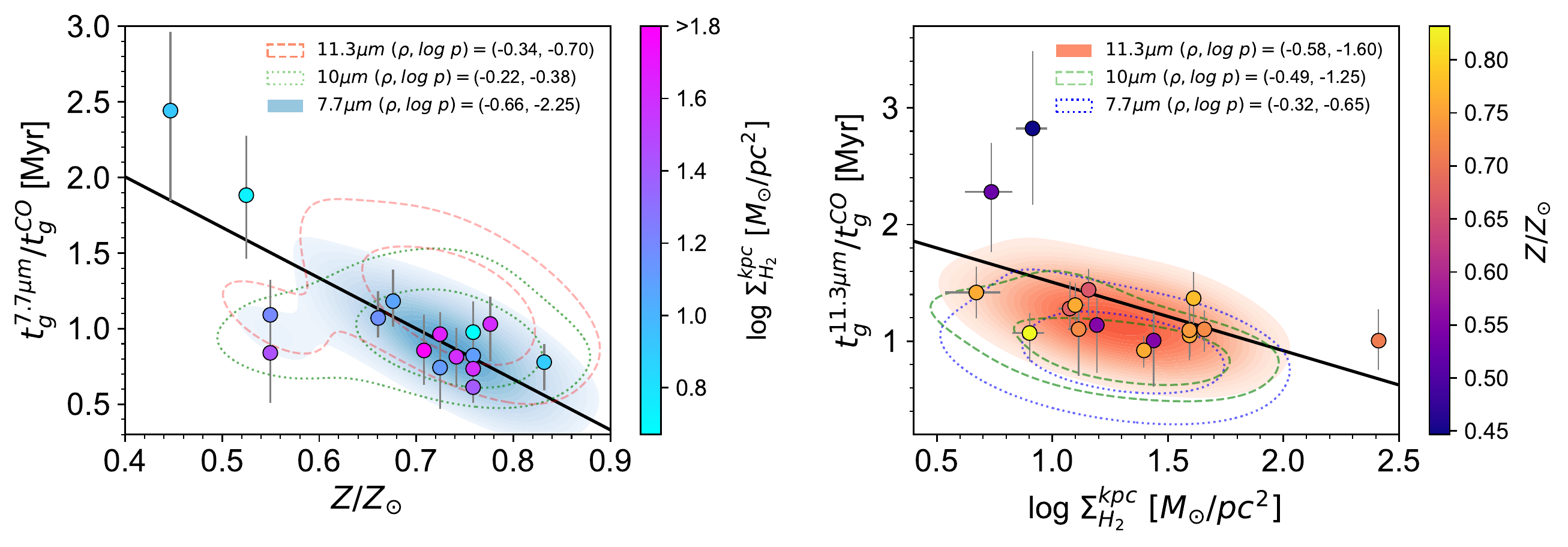}
\caption{The ratio between the compact mid-IR emitting time-scale ($t_{\rm g}^{\rm mid-IR}$) and molecular gas cloud lifetime ($t_{\rm g}^{\rm CO}$) shows negative trends with metallicity and molecular gas surface density, which are in line with theoretical expectations. Data points (circles) in the left panel, color-coded by molecular gas surface density, shows the trend between $t_{\rm g}^{\rm 7.7\mu m}/t_{\rm g}^{\rm CO}$ and metallicity ($Z/Z_{\odot}$, relative to solar), which is the tightest among mid-IR bands.  The solid line represents the best-fitting linear regression. The 1$\sigma$ distribution of our measurements in all three mid-IR bands are shown for comparison as blue, green, and red contours for 7.7, 10, and 11.3\,$\mu$m, respectively. The tightest relation is emphasized using filled contours. We also report the Spearman's correlation coefficients ($\rho$) and $p$-values in the upper right corner. The right panel is similar to the left but instead shows the negative trend between $t_{\rm g}^{\rm 11.3\mu m}/t_{\rm g}^{\rm CO}$ and kpc-scale average molecular gas surface density ($\Sigma_{\rm H_{2}}^{\rm kpc}$), where the data points (circles) are color coded by $Z/Z_{\odot}$.}\label{fig:corr_theory}
\end{figure*}

Figure~\ref{fig:hm} shows that $t_{\rm g}^{\rm mid-IR}/t_{\rm g}^{\rm CO}$ exhibits negative trends with $Z$ and gas surface densities ($\Sigma_{\rm H_{2}}^{\rm kpc}$ and $\Sigma_{\rm HI+H_{2}}^{\rm kpc}$). However, they were not characterized as statistically significant (see Section~\ref{sssec:sig}). In Figure~\ref{fig:corr_theory}, we show how the $t_{\rm g}^{\rm mid-IR}/t_{\rm g}^{\rm CO}$ varies as a function of metallicity and molecular gas surface density. Among mid-IR bands, $t_{\rm g}^{\rm 7.7\mu m}/t_{\rm g}^{\rm CO}$ shows the tightest relation with metallicity with a correlation coefficient ($\rho$) of $-0.7$, while $t_{\rm g}^{\rm 11.3\mu m}/t_{\rm g}^{\rm CO}$ best correlates with molecular gas surface density with $\rho=-0.6$. These observed relations appear to agree with theoretical expectations, where the ratio decreases with increasing metallicity and molecular gas surface density. In high-metallicity and high-density environments, the compact mid-IR emitting phase is similar to the molecular gas cloud lifetime traced with CO, with their ratio close to 1. This suggests that, in these environments, the compact mid-IR emission primarily comes from molecular gas clouds, with their entire extent being bright in CO. It implies that the conversion from cold neutral medium (traced only with compact mid-IR) to molecular gas clouds with stable CO (traced with mid-IR and CO) is rapid. However, in low-metallicity and sparse ISM environment, which are likely to be atomic gas-dominated, the assembly of clouds can take a while and, similarly, it takes a while for the column density to reach sufficient values for CO to be stable against photodissociation (see \citealp{clark12}), leading to a greater difference between the compact mid-IR and CO-bright molecular gas cloud phases. The ratio $t_{\rm g}^{\rm 11.3\mu m}/t_{\rm g}^{\rm CO}$ exceeds 2 in the two most metal-poor galaxies (NGC\,5068 and NGC\,2835). However, we again note that CO flux completeness is also lower for lower-metallicity and low-mass galaxies. This raises the possibility that the observed trends may be biased by the shallow CO observations, potentially missing small and faint CO clouds in these galaxies. Such incompleteness would lead to an underestimation of the CO-emitting time-scale and therefore an overestimation of $t_{\rm g}^{\rm 11.3\mu m}/t_{\rm g}^{\rm CO}$. 

To assess the impact of CO sensitivity, we first performed a linear regression between $t_{\rm g}^{\rm mid-IR}/t_{\rm g}^{\rm CO}$ and CO completeness ($c$). We then used the residuals to represent the time-scale ratio after removing the dependence on completeness. We find that the correlation between the residuals of $t_{\rm g}^{\rm 7.7\mu m}/t_{\rm g}^{\rm CO}$ and metallicity remains intact with $\rho=-0.6$ and $log\,p=-1.8$. In contrast, the trend between $t_{\rm g}^{\rm 11.3\mu m}/t_{\rm g}^{\rm CO}$ and $\Sigma_{\rm H_{2}}^{\rm kpc}$ disappears after accounting for the effect of completeness, which tends to be higher in galaxies with higher molecular gas surface densities. However, this approach assumes a linear dependence between $t_{\rm g}^{\rm mid-IR}/t_{\rm g}^{\rm CO}$ and CO flux completeness, and limited by a small sample. As such, the result should be interpreted with caution. When the same linear regression based correction is applied to the correlation between $t_{\rm g}^{\rm mid-IR}/t_{\rm g}^{\rm CO}$ and spatial resolution, the correlation between the residuals of $t_{\rm g}^{\rm 7.7\mu m}/t_{\rm g}^{\rm CO}$ and metallicity remains comparably tight with $\rho=-0.5$. The trend between $t_{\rm g}^{\rm 11.3\mu m}/t_{\rm g}^{\rm CO}$ and $\Sigma_{\rm H_{2}}^{\rm kpc}$ also persists but becomes weaker with $\rho=-0.3$.

\begin{figure*}
\includegraphics[scale=0.68]{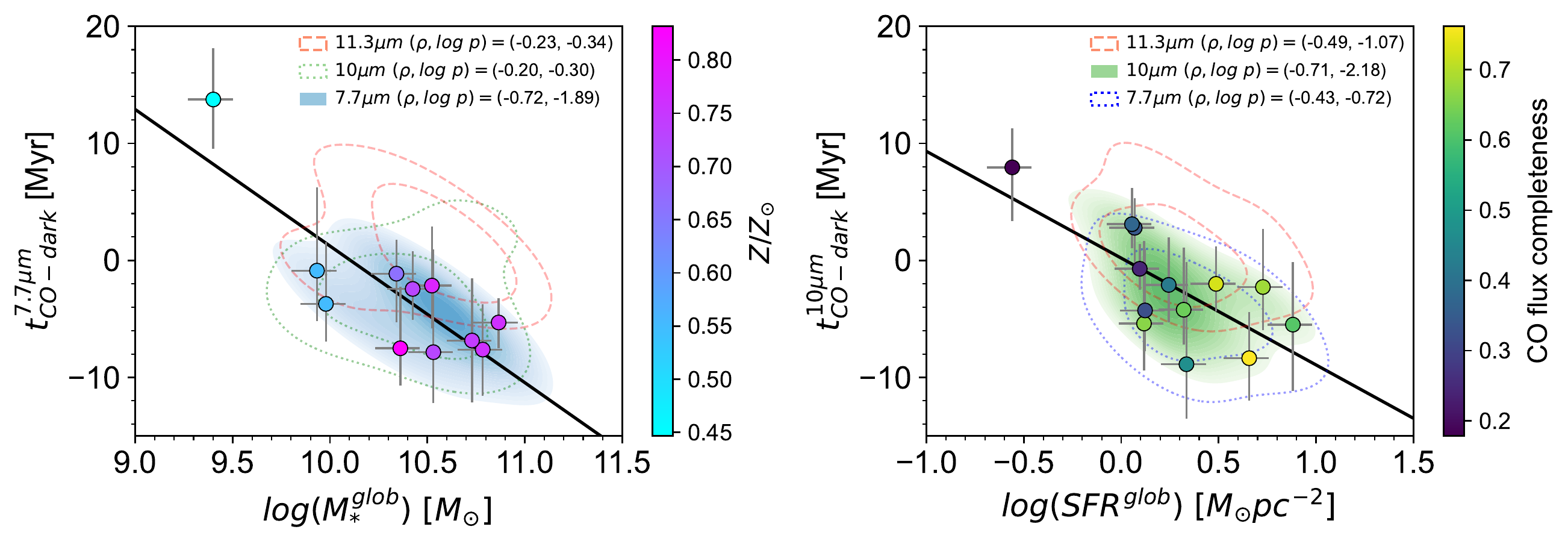}
\caption{Similar to Figure~\ref{fig:corr_theory}, we show a trends between the duration of the mid-IR emitting phase dark in in CO (and H$\alpha$) emission ($t_{\rm CO-dark}^{mid-IR}$) and the galaxy stellar mass ($M_{*}^{\rm glob}$ left) and total SFR ($SFR^{glob}$; right). Data points are $t_{\rm CO-dark}^{7.7\,\mu m}$ and $t_{\rm CO-dark}^{10\,\mu m}$ in the left and right panel, which are colored by the metallicity ($Z/Z_{\odot}$) and the completeness of CO flux observations ($c$), respectively. The correlation indicates that galaxies with higher metallicity and higher SFR has a negligible mid-IR emitting gas cloud phase (before the onset of star formation) that is not traced by CO, with $t_{\rm CO-dark}^{mid-IR}$ being close to zero. For each mid-IR band, Spearman's correlation coefficient ($\rho$) and $p$-value are shown in the upper right corner.}\label{fig:corr_codark}
\end{figure*}

If the conversion from atomic or CO-dark molecular gas clouds to CO-bright molecular gas clouds is faster in higher density and metallicity environments, we also expect to observe similar trends with $t_{\rm CO-dark}^{\rm mid-IR}$. This $t_{\rm CO-dark}^{\rm mid-IR}$ is different from $t_{\rm g}^{\rm mid-IR}/t_{\rm g}^{\rm CO}$ in that $t_{\rm CO-dark}^{\rm mid-IR}$ only considers the duration of the compact mid-IR emitting phase that is not associated with both CO emission and H$\alpha$ while $t_{\rm g}^{\rm mid-IR}/t_{\rm g}^{\rm CO}$ includes the phase coincident with H$\alpha$ emission (feedback phase). In Figure~\ref{fig:corr_codark}, we show two tightest correlationsobserved with $t_{\rm CO-dark}^{\rm mid-IR}$. We find that $t_{\rm CO-dark}^{\rm 7.7\,\mu m}$ shows an anti-correlation with stellar mass ($M_{*}^{\rm glob}$) with $\rho=-0.7$, which agrees with our expectation as stellar mass correlates with metallicity. We also find that $t_{\rm CO-dark}^{\rm 10\,\mu m}$ decreases with increasing global SFR with $\rho=-0.7$. A higher SFR on global-scale indicates that there is more molecular gas with higher densities, possibly resulting in a shorter mid-IR neutral gas phase without associated CO emission. Biases in CO observation also affect $t_{\rm CO-dark}^{\rm 10\,\mu m}$, where we measure a shorter duration of isolated mid-IR phase, dark in CO for galaxies with higher CO flux completeness, as indicated by the color of the data points. The duration of CO-dark phase obtained using other mid-IR bands shows similar trends with $M_{*}^{\rm glob}$ and $\rm SFR^{glob}$, however are weaker. Similar to $t_{\rm g}^{\rm mid-IR}/t_{\rm g}^{\rm CO}$ above, we obtained residuals of $t_{\rm CO-dark}^{\rm 7.7\,\mu m}$ after the dependence on CO flux completeness has been removed. The correlation between the residuals of $t_{\rm CO-dark}^{\rm 7.7\,\mu m}$ and $M_{*}^{\rm glob}$ remains tight with $\rho=-0.6$, while the relation between the residuals of $t_{\rm CO-dark}^{\rm 10\,\mu m}$ and $\rm SFR^{glob}$ disappears. The spatial resolution on the other hand more strongly affects the relation between the residuals of $t_{\rm CO-dark}^{\rm 7.7\,\mu m}$ and $M_{*}^{\rm glob}$ with $\rho=-0.3$. The relation between $t_{\rm CO-dark}^{\rm 10\,\mu m}$ and $\rm SFR^{glob}$ remains comparably tight with $\rho=-0.5$.

 As discussed in Section~\ref{ssec:tg}, the LMC has a very low molecular gas fraction of $f_{\rm H_{2}}=0.09$, sparse ISM with molecular gas surface density of 2\,$M_{\odot}\rm pc^{-2}$, $\rm log\,SFR = -0.7$ \citep{jameson16, schruba19}, and 50\% of solar metallicity \citep{toribio17}. Using \textsc{Hi} and CO observations, the ratio between the neutral gas cloud and molecular gas cloud lifetimes is measured to be 5, whereas $t_{\rm CO-dark}^{\rm HI}$ is $\approx 35$\,Myr \citep{ward20_HI, ward22}. These measurements would place the LMC in the upper left corners of Figure~\ref{fig:corr_theory} and Figure~\ref{fig:corr_codark}, further supporting the trends observed here. We however note that the neutral gas cloud lifetime measured in \citet{ward20_HI} uses \textsc{Hi} observations, while mid-IR emission used here do not fully capture all the neutral ISM distribution. Indeed, when using 8\,$\mu$m observations of the LMC from \textit{Spitzer} \citep{meixner06}, we find the mid-IR emitting phase to be 15\,Myr, resulting in a ratio of 1.5 between the mid-IR emitting time-scale and molecular cloud lifetime, whereas the CO-dark mid-IR emitting cloud phase lasts $\approx 5$\,Myr (J. Kim et al. in prep.). These values are smaller compared to when \textsc{Hi} is used, however still places the LMC along the observed trend.

\begin{figure*}
\includegraphics[scale=0.63]{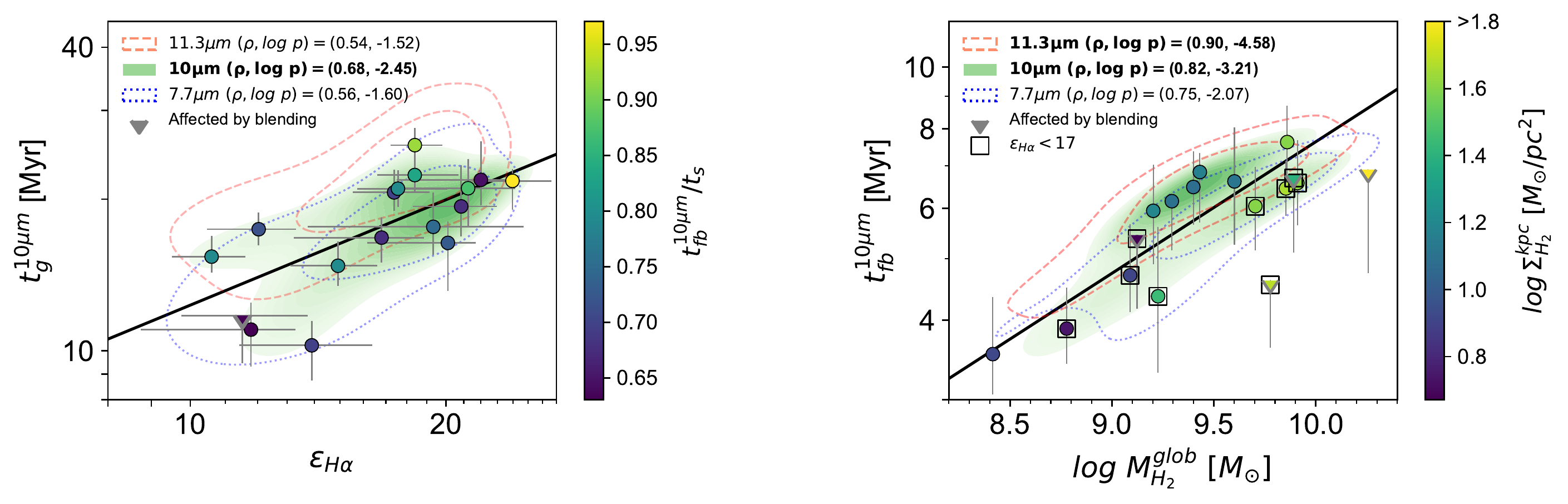}
\caption{Strong correlations identified between our measurements of time-scales and galaxy properties. The data points (circles) in the left panel shows a correlation between the 10\,$\mu$m emitting time-scale ($t_{\rm g}^{10\,\mu m}$) as function of the flux density contrast measured between peaks and galactic average in H$\alpha$ emission map ($\epsilon_{\rm H\alpha}$). The solid line represents the best-fitting linear regression. The higher the density contrast with better defined \textsc{Hii} regions, allows a more efficient heating of the ambient gas and mid-IR emitting phase to continue for a longer period during the H$\alpha$ bright phase. This is also indicated by the color of the circles, which show higher ratios of $t_{\rm fb}^{\rm 10\,\mu m}/t_{\rm s}$ for galaxies with longer mid-IR emitting phases. The outermost contours represent the 1$\sigma$ distributions of our measurements across the three mid-IR bands, shown in blue, green, and red for the 7.7, 10, and 11.3\,$\mu$m bands, respectively, with the tightest relation highlighted using filled contours.Spearman's correlation coefficient and associated $p$-value are indicated in each panel for all mid-IR bands, where significant correlations are in boldface. Triangles represent upper limits for some galaxies which are are affected by source blending (Appendix~\ref{app:robust}). They are excluded in the correlation analysis. The right panel is similar to the left panel but shows correlation between the feedback time-scale ($t_{\rm fb}^{10\,\mu m}$) and molecular gas mass ($M_{\rm H_{2}}^{glob}$). Galaxies with a shorter feedback time-scale have sparser ISM, as indicated by the color of data points (kpc-average molecular gas surface density, $\Sigma_{\rm H_{2}}^{\rm kpc}$), which facilitates the PAH and/or dust dispersal. Galaxies with $\epsilon_{\rm H\alpha}<17$ (squares) also show shorter feedback time-scales, which can be explained if PAHs disperse more quickly when \textsc{Hii} regions are not well defined with structured ISM, enabling a easier escape of ionizing photons and PAH destruction near \textsc{Hii} regions. } \label{fig:corr_tg}
\end{figure*}

\subsubsection{Identification of statistically significant correlations}\label{sssec:sig}

We search for statistically strong correlations, which are determined by their associated $p$-values smaller than an effective $p$-value ($p_{\rm eff}$), defined using the Holm-Bonferroni method (\citealp{holm79}; see also \citealp{kruijssen19b} and \citealp{kim22} for astrophysical applications). This method is used when conducting a large number of hypothesis tests simultaneously, such as correlations between multiple variables, as there is an increased risk of finding spurious significant results purely by chance. The procedure is as follows. We begin by examining whether each of our measurements (row in Figure~\ref{fig:hm}) correlates with any of the galaxy properties (columns in Figure~\ref{fig:hm}). The correlations are ranked based on increasing $p$-values. For each sorted $p$-value with a rank $i$, we compare it with an effective $p$-value ($p_{\rm eff}$). If the $p$-value is smaller than $p_{\rm eff}$, we consider the correlation to be significant. The $p_{\rm eff}$ is defined as $p_{\rm ref}/(N_{\rm corr}+1-i)$, where $p_{\rm ref}$ is the desired significance level (0.05) and $N_{\rm corr}$ is the number of independent variables being compared. In order to evaluate  $N_{\rm corr}$, we have correlated galaxy properties with each other and treat properties with a correlation coefficient higher than 0.7 as one metric. We found strong correlations between $\rm SFR^{\rm glob}$ and $M_{\rm H_{2}}^{\rm glob}$, $M_{\rm gas}^{\rm glob}$, $\Delta \rm MS$, $\Sigma_{\rm HI+H_{2}}^{\rm kpc}$, $\Sigma_{\rm SFR}^{\rm kpc}$, $\Sigma_{\rm H_{2}}^{\rm kpc}$, $\Sigma_{\rm *}^{\rm kpc}$, $P_{\rm DE}$, $c$, $M^{\rm GMC}$, and $\alpha_{\rm vir}^{\rm GMC}$. The galaxy stellar mass $M_{\rm *}$ also strongly correlates with Hubble T-type and  $\sigma^{\rm GMC}$, resulting in $N_{\rm corr} \approx 15$. We note that if we assume all the variables are independent with $N_{\rm corr}=28$, the correlations with $t_{\rm g}^{\rm mid-IR}$ and $f_{\rm diffuse}^{\rm mid-IR}$ are no longer defined as statistically significant. 

All of the significant correlations identified according to our definition are highlighted in Figure~\ref{fig:hm}. We note that these include strong correlations caused by systematic biases. For example, we find that our measurements of $\lambda$ are dependent on the resolution of the data ($l_{\rm ap, min}$) regardless of the wavelength used. These biases have already been pointed out by our previous work \citep{kim22} and imply that $\lambda$ are overestimated in galaxies with a poor resolution. In the following section, we discuss strong correlations that are likely to have physical meaning behind them. 

\begin{figure*}
\includegraphics[scale=0.65]{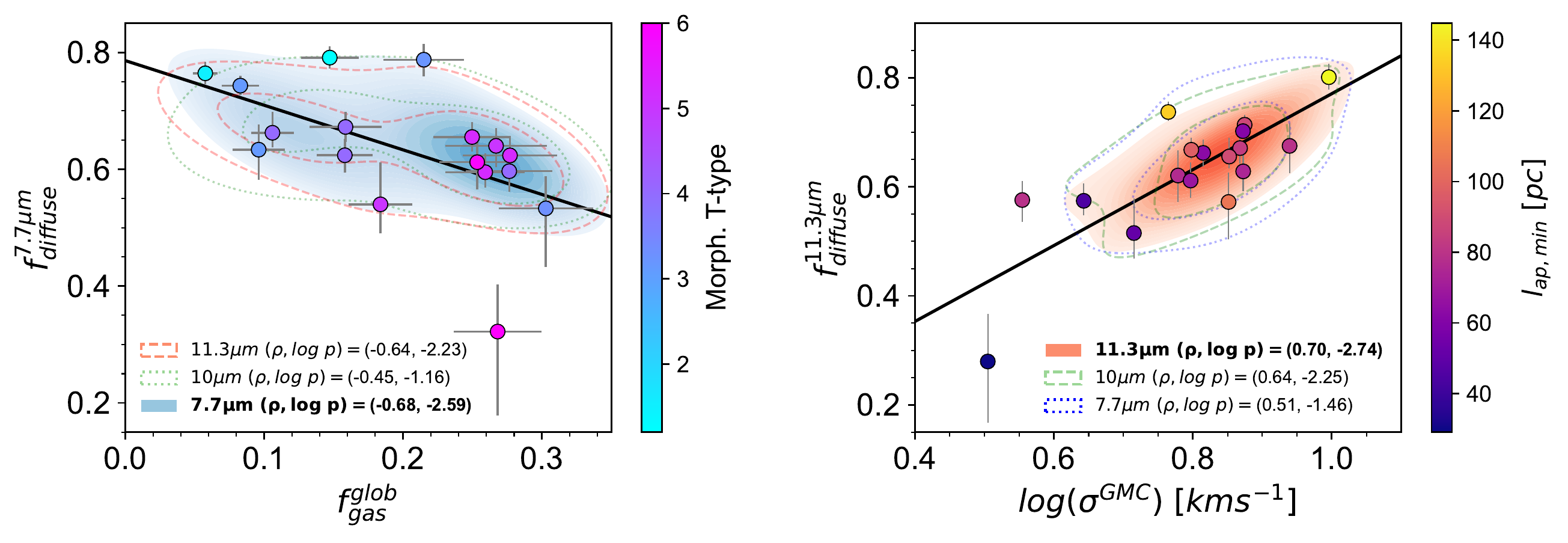}
\caption{Similar to Figure~\ref{fig:corr_tg}, we show strong correlations found with diffuse emission fractions. Left panel shows the anti-correlation between $f_{\rm diffuse}^{\rm 7.7\,\mu m}$ and gas fraction ($f_{\rm gas}^{\rm glob}$), where the data points are color coded by the Hubble morphology T-type. The correlation suggests that diffuse emission in mid-IR increases for galaxies with a low gas fraction and well defined spiral arm structure. Right panel shows the correlation between $f_{\rm diffuse}^{\rm 11.3\,\mu m}$ and the average velocity dispersion of GMCs ($\sigma^{\rm GMC}$), implying that more turbulent clouds leads to higher diffuse emission fraction in mid-IR. However, we note the possibility of resolution biasing the correlation as indicated by the color of the data point, demonstrating that diffuse emission fraction is higher for galaxies with a coarser resolution. } \label{fig:corr_fdiff}
\end{figure*}

\subsubsection{Interpreting the strong correlations}\label{sssec:corr_meaning} 

For the mid-IR emitting phase, as shown in the left panel of Figure~\ref{fig:corr_tg}, we found a strong positive correlation between $t_{\rm g}^{\rm 10\mu m}$ and $\epsilon_{\rm H\alpha}$. The $\epsilon_{\rm H\alpha}$ represents the flux density contrast of H$\alpha$ emission peaks relative to the average flux density on galactic-scale, measured on the diffuse emission filtered map. Therefore, higher density contrast implies that the flux distribution is sharper and that the \textsc{Hii} regions are better defined \citep{oey17} with less crowding and less structured background. Figure~\ref{fig:corr_tg} also shows durations of mid-IR emitting phases obtained with other mid-IR bands ($t_{\rm g}^{\rm 7.7 \mu m}$ and $t_{\rm g}^{\rm 11.3 \mu m}$). They exhibit similar trends with log\,$p\approx -1.5$, which was not small enough to be considered statistically significant.

Given that a higher density contrast in H$\alpha$ emission peak implies better defined and clean \textsc{Hii} regions, we speculate that the correlation is driven by the environments with greater density contrasts allowing a more efficient absorption of non-ionizing far-ultraviolet photons and thus heating in the ambient PAHs and dust grains surrounding the \textsc{Hii} regions. This will in turn allow the mid-IR emitting phase to continue more brightly near \textsc{Hii} regions in the ambient medium, resulting in a longer mid-IR emitting period (e.g. \citealp{jones15}). Our interpretation is also in line with observations of \citet{oey17} suggesting that \textsc{Hii} regions with a clear stratified ionization structure are associated with brighter 8$\mu$m emission compared to \textsc{Hii} regions with structured ISM and a weak transition zone between ionized and neutral gas. Better defined \textsc{Hii} regions also imply a higher spatial correspondence between dust and gas to the heating source, which would lead to hotter dust and thus brighter mid-IR emission near \textsc{Hii} regions \citep{dale07}. The data points in Figure~\ref{fig:corr_tg} are color coded by the ratio of feedback phase over H$\alpha$ emitting phase ($t_{\rm fb}^{\rm 10\mu m}/t_{\rm s}$), suggesting that galaxies with a higher density contrast and a longer mid-IR emission phase tend to have a higher $t_{\rm fb}^{\rm 10\mu m}/t_{\rm s}$ ratio. For all the mid-IR bands, the relations between $t_{\rm fb}/t_{\rm s}$ and $\epsilon_{\rm H\alpha}$ reveal suggestive positive correlations with their coefficients ranging from 0.5 to 0.7 (see Figure~\ref{fig:hm}).

In the right panel of Figure~\ref{fig:corr_tg}, we show a strong positive correlation between $t_{\rm fb}^{\rm 10\mu m}$ and $M_{\rm H_{2}}^{\rm glob}$. The correlation with $M_{\rm gas}^{\rm glob}$ is also found to be significant, however, we show only the relation with $M_{\rm  H_{2}}^{\rm glob}$ here to avoid redundancy. In our analysis with 11.3\,$\mu$m, the same correlations with $M_{\rm HI}^{\rm glob}$ and $M_{\rm gas}^{\rm glob}$ were found to be significant (see Figure~\ref{fig:hm}) with their distributions similar to 10\,$\mu$m. The $t_{\rm fb}^{\rm 11.3\mu m}$ also shows strong correlations with $\rm SFR^{\rm glob}$ and $M_{\rm HI}^{\rm glob}$. For 7.7 \,$\mu$m, the same correlations were not characterized as significant with log\,$\,p=-2.1$. 

The strong correlation between $t_{\rm fb}^{\rm 10\mu m}$ and $M_{\rm H_{2}}^{\rm glob}$ suggests that low mass galaxies have a shorter feedback time-scale. This can be attributed to the low gas surface density and therefore lower mid-plane pressure of low mass galaxies, which allows gas to be dispersed more quickly by stellar feedback. Low mass galaxies also have low gravitational potentials, enabling a faster gas ejection from the galaxy disc \citep{agertz20}. In addition, similar to the relationship between $t_{\rm g}^{10 \mu m}$ and $\epsilon_{\rm H\alpha}$, galaxies with smaller density contrast in \textsc{Hii} regions $\epsilon_{\rm H\alpha}<17$ appear to have shorter feedback time-scales. In an environment with lower density contrasts and thus not well defined \textsc{Hii} regions with structured ISM distribution, it would be easier for ionizing photons to escape, leading to a faster PAH (or dust) dispersal and shorter feedback time-scales. However, in a high density contrast environment with clean \textsc{Hii} regions, photons more efficiently illuminate the surrounding shells of material in mid-IR, resulting in a longer overlap between H$\alpha$ and mid-IR at scales of 50{-}150\,pc. As discussed in Section~\ref{ssec:tfb}, at this scale, the spatial offset between H$\alpha$ peak and the surrounding shells of ISM emitting in mid-IR emission is not spatially resolved, resulting in a long overlap time-scale between the two phases. If the spatial offset can be resolved, we expect to measure a shorter overlapping time-scale.

Finally, as shown in Figure~\ref{fig:corr_fdiff}, we found correlations with diffuse emission fraction, where $f_{\rm diffuse}^{7.7\,\mu m}$ shows a strong anti-correlation with the galaxy gas fraction ($f_{\rm gas}^{\rm glob}$) and $f_{\rm diffuse}^{11.3\,\mu m}$ correlates with the average velocity dispersion of GMCs ($\sigma^{\rm GMC}$). The anti-correlation with $f_{\rm gas}^{\rm glob}$, indicates that galaxies with higher gas fraction, which tend to have a more irregular galaxy morphology (color of data points), have a lower diffuse emission fraction measured in 7.7\,$\mu$m. Other mid-IR bands also exhibit similar trends, where $f_{\rm diffuse}^{11.3\,\mu m}$ is strongly anti-correlated with morphology T-type (see Figure~\ref{fig:hm}). This trend may arise because mid-IR emission from spurs within spiral arms are filtered out and excluded from our analysis, except for the bright knots located within them. This results in a higher diffuse emission fraction for galaxies with a distinct spiral arm structure. The correlation between $f_{\rm diffuse}^{11.3\,\mu m}$ and $\sigma^{\rm GMC}$, can be explained if more turbulent clouds allow photons to travel further outside the star-forming region, resulting in a larger fraction of diffuse emission. We however note that a higher diffuse emission fraction is measured for some galaxies with a coarser resolution, as indicated by the color, potentially biasing the correlation.

\begin{figure*}
\includegraphics[scale=0.70]{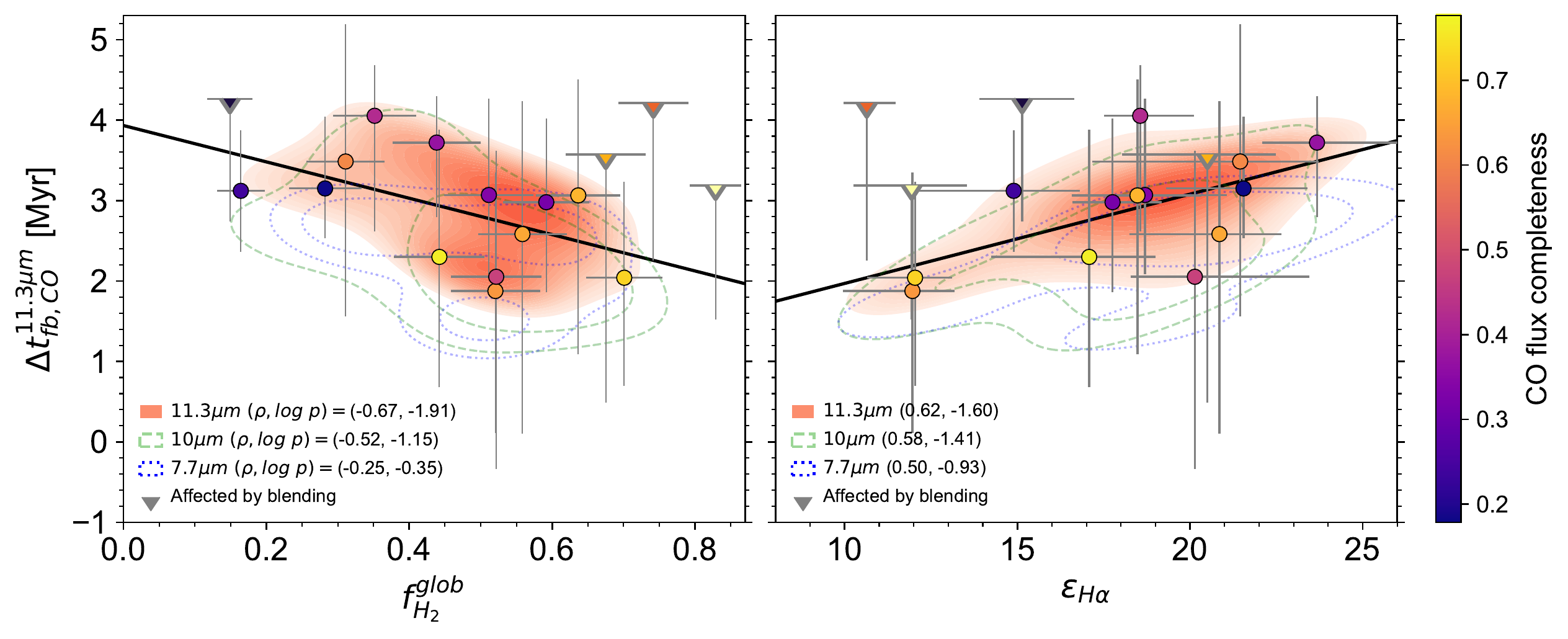}
\caption{Similar to Figure~\ref{fig:corr_tg}, we show suggestive trends of the difference in feedback-time-scales obtained with mid-IR and CO emission ($\Delta t_{\rm fb, CO}^{11.3\, \mu m}$), as a function of the molecular gas fraction ($f_{\rm H_{2}}^{\rm glob}$; left panel) and the density contrast measured in H$\alpha$ emission map ($\epsilon_{\rm H\alpha}$; right panel). The data points are colored based on the CO flux completeness.} \label{fig:corr_dtfb}
\end{figure*}

\subsubsection{Trends with the difference in the feedback time-scale measured with mid-IR versus CO }\label{sssec:dtfb}
As discussed in Section~\ref{ssec:tfb}, across all the galaxies and mid-IR bands, we find that the H$\alpha$ overlap time-scale measured with mid-IR is longer than that obtained using CO emission as the gas tracer. While not characterized as statistically significant according to our definition, in Figure~\ref{fig:corr_dtfb}, we show the two strongest correlations obtained with the difference between the two feedback time-scales ($\Delta t_{\rm fb, CO}^{\rm mid-IR}$). We find that $\Delta t_{\rm fb, CO}^{11.3\, \mu m}$ shows a suggestive trend with the molecular gas fraction ($f_{\rm H_{2}}^{\rm glob}$) and flux density contrast measured in H$\alpha$ map ($\epsilon_{\rm H\alpha}$) with log\,$p<-1.5$. In other mid-IR bands, both relations become weaker. 

The $\Delta t_{\rm fb, CO}^{11.3\, \mu m}$ is longer in galaxies with a lower molecular gas fraction, which can be explained if, in atomic gas-dominated galaxies, CO emission goes under the detection limit faster during the cloud dispersal compared to the mid-IR emission. The CO flux completeness is also low in these galaxies as indicated by the color of the points. Interestingly, we find a longer $\Delta t_{\rm fb, CO}^{\rm 11.3\,\mu m}$ for galaxies with a higher $\epsilon_{\rm H\alpha}$. This is similar to the trend in the left panel of Figure~\ref{fig:corr_tg}, suggesting that photons coming from well defined \textsc{Hii} regions are more efficient in illuminating the surrounding neutral ISM material while the CO emission has already dispersed. This implies that mid-IR emission persists in star-forming regions and is related to the ISM distribution around \textsc{Hii} regions. We note that after removing the dependence of CO flux completeness on $\Delta t_{\rm fb, CO}^{\rm 11.3\,\mu m}$ via linear regression, the correlation of $\Delta t_{\rm fb, CO}^{11.3\, \mu m}$  with molecular gas fraction weakens from $\rho=-0.7$ to $\rho=-0.3$ and the correlation with $\epsilon_{\rm H\alpha}$ weakens from $\rho=0.6$ to $\rho=0.4$. The $\Delta t_{\rm fb, CO}^{\rm 11.3\,\mu m}$ does not show trends with resolution (Figure~\ref{fig:hm} and the observed trends remain unchanged after removing any potential resolution dependence using the same linear regression-based approach. 

\begin{figure*}
\centering
\includegraphics[scale=0.5]{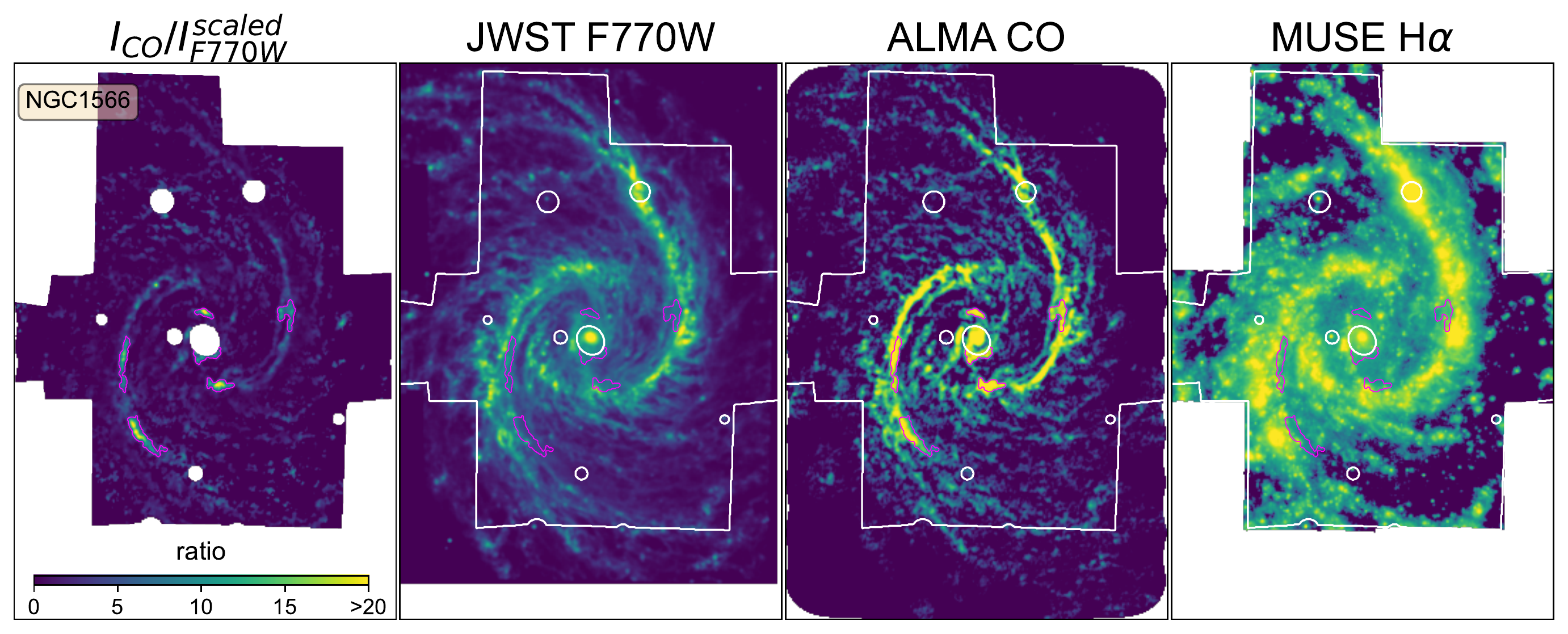}
\includegraphics[scale=0.5]{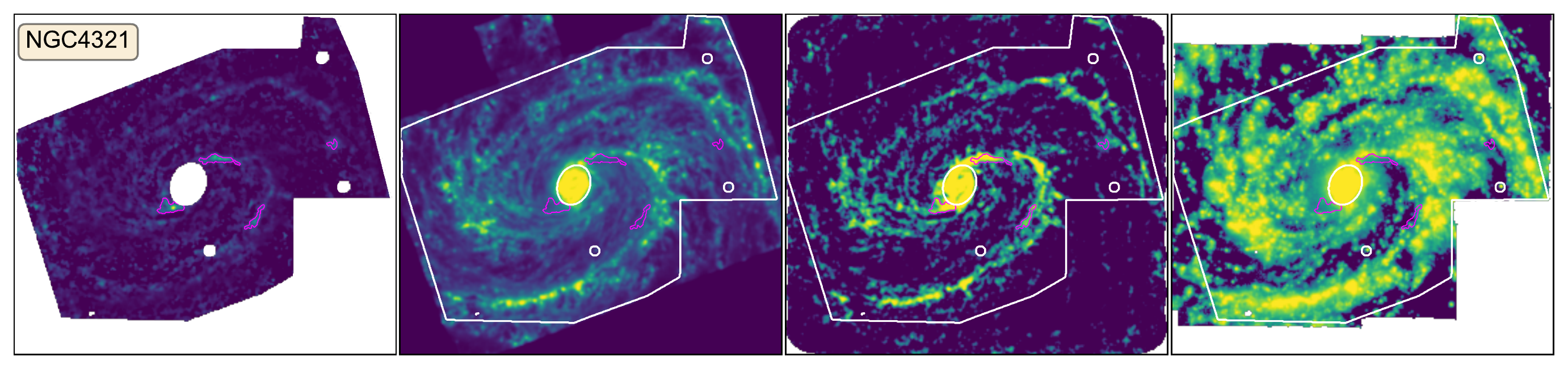}
\includegraphics[scale=0.5]{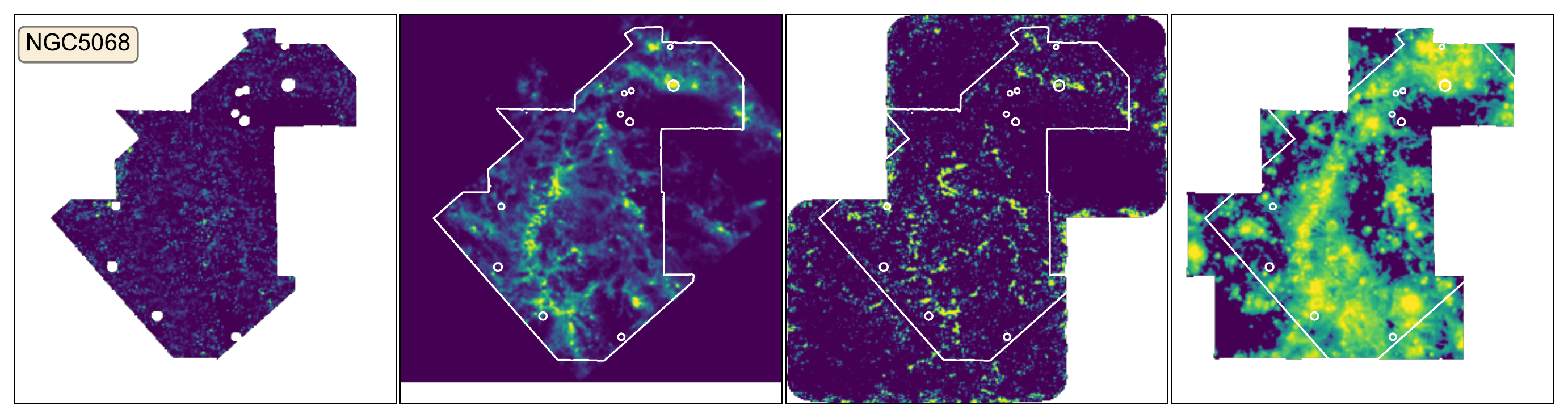}
\includegraphics[scale=0.5]{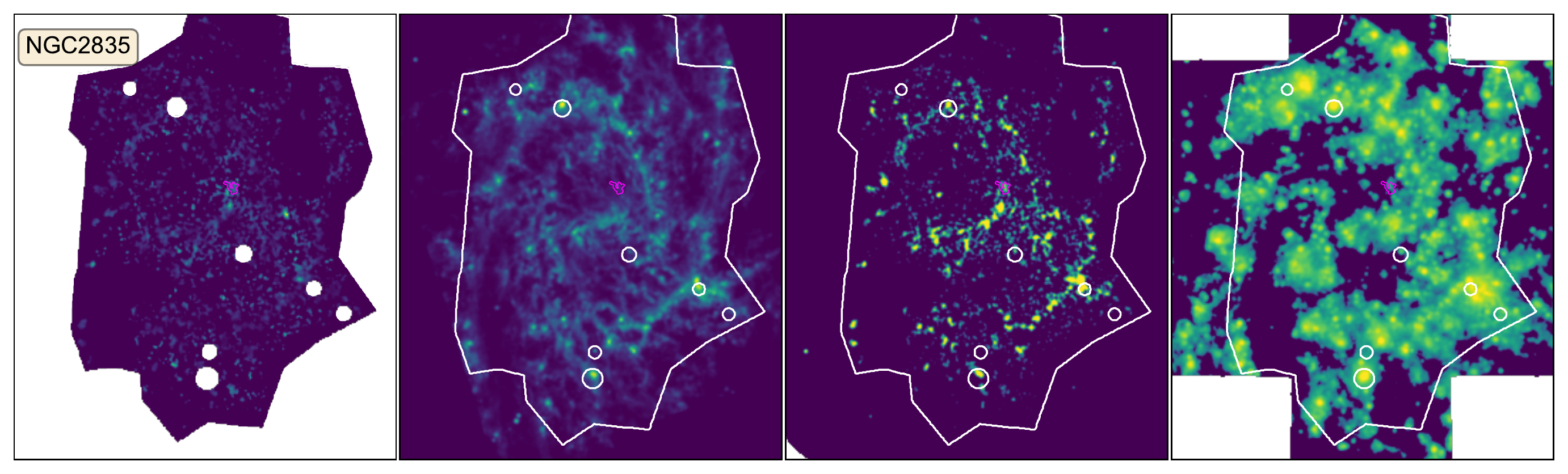}
\caption{Galaxies with the most negative $t_{\rm CO-dark}^{7.7\,\mu m}$ have molecular gas ridges with unusually high CO-to-PAH flux ratios (magenta in top two rows; NGC\,1566 and NGC\,4321), indicating that the tight correlation between F770W and CO flux is broken. This leads to an isolated CO emitting phase before the onset of mid-IR emitting phase. The first column shows the ratio of CO flux ($I_{\rm CO}$) to that expected from F770W flux ($I_{\rm F770W}^{\rm scaled}$), using the correlation from \citet{chown24}. The ridges with flux ratio higher than 5 and bigger than 10 times the $N_{\rm pix}$ are highlighted in magenta. The solid white polygon indicates our analyses region and circle is the masked emission peaks (see Section~\ref{sec:method}). We also show F770W, CO and H$\alpha$ observations in each column. For comparison, galaxies with the most positive $t_{\rm CO-dark}^{7.7\,\mu m}$ are shown in the two bottom rows, where we do not identify significant ridges with unusual flux ratios. } \label{fig:weird}
\end{figure*}

\subsection{Galaxies with decorrelation between CO and mid-IR emission}\label{ssec:copahdecorr}

As shown in Figure~\ref{fig:timeline}, the mid-IR emitting phase typically spans the entire CO emitting phase in most galaxies. This is consistent with expectations, as molecular gas form from compact neutral ISM structures \citep[e.g.,][]{clark12}, before star formation takes place. However, we found some exceptions that contradict physical expectations, NGC\,1566 and NGC\,4321, where CO emitting phase begins earlier than the mid-IR emitting phase, particularly when 7.7\,$\mu$m is used. This implies that there is a substantial decorrelation between CO and mid-IR emission in these galaxies, resulting in a negative $t_{\rm CO-dark}^{\rm mid-IR}$. 

Figure~\ref{fig:weird} shows the ratio of CO flux ($I_{\rm CO}$) and that expected from F770W emission ($I_{\rm F770W}^{\rm scaled}$), based on the correlation from \citet{chown24}. Mid-IR observations used in Figure~\ref{fig:weird} are described in Section~\ref{sec:obs}, whereas the CO data is from PHANGS-ALMA \citep{leroy21_survey}. We note that these are not diffuse emission filtered. Contours highlight ridges with flux ratios ($I_{\rm CO}/I_{\rm F770W}^{\rm scaled}$) higher than 5 and with their size at least 10 times larger than the minimum emission peak size, $N_{\rm pix, min}$ (see Table~\ref{tab:input}). Notably, NGC\,1566 and NGC\,4321, which have the most negative $t_{\rm CO-dark}^{\rm 7.7\,\mu m}$ ($-7.6_{-4.0}^{+3.9} $ and a range from $-11.9$ to $-3.4$, respectively) show long ridges that deviate from the tight correlation observed between F770W and CO flux \citep{chown24}. These regions, with unusually high CO-to-PAH flux ratios, do not coincide with \textsc{Hii} regions, indicating that the lack of PAH emission is not due to massive stars dissociating PAHs \citep{egorov23, sutter24}. They are also preferentially located in dynamically complex areas, like concave parts of spiral arms, bar ends, and central rings. These two galaxies (NGC\,1566 and NGC\,4321) and NGC\,1365 are the only galaxies in our sample that exhibit these long ridges with unusual flux ratios, while other galaxies show none or only one or two small regions. For comparison, we also show NGC\,5068 and NGC\,2835, which are two galaxies with the longest $t_{\rm CO-dark}^{\rm mid-IR}$ and follow the observed correlation between CO and mid-IR well without significant outliers.

There are multiple potential explanations for the long ridges with anomalous CO-to-PAH ratios. One possibility is a very low UV radiation field, which could suppress PAH excitation. However, given their location, it seems unlikely that these ridges are particularly in well-shielded regions. Another explanation could be that CO emissivity is substantially higher in these ridges, possibly due to large-scale streaming motions changing the opacity \citep{teng23}, as these ridges are situated along the bar ends and edges of spiral arms. Lastly, it could be that PAH emission in the $7.7{–}11.3$\,$\mu$m bands has been suppressed, potentially due to PAH dissociation from large-scale shocks, or because PAHs are growing in size or even sticking onto other dust grains via accretion and coagulation \citep{draine21, hensley23}. To investigate this, we have used stellar continuum-subtracted 3.3\,$\mu$m imaging from H. Koziol et al. (in preparation) and \citet{sandstrom23_33pah} to measure the 3.3\,$\mu$m/11.3\,$\mu$m and 11.3\,$\mu$m/7.7\,$\mu$m flux ratios. We found that PAHs in these ridges appear to be larger (lower 3.3\,$\mu$m/11.3\,$\mu$m ratio) and more neutral (higher 11.3\,$\mu$m/7.7\,$\mu$m ratio), which supports the idea of coagulation. However, these ratios are degenerate with PAH properties, and further spectroscopic observations such as those covering $\rm H_{2}$ rotational and rot-vibrational lines, shock tracers like [Fe II], and various PAH features are needed to confirm these speculations.

\begin{figure*}
\centering
\includegraphics[scale=0.6]{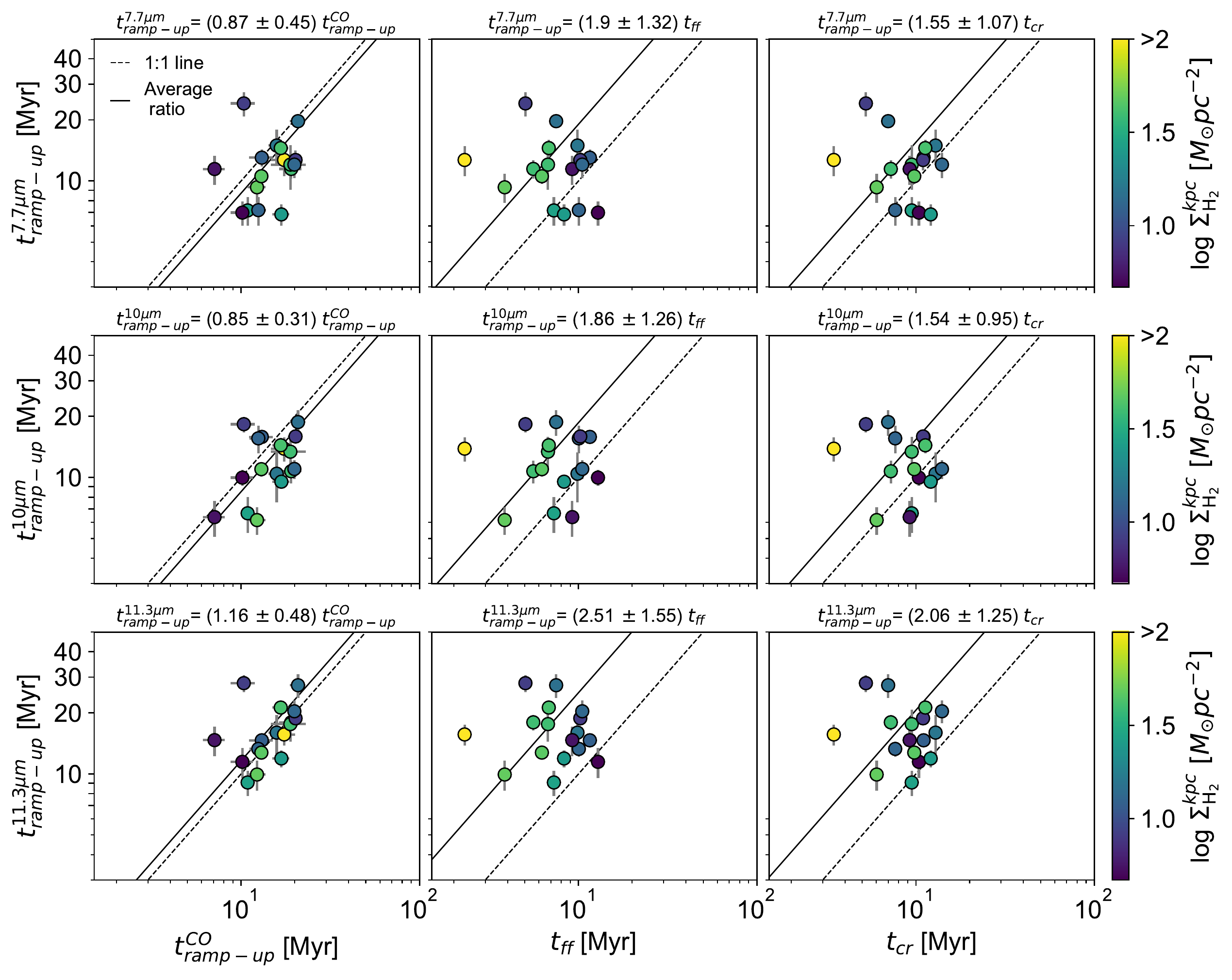}
\caption{The time from neutral gas cloud formation to the onset of star formation traced with mid-IR (ramp up time-scale; $t_{\rm ramp up}^{\rm mid-IR}$) is compared to that measured with CO ($t_{\rm ramp up}^{\rm CO}$), as well as to analytical predictions from \citet{sun22}, which are related to the dynamics within clouds (GMC free-fall and turbulence crossing time-scales; $t_{\rm ff}$ and $t_{\rm cr}$, respectively). Data points are color-coded by molecular gas surface density ($\Sigma_{\rm H_{2}}^{\rm kpc}$). The average and 1$\sigma$ range of the ratio of two time-scales are shown in each panel. The solid line shows the average ratio between the time-scales and a one-to-one relation is shown in dashed line.  We find the best agreement between $t_{\rm ramp up}^{\rm mid-IR}$ and $t_{\rm ramp up}^{\rm CO}$, whereas the ramp-up time-scale measured with mid-IR is somewhat longer than $t_{\rm ff}$ and $t_{\rm cr}$. } \label{fig:tcomp}
\end{figure*}

\begin{figure}
\centering
\includegraphics[scale=0.65]{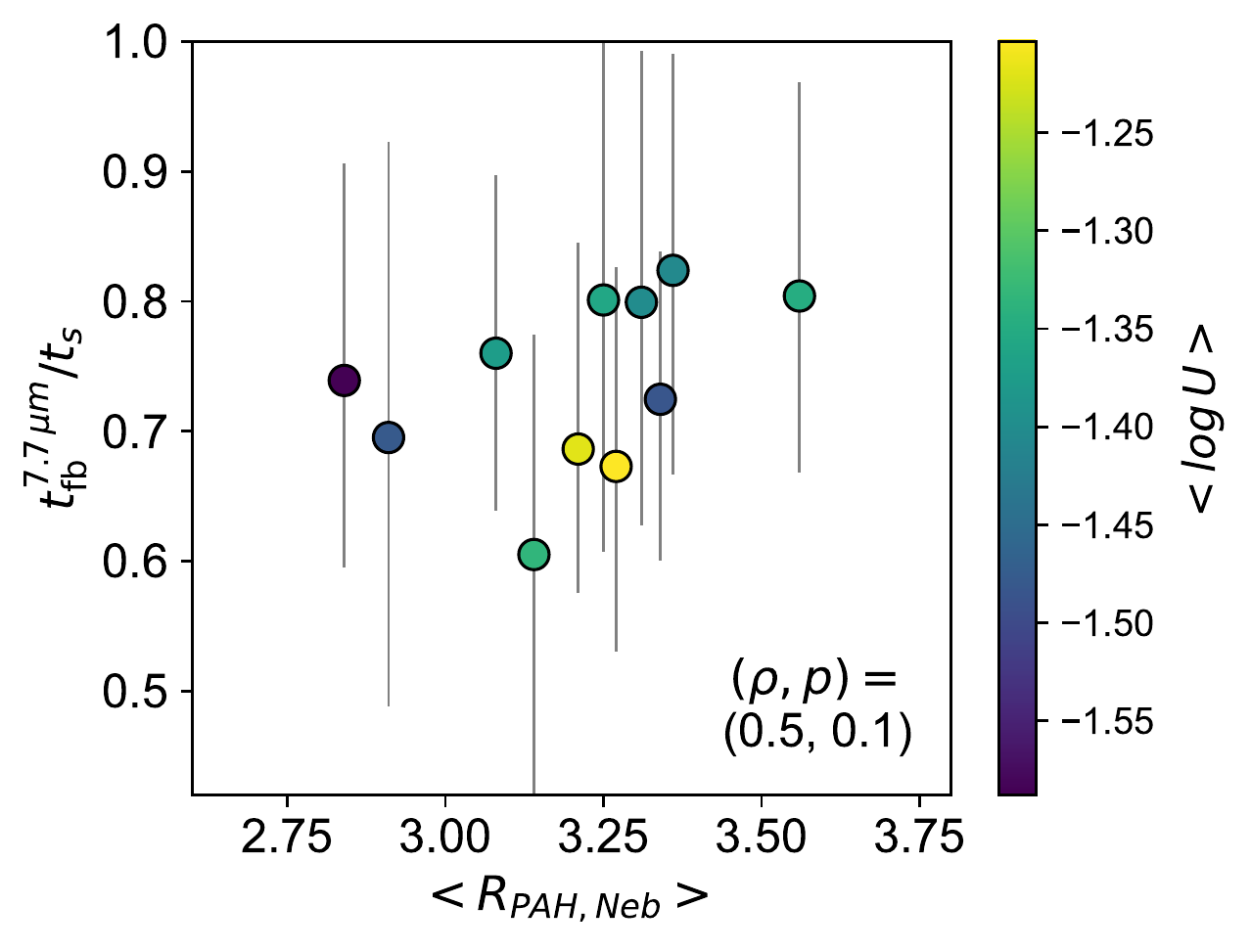}
\caption{The fraction of the H$\alpha$ emitting phase associated with 7.7\,$\mu$m emission ($t_{\rm fb}^{7.7\,\mu m}/t_{s}$) as a function of the average PAH abundance in \textsc{Hii} regions \citep[$<R_{\rm PAH, Neb}>$][]{sutter24}. Data points are color-coded by the average ionization parameter ($<log\,U>$), which is the ratio between ionizing photon flux and the hydrogen density in \textsc{Hii} regions from \citet{GROVES_HIICAT}. Spearman's correlation coefficient and the associated p-value are also indicated. The 7.7\,$\mu$m shows the strongest correlations while other bands show weaker relations.} \label{fig:rpah}
\end{figure}

\subsection{Comparison with characteristic time-scales}\label{ssec:tcomp}

The measured mid-IR emitting phase can be divided into two stages: the quiescent ramp-up phase, associated with mid-IR emission before the onset of massive star formation detected in H$\alpha$ ($t_{\rm ramp-up}^{\rm mid-IR}=t_{\rm g}^{\rm mid-IR}-t_{\rm fb}^{\rm mid-IR}$) and the mid-IR and H$\alpha$ emission overlapping phase ($t_{\rm fb}^{\rm mid-IR}$). Using CO observations, the ramp-up period has been measured to be $\sim 5-20$\,Myr and is comparable to the gravitational free-fall or turbulence crossing time-scales of GMCs \citep{corbelli17, kim21, kim23, schinnerer24}. In Figure~\ref{fig:tcomp}, we compare the duration of the ramp up phase measured with mid-IR to that constrained with CO ($t_{\rm ramp-up}^{\rm CO}$), as well as analytical predictions of GMC free-fall and turbulence crossing times ($t_{\rm ff}$ and $t_{\rm cr}$), which are mass-weighted harmonic averages from \citet{sun22}. The compact mid-IR emitting gas clouds experience a ramp-up period of $5-30$\,Myr, showing the best agreement with $t_{\rm ramp-up}^{\rm CO}$ measured in \citet{kim22}. Gas clouds with lower molecular gas surface densities, indicated by the color of the data points, show a good agreement between $t_{\rm ramp-up}^{\rm mid-IR}$ and the GMC free-fall time-scale. However, the overall agreement between $t_{\rm ramp-up}^{\rm mid-IR}$ and $t_{\rm cr}$ is better than with $t_{\rm ff}$.

\subsection{PAH abundance and feedback duration}\label{ssec:rpah}

For the feedback time-scale, we find that mid-IR emission remains strongly associated with H$\alpha$ emission, with its duration covering 60{-}90\% of the total H$\alpha$ emitting phase. This overlap period is likely to be dependent on various factors, including PAH destruction by ionizing photons, ISM heating, and reshaping of the surrounding dust reservoir. While PAH destruction within \textsc{Hii} regions would shorten the feedback phase, PAHs would glow more brightly near areas where massive stars have recently formed, prolonging the measured time-scale. 

In Figure~\ref{fig:rpah}, we show a relationship between the fraction of H$\alpha$ emitting phase associated with 7.7\,$\mu$m emission ($t_{\rm fb}^{7.7\,\mu m}/t_{s}$) and the average PAH abundance of \textsc{Hii} regions ($R_{\rm PAH, Neb}$) from \citet{sutter24}. Among other mid-IR bands, 7.7\,$\mu$m shows the strongest relation. The $R_{\rm PAH, Neb}$ is estimated using a combination of JWST filters ([F770W+F1130W]/F2100W), which are known to trace the fraction dust in the form of PAHs by mass \citep{chastenet23, egorov23}. We find a moderate positive relationship between $t_{\rm fb}^{7.7\,\mu m}/t_{s}$ and $R_{\rm PAH, Neb}$, indicating that a lack of PAHs leads to a shorter overlap between mid-IR and H$\alpha$. The data points are color-coded according to the ionization parameter, which is estimated in \citet{GROVES_HIICAT} using the prescription of \citet{diaz91}. This ionization parameter is related to the ratio of the density of ionizing photons to the number density of hydrogen atoms in the gas. We use H$\alpha$ luminosity weighted averages. For the two galaxies with higher ionization parameters, the overlap period appears to be shorter. This could be due to the more intense ionizing radiation field destroying PAHs, rather than exciting them and allowing PAHs to illuminate more brightly near \textsc{Hii} regions. However, we note that we are only comparing a single average value for each galaxy, while there are likely wide variations in both PAH abundance and the radiation field within each galaxy. For example, \citet[][in preperation]{egorov23} finds a strong anti-correlation between PAH abundances and ionization parameter for individual \textsc{Hii} regions. In future studies, it would be valuable to explore the distribution of these parameters, rather than relying on single average values. By considering the full range of parameter values, such as the variation in PAH abundances and ionization parameters across different regions within each galaxy, we would gain a more comprehensive understanding of the processes at play.

\section{Conclusions}\label{sec:summ}
Leveraging PHANGS-JWST Cycle~1 and PHANGS-MUSE observations, we present time-scales of PAH (7.7\,$\mu$m and 11.3\,$\mu$m) and dust continuum (10\,$\mu$m) emission of gas clouds for 17 galaxy disks (excluding centers), by applying the statistical method developed by \citet{kruijssen14} and \citet{kruijssen18}. Assuming that mid-IR emission (7.7-11.3\,$\mu$m) traces the gas column density, this enables us to translate relative spatial distributions of mid-IR and H$\alpha$ emission into their underlying time-scales, evolving from the mid-IR emitting gas cloud phase to exposed young stellar \textsc{Hii} regions. For each mid-IR band, we provide constraints on the mid-IR emitting time-scale ($t_{\rm g}^{\rm mid-IR}$), the PAH and/or dust feedback time-scale (the period during which mid-IR and H$\alpha$ are coincident; $t_{\rm fb}^{\rm mid-IR}$), and the average distance between independent star-forming regions ($\lambda^{\rm mid-IR}$). We also determine additional physical quantities such as the fraction of mid-IR emitting phase with associated H$\alpha$ ($t_{\rm fb}^{\rm mid-IR}/t_{\rm g}^{\rm mid-IR}$), fraction of the H$\alpha$ emitting phase with associated mid-IR emission ($t_{\rm fb}^{\rm mid-IR}/t_{\rm s}$), feedback time-scale difference between mid-IR versus CO ($\Delta t_{\rm fb, CO}^{\rm mid-IR}$), duration of the mid-IR emitting gas cloud phase dark in CO (and H$\alpha$) emission ($t_{\rm CO-dark}^{\rm mid-IR}$), and diffuse emission fractions ($f_{\rm diffuse}^{\rm mid-IR}$ and $f_{\rm diffuse}^{\rm H\alpha}$). We have correlated these measurements with global galaxy and average GMC properties, as well as observational biases. We found several statistically significant correlations, allowing us to infer physical mechanisms that are responsible for the mid-IR emission. 

Our key findings are as follows:
\begin{enumerate}
    \item Across our sample, we find that the mid-IR emitting time-scale ranges 10{-}30\,Myr across all the mid-IR bands. The average and 1$\sigma$ range are $18\pm5$, $18\pm4$, and $23\pm5$\,Myr for the 7.7, 10, and 11.3\,$\mu$m, respectively. On average, the mid-IR emitting time-scales of gas clouds are 1 to 1.3 times the GMC lifetime measured in \citet{kim22}, encompassing the CO-emitting phase in most galaxies. The duration of the mid-IR emitting gas cloud phase not detected in CO (and H$\alpha$) emission is nearly nonexistent with an average and 1$\sigma$ range of $t_{\rm CO-dark}^{\rm mid-IR}$ being $-3\pm6$, $-2\pm5$, and $3\pm5$\,Myr for the 7.7\,$\mu$m, 10\,$\mu$m, and 11.3\,$\mu$m, respectively. This very short duration is most likely due to the fact that our observations are focused on the central part of the galaxy, where most of the gas is already in the molecular gas phase showing a tight correlation between mid-IR and CO emission \citep{chown24}. 
    \item For all the mid-IR bands, the H$\alpha$ overlap phase or feedback time-scales for PAH and/or dust grain dispersal ranges from 3 to 7\,Myr across our sample, with an average and 1$\sigma$ range of $6\pm 1$\,Myr. This constitutes, on average, 30\% of the mid-IR emitting time-scale. The feedback time-scale covers a significant fraction of the H$\alpha$ emitting phase with an average of $t_{\rm fb}/t_{\rm s}=70-80\%$ across mid-IR bands. We find that the feedback time-scale with mid-IR is always longer than the time-scale obtained with CO, with an average difference of $\Delta t_{\rm fb, CO}=2-3$\,Myr, across all mid-IR bands. The difference between mid-IR and CO is likely due to the fact that PAH and dust continuum emission is enhanced near \textsc{Hii} regions by the intense radiation from young stars, while CO is dispersed quickly. Star-forming regions undergoing independent evolution from gas to stars are separated by 110-130\,pc on average. 
    
    \item The diffuse emission fraction in mid-IR observations ranges 30-70\%, with an average and a standard deviation of $60\pm 10\%$. As for the H$\alpha$ maps, we measure diffuse emission fraction to range from 40 to 60\%. These fractions show a good agreement with the fraction of mid-IR and ionized gas emission outside \textsc{Hii} regions \citep{belfiore22, pathak24}. 

    \item We find that the ratio between mid-IR emitting time-scale and molecular gas cloud lifetime ($t_{\rm g}^{\rm mid-IR}/t_{\rm g}^{\rm CO}$) decreases with increasing metallicity and molecular gas surface density, in agreement with theoretical expectations where the conversion from atomic to molecular gas is expected to be faster in such environments. Similarly, CO-dark mid-IR emitting gas cloud phase ($t_{\rm CO-dark}^{\rm mid-IR}$) show the tightest correlations with stellar mass ($M_{*}^{\rm glob}$) and global SFR ($\rm SFR^{\rm glob}$). 

    \item We identify several statistically significant correlations between our measurements and galaxy environmental properties. For the mid-IR emitting time-scale (particularly $t_{\rm g}^{\rm 10\,\mu m}$), we find a strong positive correlation with $\epsilon_{H\alpha}$, which is the average flux density contrast measured in the H$\alpha$ map, between the peak and galactic average. The correlation can be explained if cleanly defined \textsc{Hii} regions without structured ISM (larger $\epsilon_{H\alpha}$) allow photons to heat the surrounding ISM more efficiently, enabling the mid-IR emitting phase to continue for a longer period. 

    \item The H$\alpha$ overlap time-scales obtained with 10\,$\mu$m and 11.3\,$\mu$m ($t_{\rm fb}^{\rm 10\,\mu m}$ and $t_{\rm fb}^{\rm 11.3\,\mu m}$, respectively) show a strong correlation with galaxy atomic gas mass ($M_{\rm HI}^{\rm glob}$) and the total neutral gas mass ($M_{\rm gas}^{\rm glob}$). This can be understood if the sparser ISM of low mass galaxies and their low gravitational potential facilitate a faster dispersal of PAHs and/or dust grains. 

    \item We find that diffuse emission fraction measured in 7.7\,$\mu$m ($f_{\rm diffuse}^{7.7\,\mu m}$) is higher in galaxies with a low gas fraction ($f_{\rm gas}^{\rm glob}$) and with well-defined spiral arm structures. We suspect this is due to the diffuse emission located between the arms (known as spurs), as only the star-forming knots and gas clumps within the structure are considered compact. For $f_{\rm diffuse}^{11.3\,\mu m}$, we find that galaxies with more turbulent molecular clouds, characterized by a high velocity dispersion ($\sigma^{\rm GMC}$), exhibit a higher diffuse emission fraction. 

    \item While not statistically significant, we note two tentative correlations detected with $\Delta t_{\rm fb, CO}^{11.3\,\mu m}$. The $\Delta t_{\rm fb, CO}^{11.3\,\mu m}$ becomes longer in galaxies with a low molecular gas fraction (and therefore a low CO flux completeness). This occurs because CO emission falls below the detection limit faster than the mid-IR during the feedback phase. However, another suggestive trend with $\epsilon_{\rm H\alpha}$ suggests that $\Delta t_{\rm fb, CO}^{11.3\,\mu m}$ may have physical implications as well. Specifically, $\Delta t_{\rm fb, CO}^{11.3\,\mu m}$ increases with higher $\epsilon_{\rm H\alpha}$. This may be interpreted as photons from star-forming regions continuing to illuminate the surrounding ISM even after CO has dissociated in regions with a high $\epsilon_{\rm H\alpha}$.

    \item We identified galaxies (such as NGC 1566 and NGC 4321) with long CO-bright ridges where CO and mid-IR emission decorrelates with unusually high CO-to-PAH flux ratios. In these galaxies, there is an isolated CO emitting phase before the mid-IR emission phase begins, leading to a negative $t_{\rm CO-dark}^{\rm mid-IR}$. 
    
    \item Finally, the mid-IR emitting phase is divided into two stages: the quiescent ramp-up phase before massive star formation and the overlapping phase with mid-IR and H$\alpha$ emission. The ramp-up period, measured with mid-IR, is found to be $\sim 5-30$,Myr, comparable to the ramp-up period measured with CO \citep{kim22}. For gas clouds in a low density environment, ramp-up periods, measured with mid-IR, show a good agreement with GMC free-fall time-scale \citep{sun22}. 
\end{enumerate}

Building on our previous measurements of the GMC lifecycle across 54 galaxies, we have extended the timeline for a sub-sample of 17 galaxies by incorporating mid-IR (PAH and dust continuum) emission as a probe for gas distribution. In the future, we aim to expand this analysis to the full 54 galaxies using PHANGS-JWST Cycle~2 data. The small difference observed between the mid-IR emitting time-scale of gas clouds and the molecular gas cloud lifetimes is likely due to the fact that most of the gas in our sample is in the CO-bright molecular phase. Additionally, we find that radiation from star-forming regions as well as the shape of the ISM around the \textsc{Hii} regions plays a key role in mid-IR emission. Extending this analysis to a wider range of environments with low metallicity and low gas surface density would help determine whether the time-scales measured with mid-IR and CO differ.

\section{Acknowledgments}
We thank the anonymous referee for helpful comments, which improved the quality of the manuscript. This work was carried out as part of the PHANGS collaboration. This work is based on observations made with the NASA/ESA/CSA JWST. Support for program \#2107 was provided by NASA through a grant from the Space Telescope Science Institute, which is operated by the Association of Universities for Research in Astronomy, Inc., under NASA contract NAS 5-03127. 

J.K. is supported by a Kavli Fellowship at the Kavli Institute for Particle Astrophysics and Cosmology (KIPAC). 

MC and LR gratefully acknowledge funding from the DFG through an Emmy Noether Research Group (grant number CH2137/1-1).
COOL Research DAO \citep{cool_whitepaper} is a Decentralized Autonomous Organization supporting research in astrophysics aimed at uncovering our cosmic origins.

KK gratefully acknowledges funding from the Deutsche Forschungsgemeinschaft (DFG, German Research Foundation) in the form of an Emmy Noether Research Group (grant number KR4598/2-1, PI Kreckel) and the European Research Council’s starting grant ERC StG-101077573 (``ISM-METALS").

RSK and SCOG acknowledge financial support from the European Research Council via the ERC Synergy Grant ``ECOGAL'' (project ID 855130),  from the German Excellence Strategy via the Heidelberg Cluster of Excellence (EXC 2181 - 390900948) ``STRUCTURES'', and from the German Ministry for Economic Affairs and Climate Action in project ``MAINN'' (funding ID 50OO2206). RSK also thanks the 2024/25 Class of Radcliffe Fellows for highly interesting and stimulating discussions.

OE acknowledges funding from the Deutsche Forschungsgemeinschaft (DFG, German Research Foundation) -- project-ID 541068876.

KG is supported by the Australian Research Council through the Discovery Early Career Researcher Award (DECRA) Fellowship (project number DE220100766) funded by the Australian Government. 

IP acknowledges funding by the European Research Council through ERC-AdG SPECMAP-CGM, GA 101020943.

JS acknowledges support by the National Aeronautics and Space Administration (NASA) through the NASA Hubble Fellowship grant HST-HF2-51544 awarded by the Space Telescope Science Institute (STScI), which is operated by the Association of Universities for Research in Astronomy, Inc., under contract NAS~5-26555.

EWR acknowledges the support of the Natural Sciences and Engineering Research Council of Canada (NSERC), funding reference number RGPIN-2022-03499 and the support of the Canadian Space Agency (23JWGO2A07).

This paper makes use of the following ALMA data, which have been processed as part of the PHANGS--ALMA survey: \\
\noindent ADS/JAO.ALMA\#2012.1.00650.S, \linebreak 
ADS/JAO.ALMA\#2013.1.01161.S, \linebreak 
ADS/JAO.ALMA\#2015.1.00925.S, \linebreak 
ADS/JAO.ALMA\#2015.1.00956.S, \linebreak 
ADS/JAO.ALMA\#2017.1.00392.S, \linebreak 
ADS/JAO.ALMA\#2017.1.00886.L, \linebreak 
ADS/JAO.ALMA\#2018.1.01651.S. \linebreak 
ALMA is a partnership of ESO (representing its member states), NSF (USA), and NINS (Japan), together with NRC (Canada), NSC and ASIAA (Taiwan), and KASI (Republic of Korea), in cooperation with the Republic of Chile. The Joint ALMA Observatory is operated by ESO, AUI/NRAO, and NAOJ. The National Radio Astronomy Observatory is a facility of the National Science Foundation operated under cooperative agreement by Associated Universities, Inc.


%

\vspace{5mm}
\facilities{JWST, VLT/MUSE, ALMA}


\software{Astropy \citep{astropy:2013, astropy:2018, astropy:2022}, Clumpfind \citep{williams94}, Heisenberg \citep{kruijssen18}, Matplotlib \citep{Hunter07}, SciPy \citep{scipy}, seaborn \citep{seaborn}, pandas \citep{pandas}}



\appendix

\section{Full sample three color images}\label{app:images}

From Figures~\ref{fig:obs_1300} to \ref{fig:obs_5068}, we show three color composite images as well as distributions of identified emission peaks across our galaxy sample. We exclude NGC\,0628 here, which is shown in Figure~\ref{fig:obs}.

\begin{figure*}
\includegraphics[width=\textwidth]{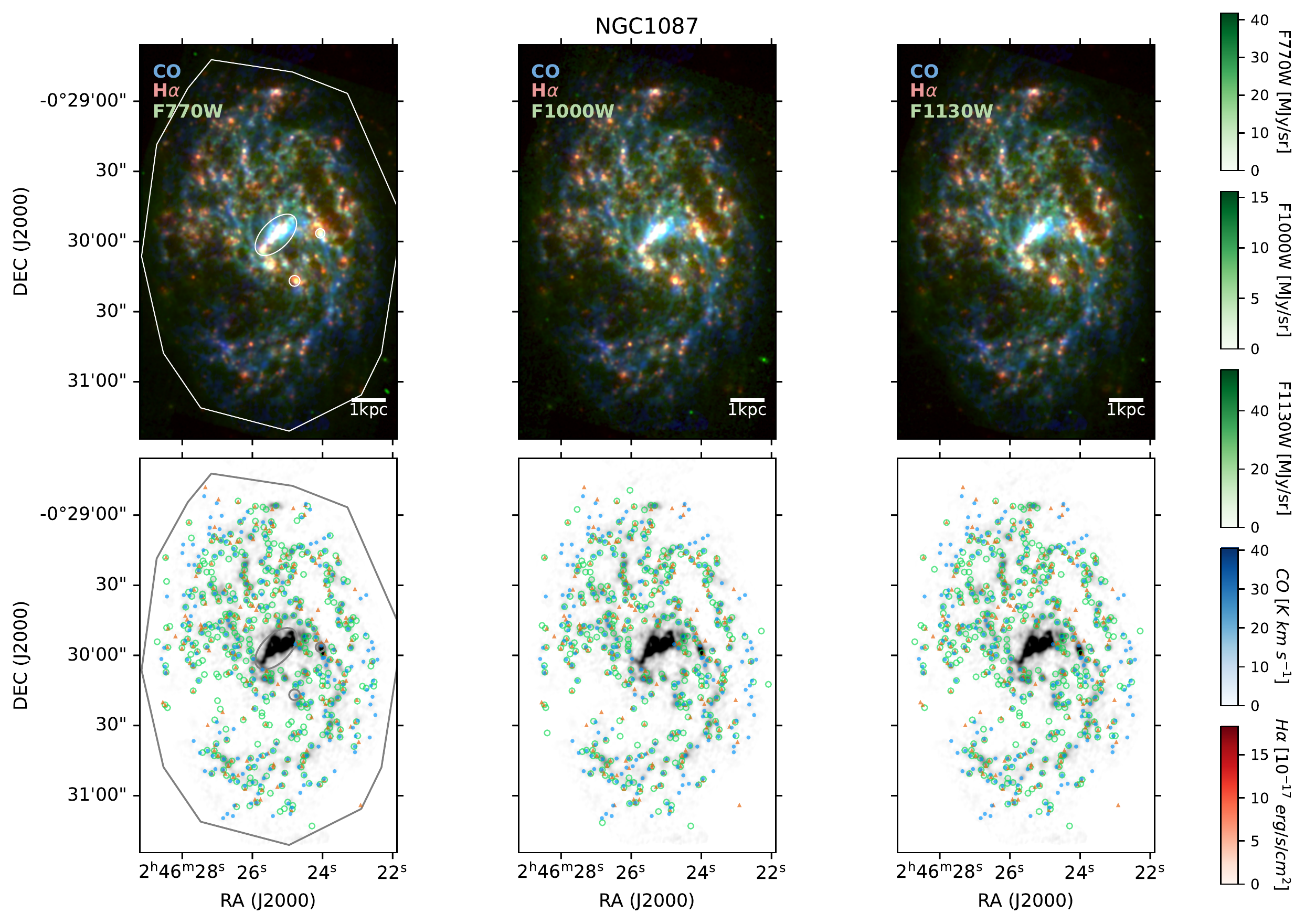}
\caption{\textit{Top:} Composite three color images of NGC\,1087 created using CO (blue), H$\alpha$ (red), and mid-IR (green) observations. Each panel, from left to right, represents mid-IR emission of F770W, F1000W, and F1130W, respectively. The mid-IR observations have been convolved and regridded to match the coarser resolution and pixel grid of of H$\alpha$ observations (see Section~\ref{ssec:post}). For visualization purposes, a power-law brightness scale with gamma correction ($\gamma=2$) has been applied in the top panels. The color bars on the right reflect the true flux ranges in each observation. The left panel highlights the area analyzed, outlined by a polygon. It excludes the crowded galaxy center (ellipse) as well as artifacts and exceptionally bright peaks (circles). \textit{Bottom:} Locations of identified H$\alpha$ (orange triangles), CO (blue filled circles), and mid-IR (green open circles) emission peaks (see Section~\ref{sec:method}) are overlaid on CO map, shown in grayscale with a linear brightness scale. Again, from left to right, the mid-IR emission peaks correspond to peaks identified in F770W, F1000W, and F1130W maps respectively. The CO grayscale image uses the same intensity range as the CO emission in the top panels shown in blue, with range of flux indicated in the color bar on the right.} \label{fig:obs_1087}
\end{figure*}

\begin{figure*}
\includegraphics[width=\textwidth]{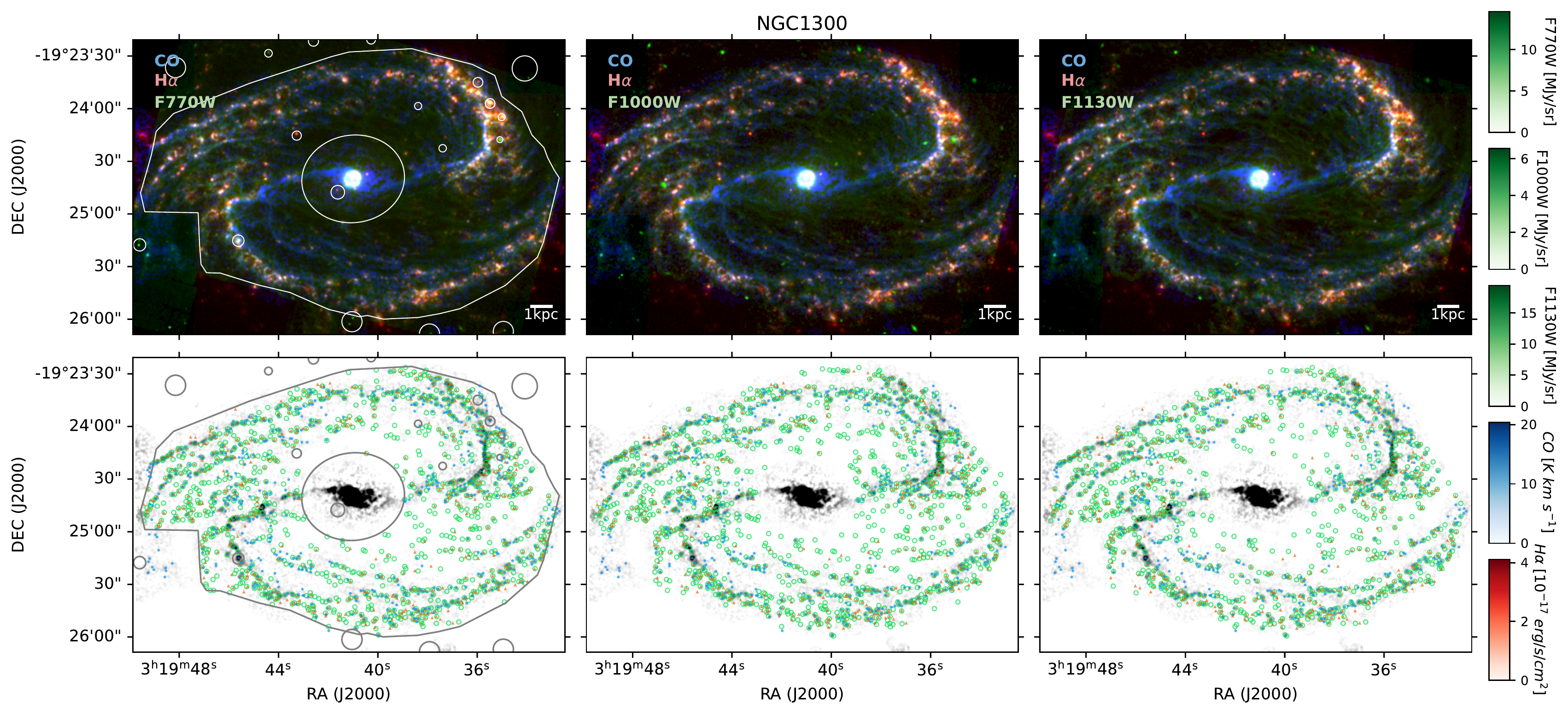}
\caption{Similar to Figure~\ref{fig:obs_1087}, showing observations and identified emission peaks of NGC\,1300.} \label{fig:obs_1300}
\end{figure*}

\begin{figure*}
\includegraphics[width=\textwidth]{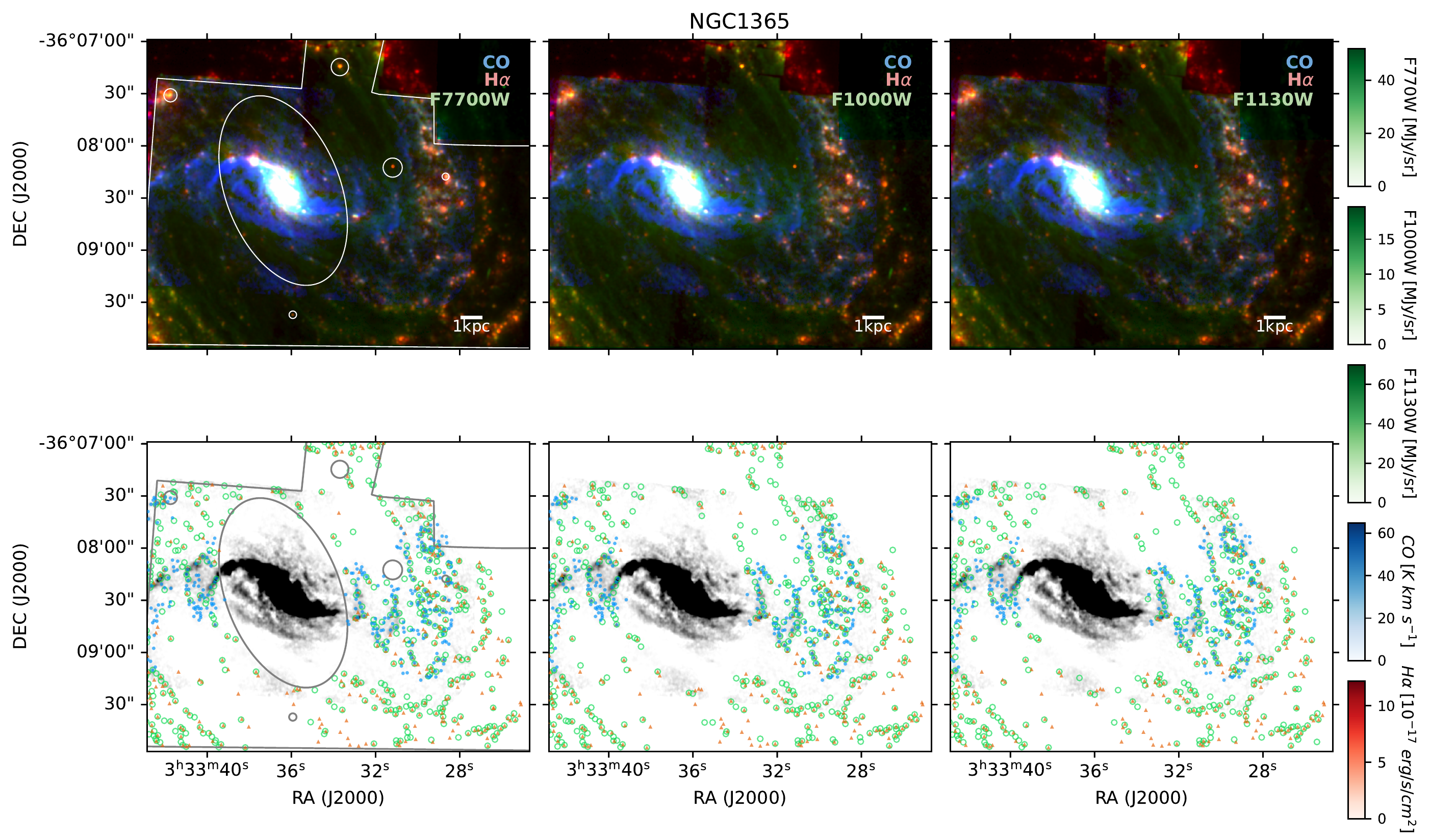}
\caption{Similar to Figure~\ref{fig:obs_1087}, showing observations and identified emission peaks of NGC\,1365. We note that for this galaxy the CO observation has a smaller FoV. } \label{fig:obs_1365}
\end{figure*}

\begin{figure*}
\includegraphics[width=\textwidth]{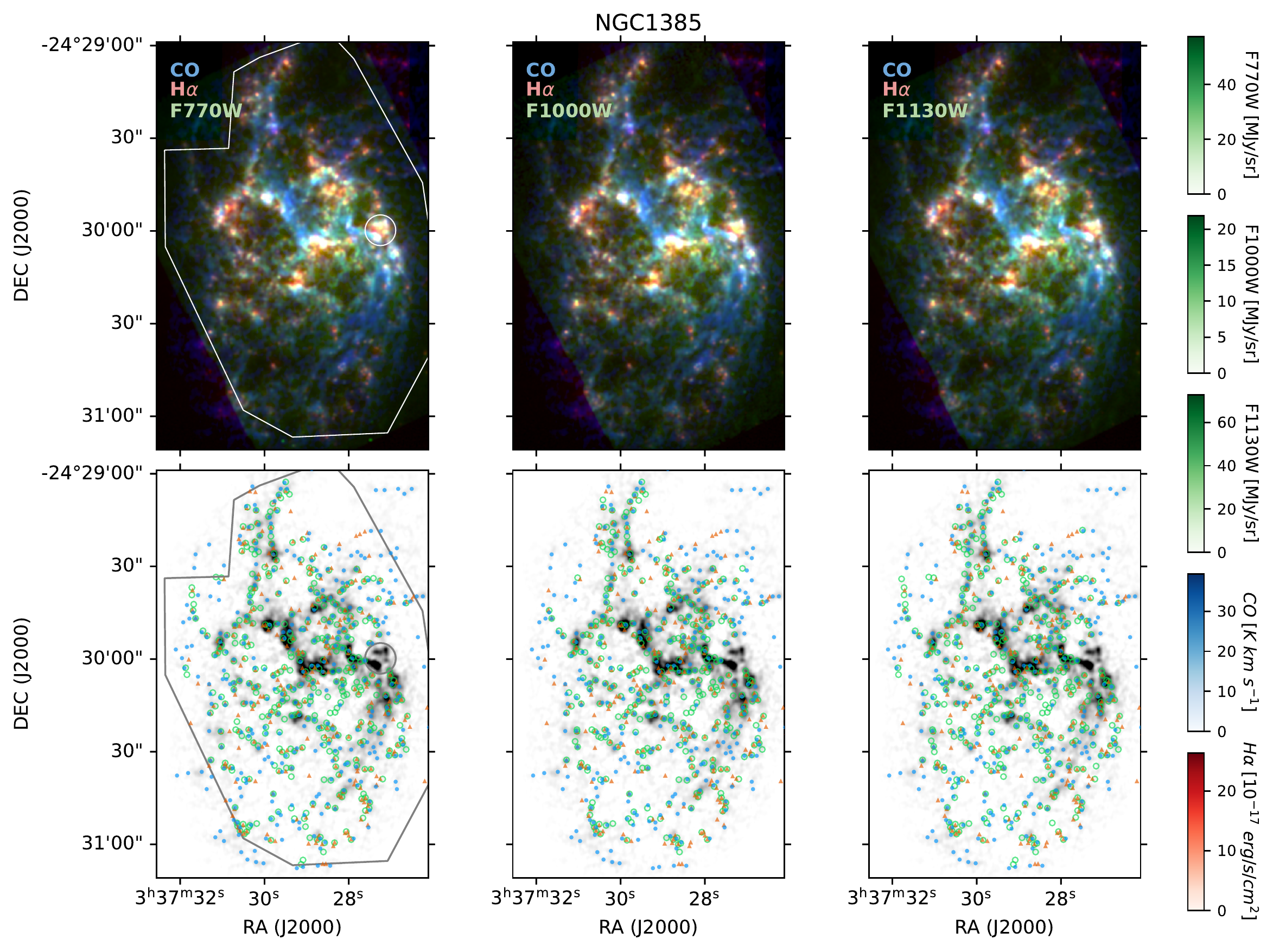}
\caption{Similar to Figure~\ref{fig:obs_1087}, showing observations and identified emission peaks of NGC\,1385.} \label{fig:obs_others}
\end{figure*}

\begin{figure*}
\includegraphics[width=\textwidth]{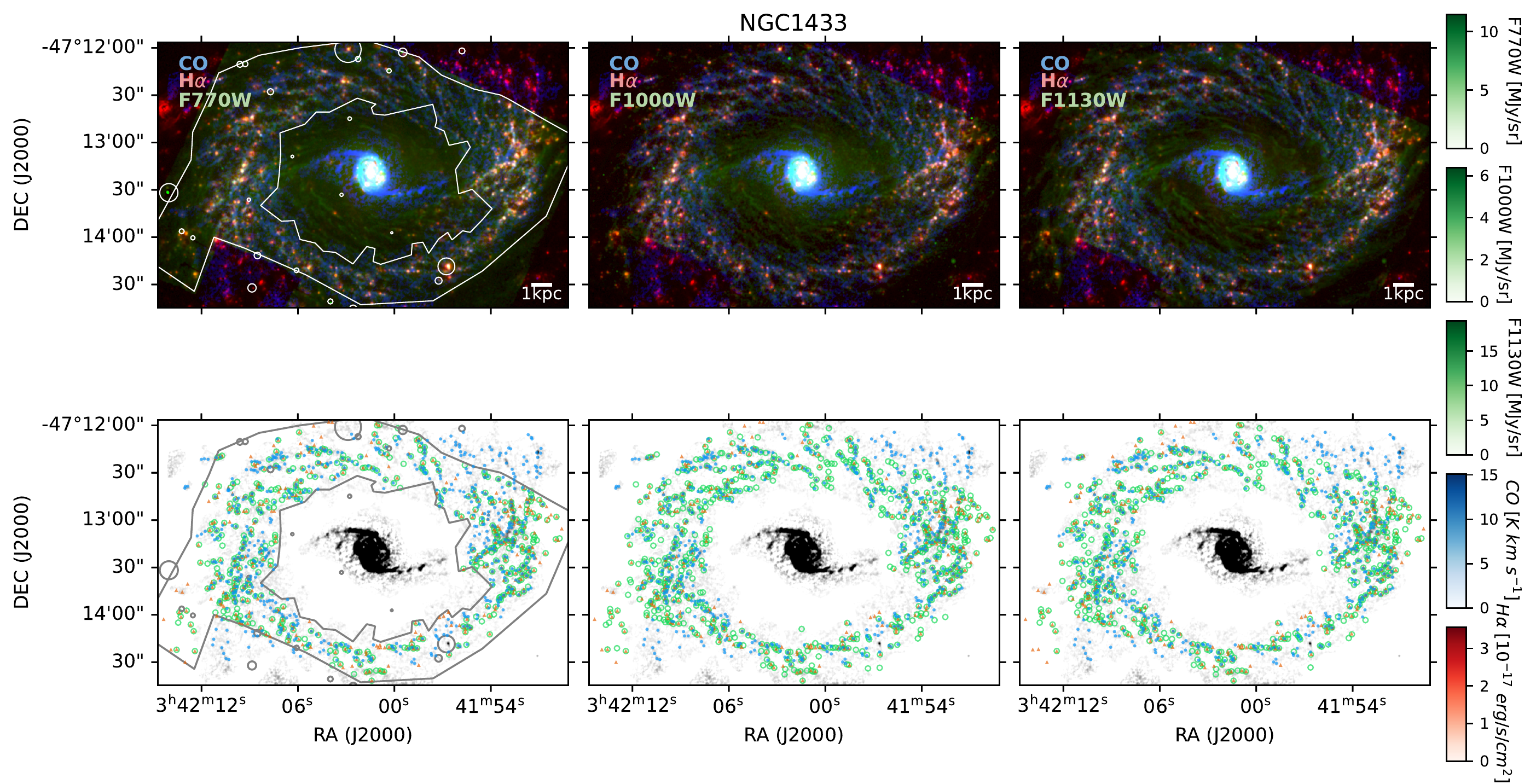}
\caption{Similar to Figure~\ref{fig:obs_1087}, showing observations and identified emission peaks of NGC\,1433.} \label{fig:obs_others}
\end{figure*}

\begin{figure*}
\includegraphics[width=\textwidth]{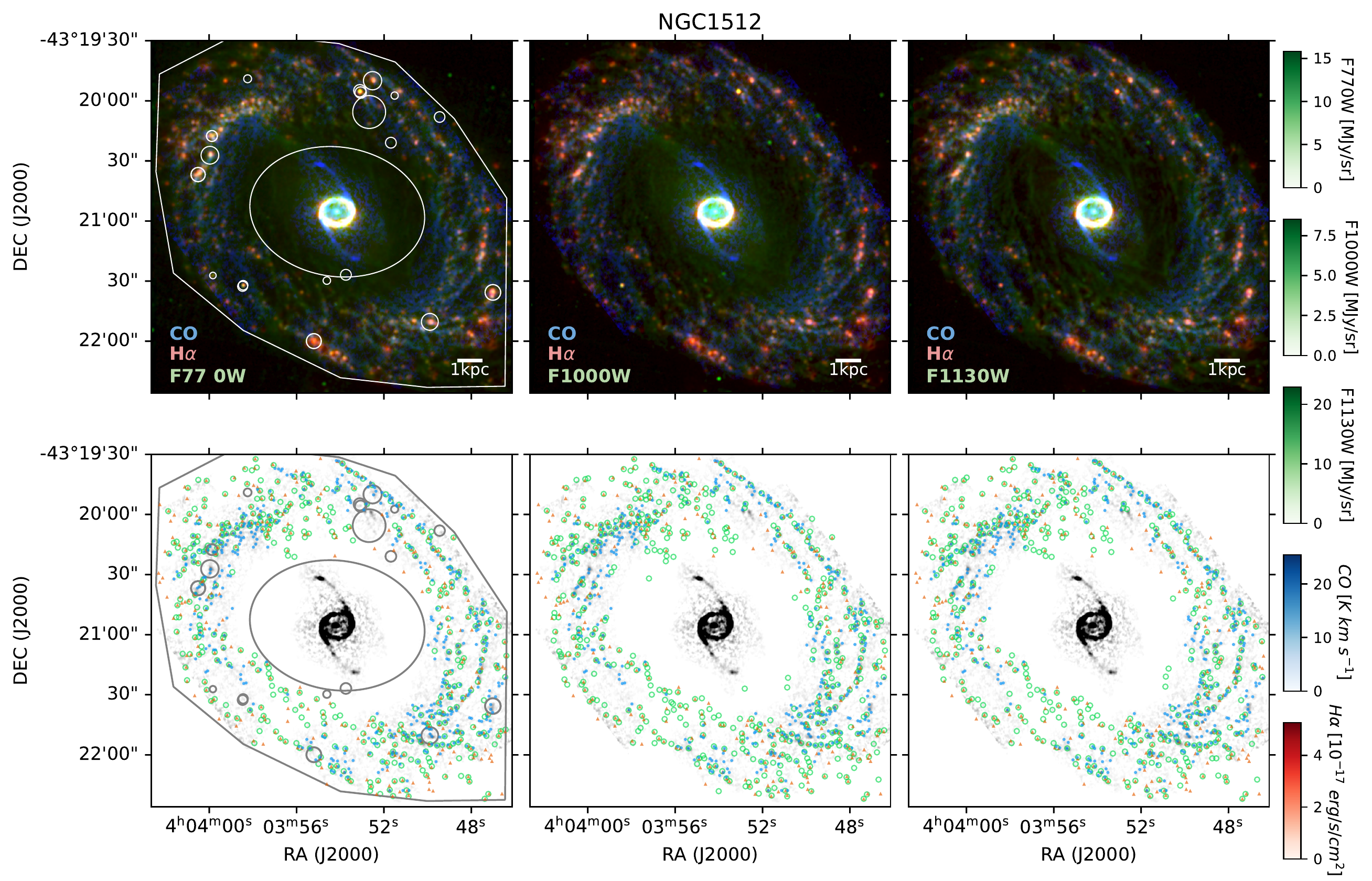}
\caption{Similar to Figure~\ref{fig:obs_1087}, showing observations and identified emission peaks of NGC\,1512.} \label{fig:obs_others}
\end{figure*}

\begin{figure*}
\includegraphics[width=\textwidth]{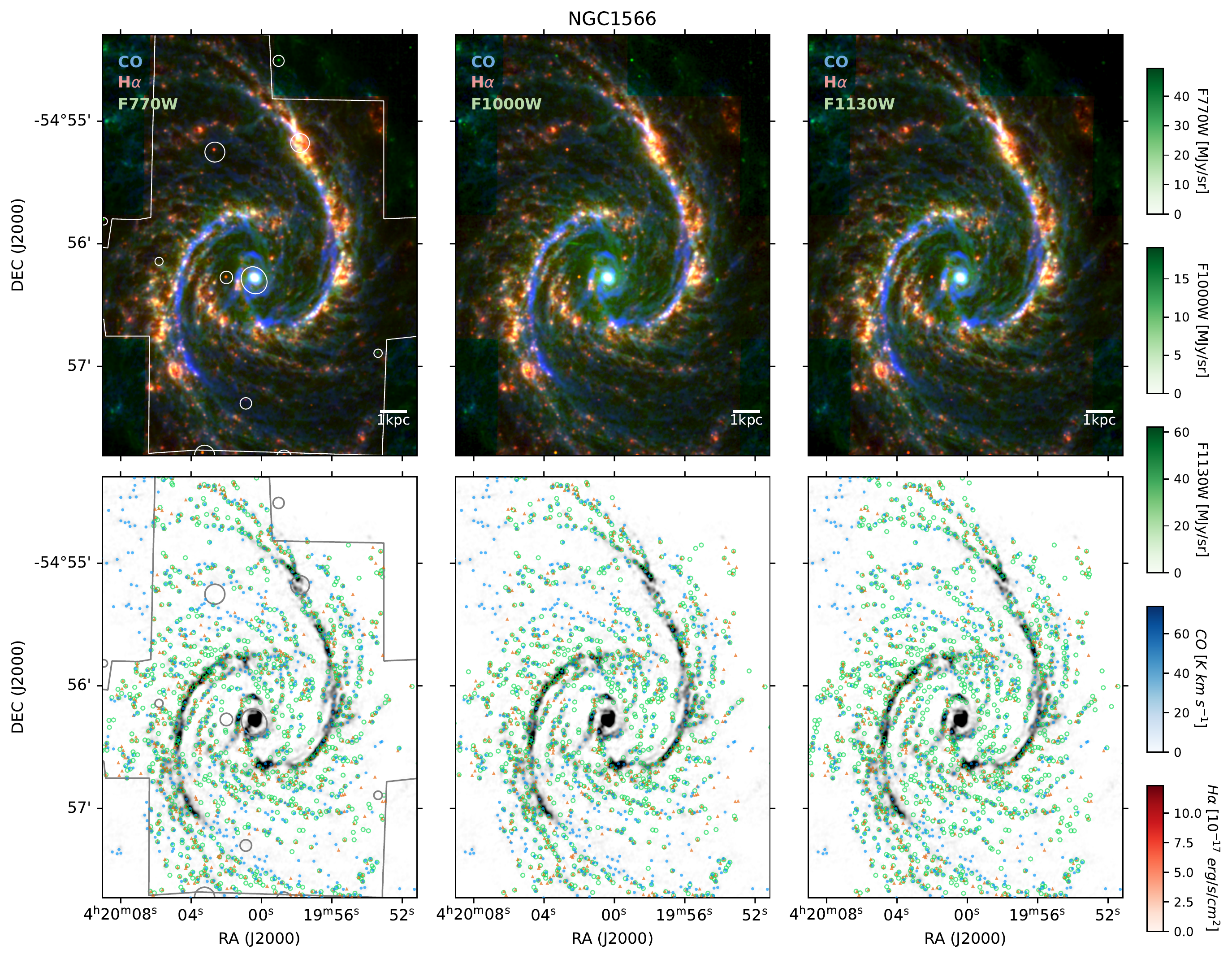}
\caption{Similar to Figure~\ref{fig:obs_1087}, showing observations and identified emission peaks of NGC\,1566.} \label{fig:obs_others}
\end{figure*}

\begin{figure*}
\includegraphics[width=\textwidth]{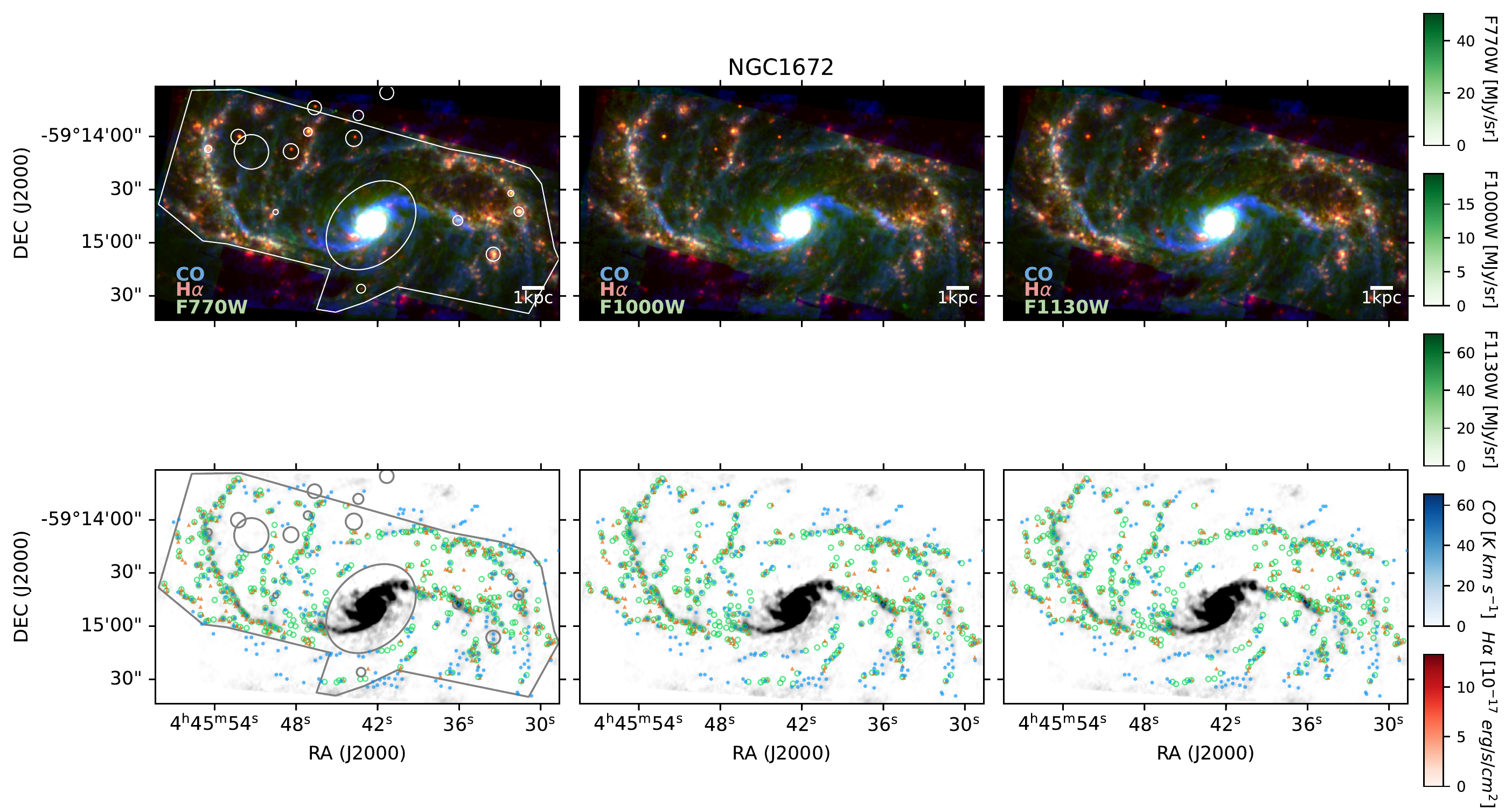}
\caption{Similar to Figure~\ref{fig:obs_1087}, showing observations and identified emission peaks of NGC\,1672.} \label{fig:obs_others}
\end{figure*}

\begin{figure*}
\includegraphics[width=\textwidth]{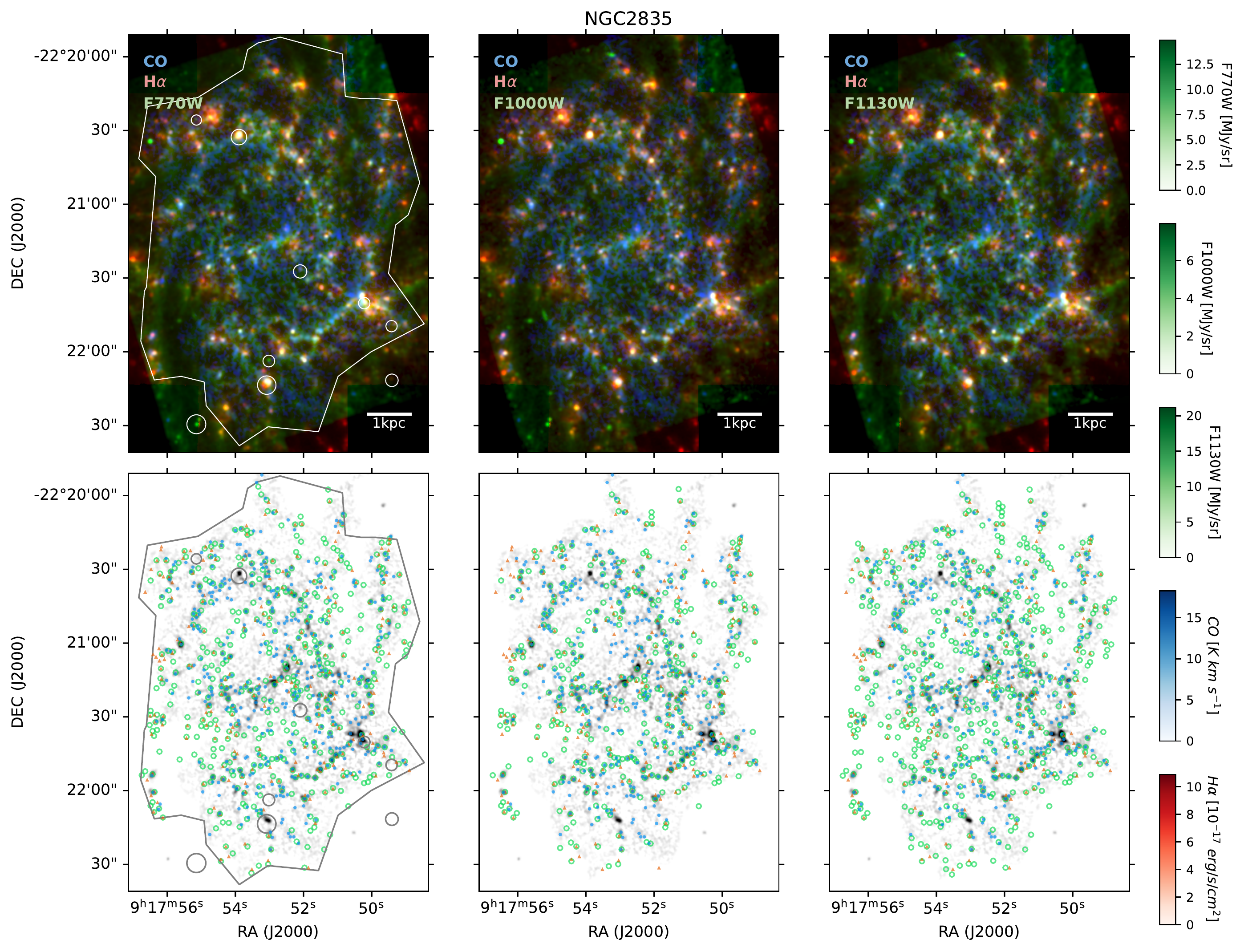}
\caption{Similar to Figure~\ref{fig:obs_1087}, showing observations and identified emission peaks of NGC\,2835.} \label{fig:obs_others}
\end{figure*}

\begin{figure*}
\includegraphics[width=\textwidth]{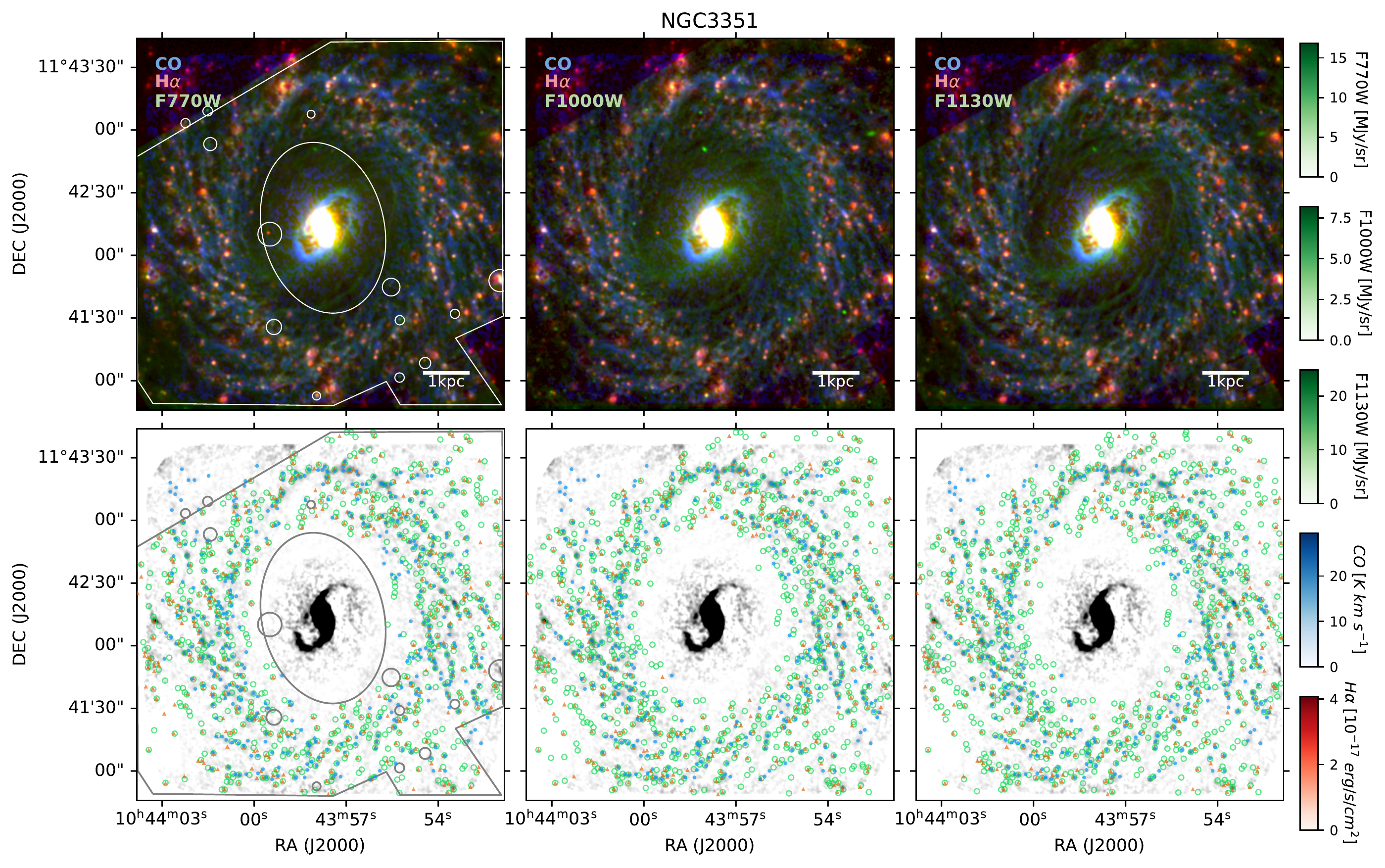}
\caption{Similar to Figure~\ref{fig:obs_1087}, showing observations and identified emission peaks of NGC\,3351.} \label{fig:obs_others}
\end{figure*}

\begin{figure*}
\includegraphics[width=\textwidth]{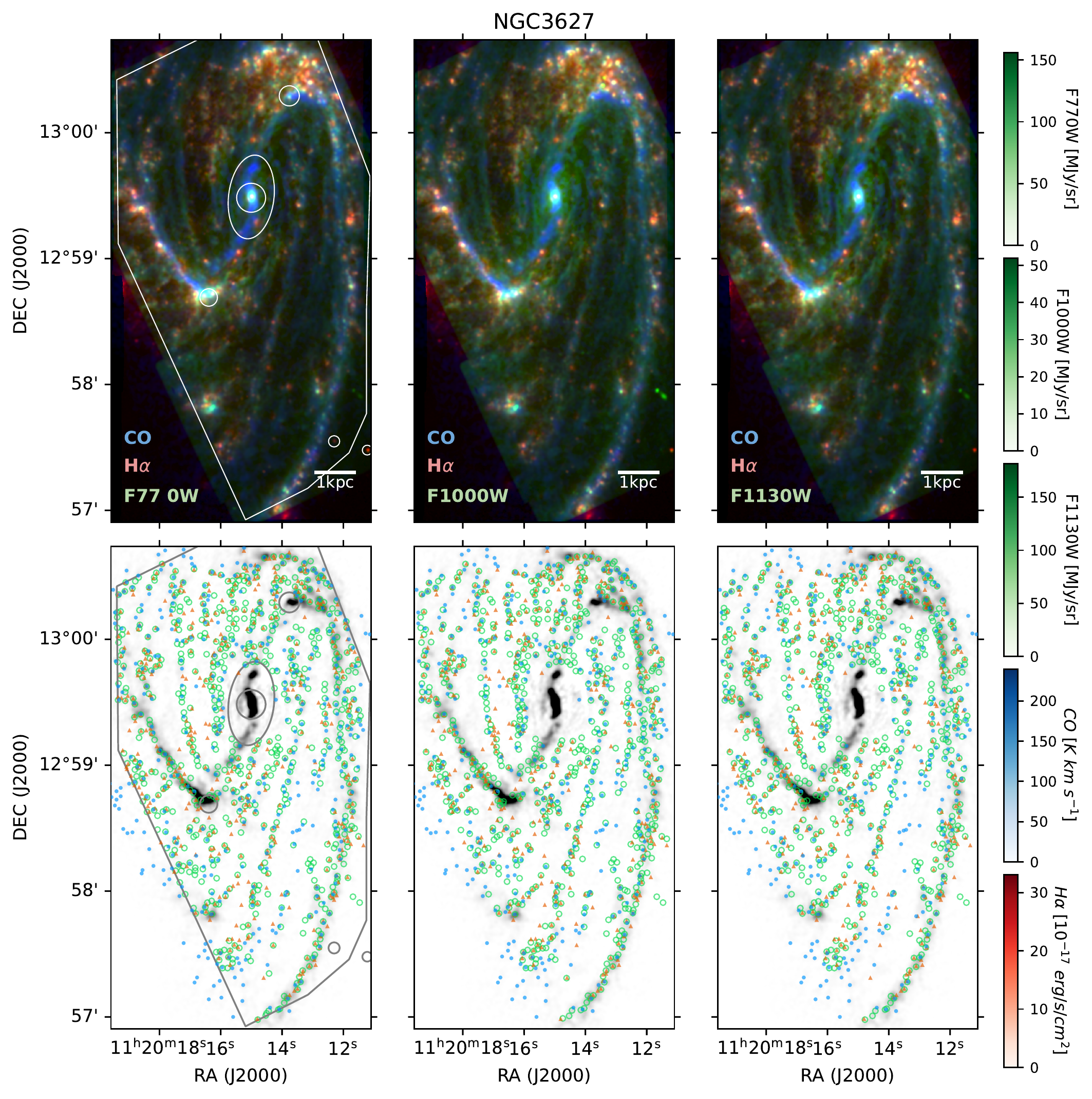}
\caption{Similar to Figure~\ref{fig:obs_1087}, showing observations and identified emission peaks of NGC\,3627.} \label{fig:obs_others}
\end{figure*}

\begin{figure*}
\includegraphics[width=\textwidth]{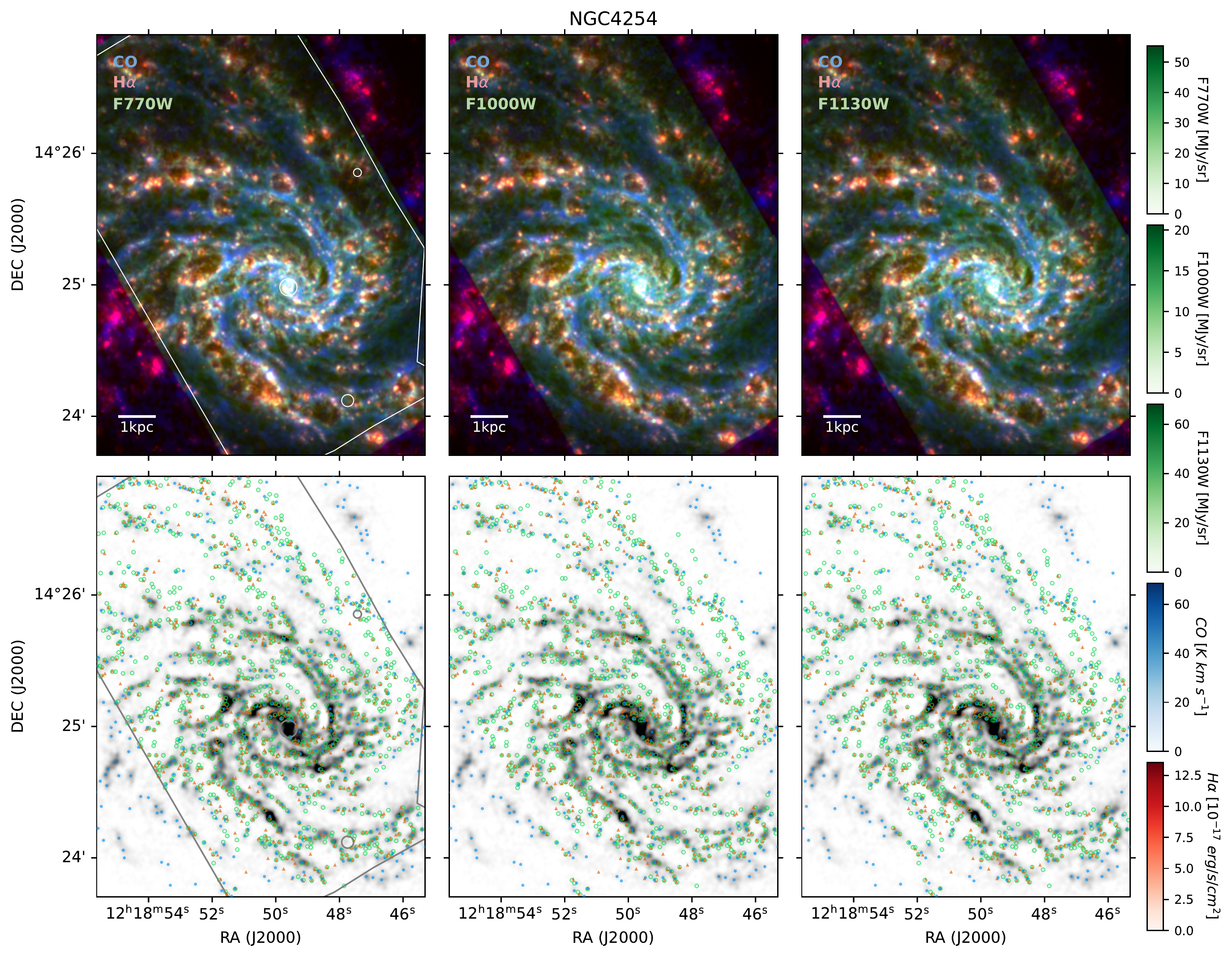}
\caption{Similar to Figure~\ref{fig:obs_1087}, showing observations and identified emission peaks of NGC\,4254.} \label{fig:obs_others}
\end{figure*}

\begin{figure*}
\includegraphics[width=\textwidth]{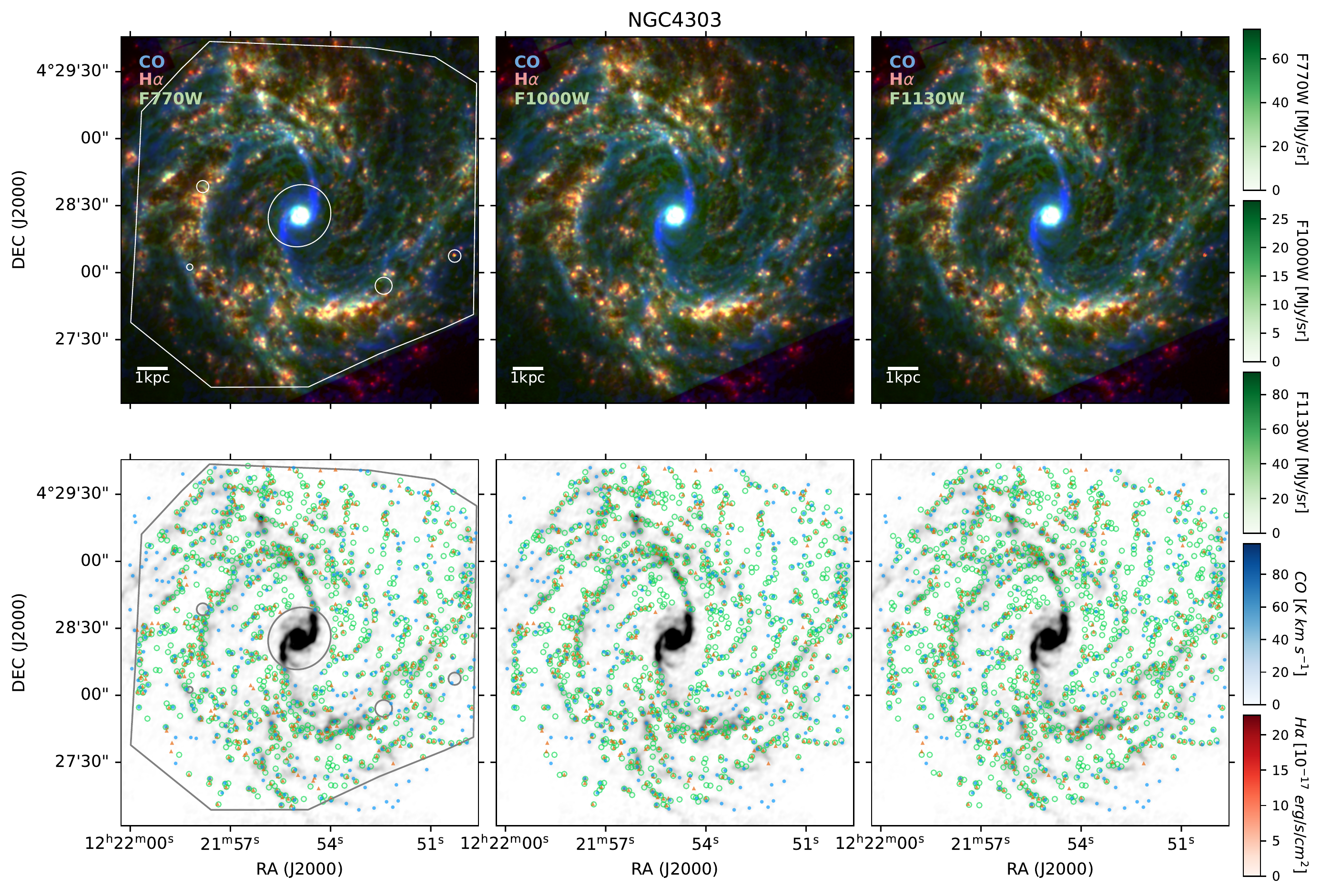}
\caption{Similar to Figure~\ref{fig:obs_1087}, showing observations and identified emission peaks of NGC\,4303.} \label{fig:obs_others}
\end{figure*}

\begin{figure*}
\includegraphics[width=\textwidth]{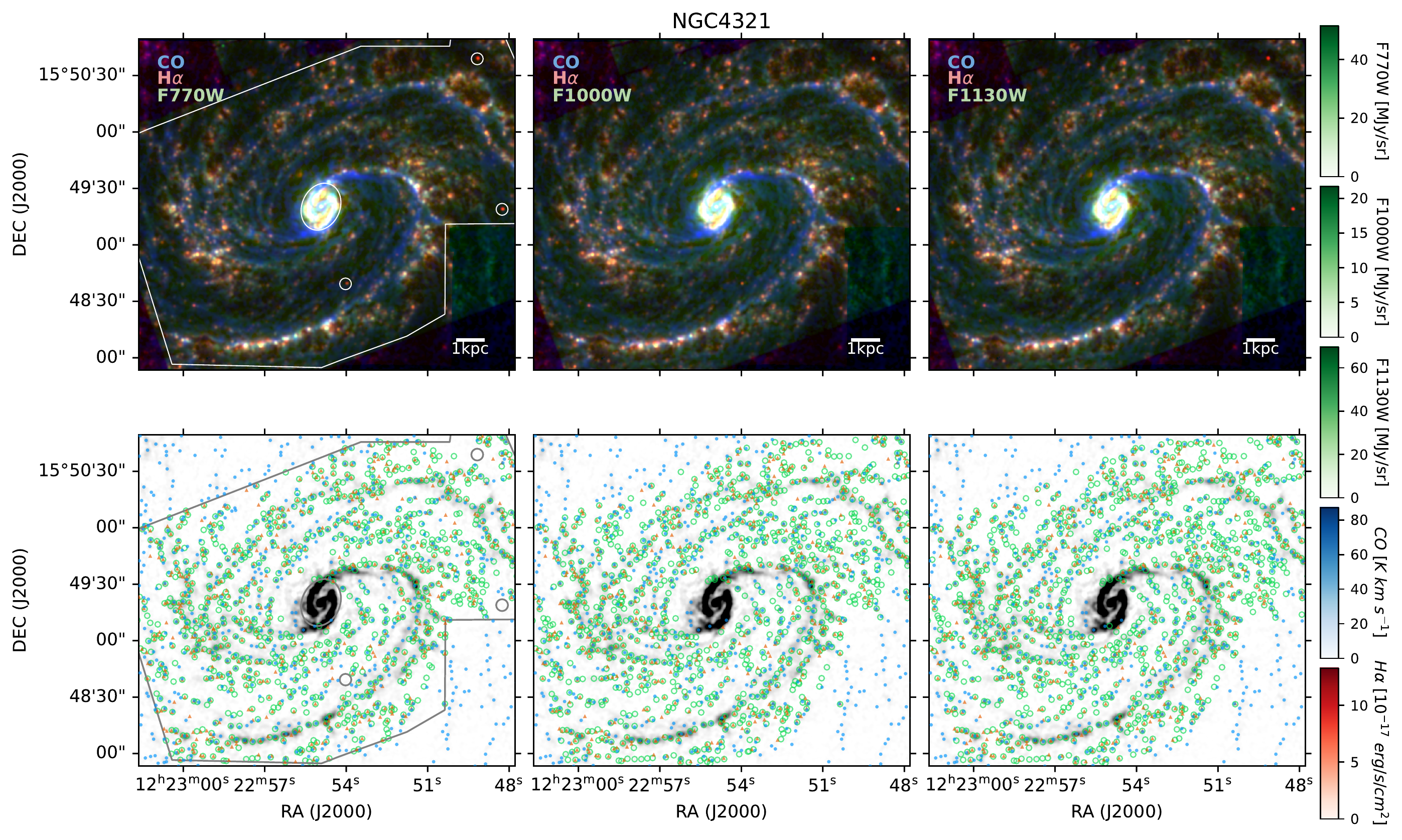}
\caption{Similar to Figure~\ref{fig:obs_1087}, showing observations and identified emission peaks of NGC\,4321.} \label{fig:obs_others}
\end{figure*}

\begin{figure*}
\includegraphics[width=\textwidth]{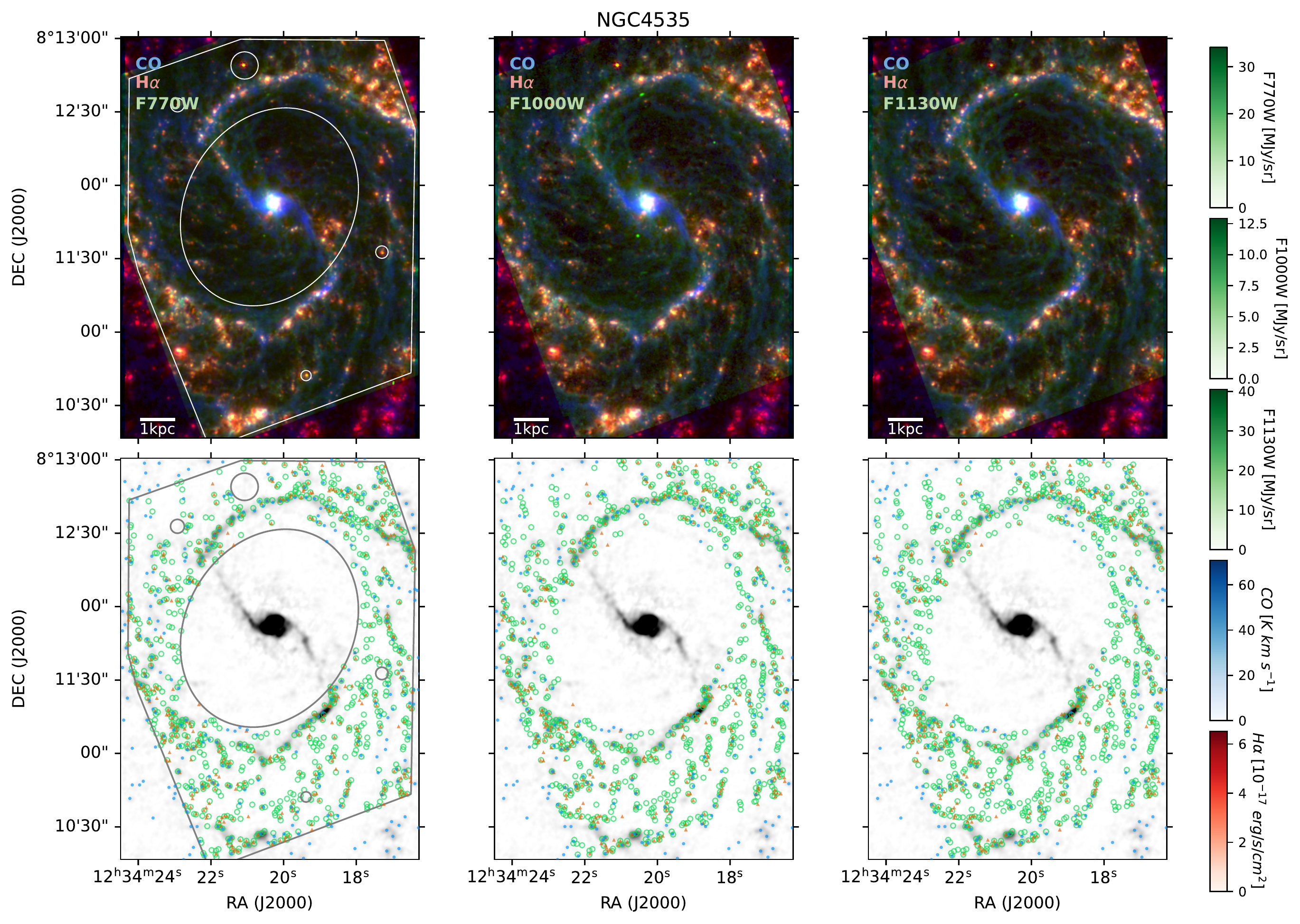}
\caption{Similar to Figure~\ref{fig:obs_1087}, showing observations and identified emission peaks of NGC\,4535.} \label{fig:obs_others}
\end{figure*}

\begin{figure*}
\includegraphics[width=\textwidth]{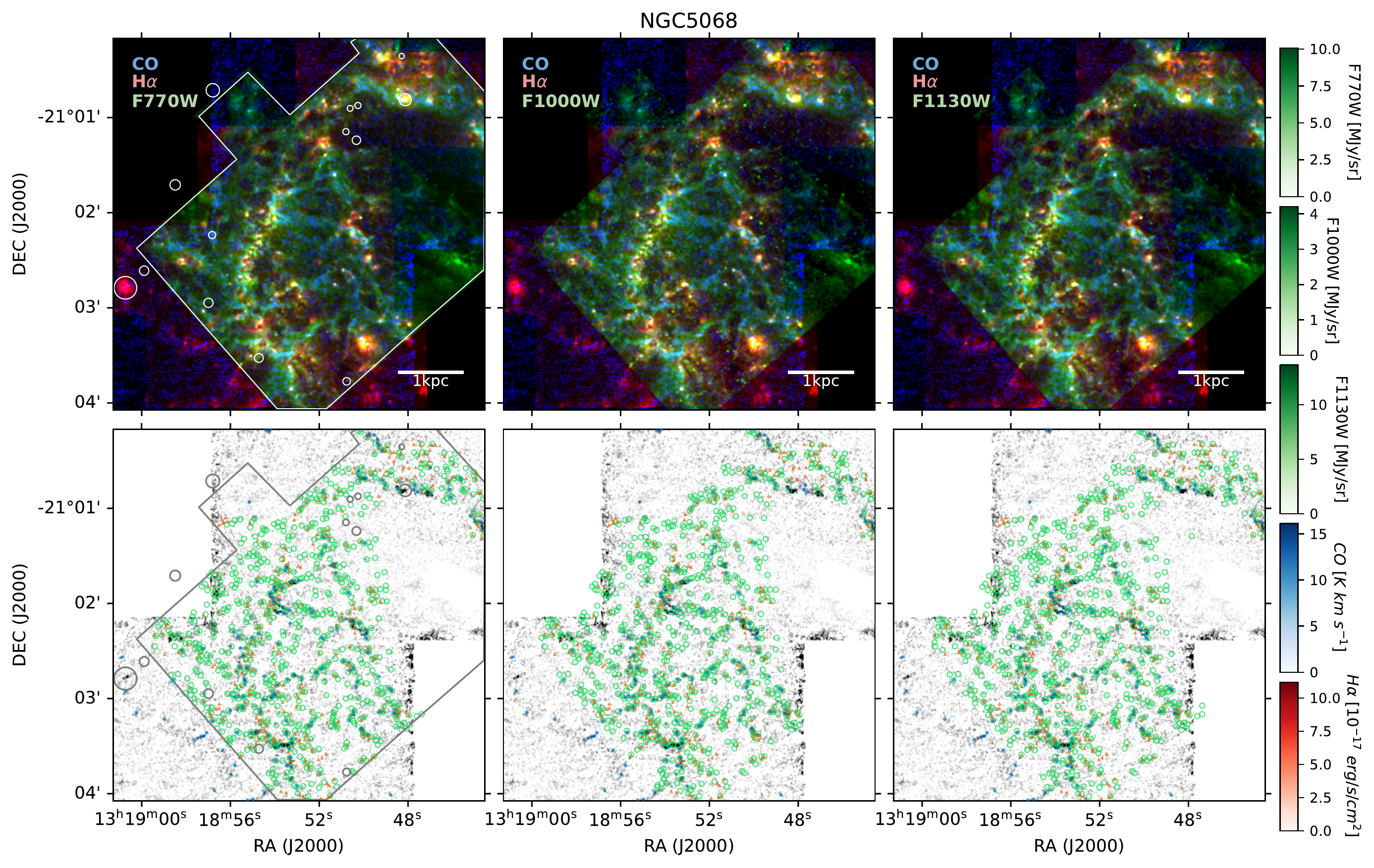}
\caption{Similar to Figure~\ref{fig:obs_1087}, showing observations and identified emission peaks of NGC\,5068.} \label{fig:obs_5068}
\end{figure*}

\section{Accuracy of our measurements}\label{app:robust}
\citet[][Section 4.4]{kruijssen18} lists a set of requirements that our measurements have to satisfy in order to be considered accurate within 30\% error. We verify whether these are fulfilled for our constrained parameters ($t_{\rm g}$, $t_{\rm fb}$, and $\lambda$). 

\begin{figure}
\includegraphics[scale=0.9]{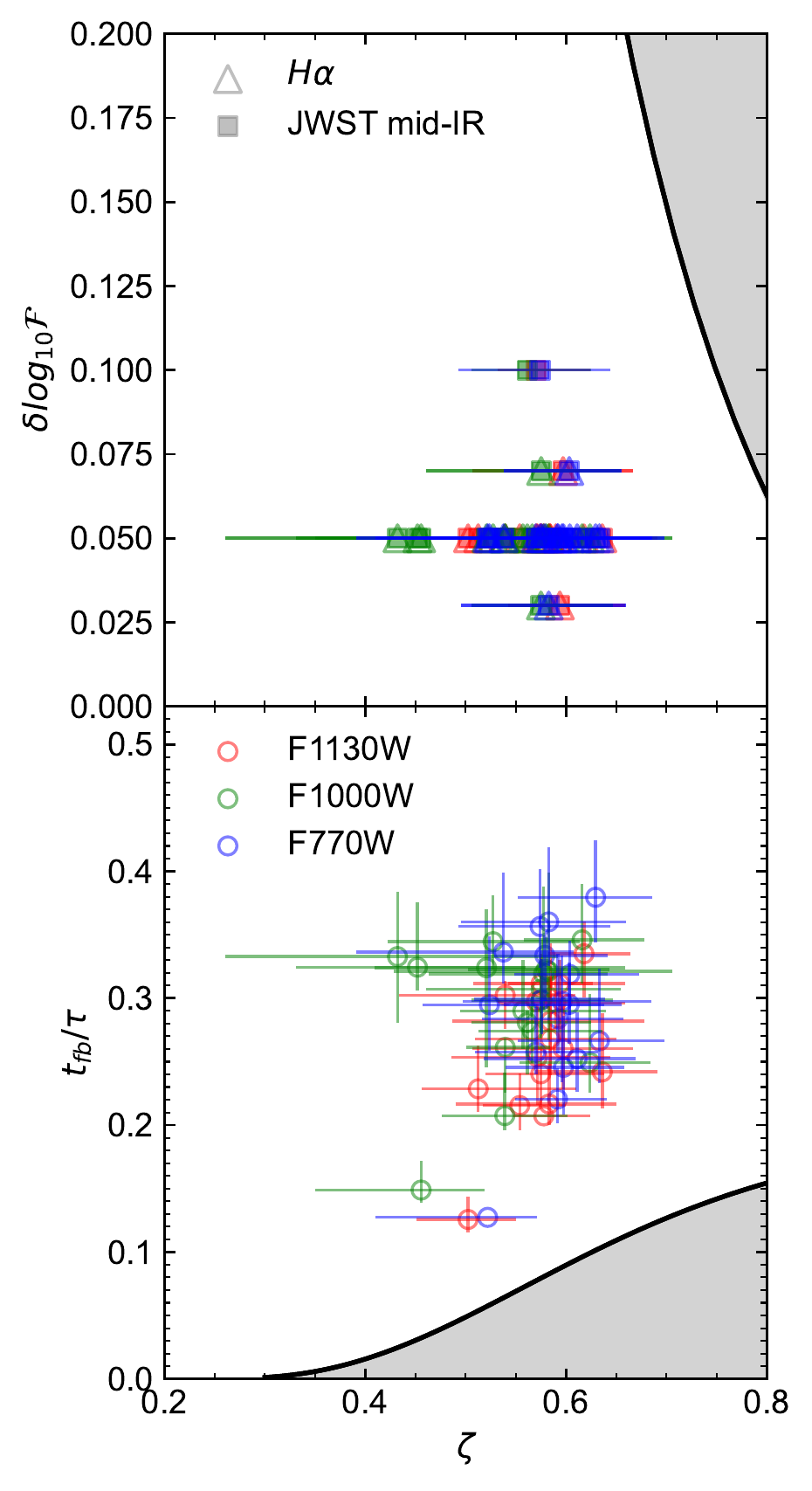}
\caption{\textit{Top:} The flux contrast ($\rm{\delta}log_{10}\mathcal{F}$) adopted during the peak identification for mid-IR (square) and H$\alpha$ (triangles) observations shown as a function of the average filling factor ($\zeta$). Across our analysis with each mid-IR band, shown as 7.7\,$\mu$m (blue), 10\,$\mu$m (green), and 11.3\,$\mu$m (red), the adopted flux contrast is well below the upper limit (gray region). \textit{Bottom}: Measured $t_{\rm fb}/\tau$ as a function of $\zeta$. In both panels, the gray region indicates the parameter space where the high filling factor biases our measurements of the $t_{\rm fb}$ \citep{kruijssen18}. } \label{fig:blending}
\end{figure}

\begin{enumerate}
    \item The mid-IR emitting time-scale ($t_{\rm g}$) and the H$\alpha$ emitting time-scale ($t_{\rm s}$) should differ by no more than one order of magnitude. This is satisfied in all galaxies across all mid-IR bands with $\lvert \log(t_{\rm g}/t_{\rm s})\rvert<0.77$.
    \item Individual star-forming regions should be sufficiently resolved with $\lambda>1.5\,l_{\rm ap, min}$. Across all mid-IR bands, this is not satisfied in NGC\,1365, NGC\,1512, and NGC\,4321. Only in the analysis with 7.7\,$\mu$m, two additional galaxies (NGC\,1300  NGC\,2835) suffer from insufficient resolution. In these galaxies, the ratio  $\lambda/l_{\rm ap, min}$ ranges from 1.3 to 1.4, indicating that only $t_{\rm g}$ is robust while $t_{\rm fb}$ and $\lambda$ are upper limits.
    \item We verify that the number of identified emission peaks in both mid-IR and H$\alpha$ maps is above 35 (see also Figure~\ref{fig:obs}, as well as Figures \ref{fig:obs_1087} to \ref{fig:obs_5068}). 
    \item The measured mid-IR-to-H$\alpha$ flux ratio on small-scale with apertures placed on mid-IR (H$\alpha$) peaks should not be below (above) the galactic average value. This is satisfied as shown in Figure~\ref{fig:tuningforks}. 
    \item The global star formation rate of the analyzed region should not fluctuate by more than 0.2\,$dex$ during the last evolutionary cycle ($\tau=t_{\rm g}+t_{\rm s}-t_{\rm fb}$), when averaged over a time period of $t_{\rm g}$ or $t_{\rm s}$. Using the SFR history measured by \citet{pessa23}, we verify that this condition is satisfied. 
    \item Regions undergoing evolution from gas to stars should be detected in both the mid-IR and the H$\alpha$ map at some point of their lifecycle. This requirement can be verified by examining whether the sensitivity in mid-IR observations matches that of H$\alpha$ observations. According to \citet{leroy23}, the 1$\sigma$ noise at 1.\arcsec7 of JWST mid-IR emission ranges from 0.2 to 0.5\,$M_{\odot}\rm pc^{-2}$, obtained using a median CO-to-mid-IR ratio and a typical CO-to-$\rm H_{2}$ conversion factor. We first estimate the minimum gas cloud mass corresponding to a typical 5$\sigma$ sensitivity by multiplying the representative 1$\sigma$ noise (0.3\,$M_{\odot}\rm pc^{-2}$ across mid-IR bands) by 5 and the area corresponding to the size of 1.\arcsec7 at a typical distance of 15\,Mpc. This yields a 5$\sigma$ detection threshold corresponding to a gas cloud mass of $\sim 1.8\times10^4\,M_{\odot}$. We then multiply this by the mean star formation efficiency per cloud lifecycle as constrained within our method (3\%; \citealp{kim22}) to obtain the expected minimum star-forming region mass ($\sim 500\,M_{\odot}$) from mid-IR observations. The minimum mass is compared to the minimum mass of the stellar population required to generate the typical 5$\sigma$ sensitivity of the H$\alpha$ map ($\sim 8\times 10^{37}\,{\rm erg\,s^{-1}kpc}^{-2}$) on the scales of star-forming regions ($\lambda\sim 120$\,pc). We obtain the initial mass using the \textsc{Starburst99} model with an assumption that star formation was instantaneous 5\,Myr ago. We find that the minimum mass from mid-IR ($\sim 500\,M_{\odot}$) matches well with that from H$\alpha$ ($\sim 600\,M_{\odot}$). 

    \item When regions are crowded, a small enough flux contrast ($\rm{\delta}log_{10}\mathcal{F}$) needs to be adopted during the peak identification process to distinguish adjacent peaks. \citet{kruijssen18} provides upper limits for $\rm{\delta}log_{10}\mathcal{F}$ as a function of $\zeta$, which represents the filling factor of emission peaks. The $\zeta$ is defined as $2r/\lambda$, where the $r$ is the mean radius of gas or SFR tracer peaks. We obtain the average $\zeta$ by weighting the $\zeta$ for each gas and SFR tracer with their associated time-scales. Figure~\ref{fig:blending} shows that the adopted $\rm{\delta}log_{10}\mathcal{F}$ is well below the upper limit. 

    \item Galaxies with a high filling factor ($\zeta$) might be incorrectly constrained to have a longer feedback time-scale. In Figure~\ref{fig:blending}, we compare the ratio between the feedback time-scale ($t_{\rm fb}$) and the total duration of the evolutionary cycle ($\tau=t_{\rm g}+t_{\rm s}-t_{\rm fb}$) as a function of $\zeta$ to the analytical predictions of \citet{kruijssen18},outlining the region where the constrained $t_{\rm fb}$ should be considered to be an upper limit due to the high filling factor. We find that none of our measurements fall into this region. 

    \item As shown in the bottom panel of Figure~\ref{fig:blending}, we ensure that the $t_{\rm fb}/\tau$ is above 0.05 and below 0.95 in all of our measurements.

    \item Similarly to requirement (5) above, during the last evolutionary cycle ($\tau$), the SFR should not vary by more than 0.2${\rm dex}$ when averaged over a time period of $t_{\rm fb}$. This is satisfied using the same argument as above.

    \item After masking the dense central regions of the galaxy, we ensure that visual inspection does not identify areas with substantial blending.

\end{enumerate}

To summarize, we find that almost all of the the requirements are satisfied except for the condition (2). This is not satisfied for NGC\,1365, NGC\,1512, and NGC\,4321 across all three mid-IR bands. NGC\,1300 and NGC2835 also fail to satisfy this condition for our measurements with 7.7\,$\mu$m as the gas tracer. This implies that the measured $t_{\rm fb}$ and $\lambda$ are upper limits for this subset of galaxies as indicated in Table~\ref{tab:result}. 

\section{Impact of masking extremely bright peaks}\label{app:masking}

Our method uses flux averages and its distributions to determine the mid-IR-to-H$\alpha$ flux ratios and their associated uncertainties. Therefore, exceedingly bright emission peaks that does not represent the overall cloud and \textsc{Hii} region population can potentially bias our measured time-scales. For example, in the LMC, when 30 Doradus is included in the analysis, constituting $\sim$20\% of the total H$\alpha$ emission, \citet{ward22} have found that the 1$\sigma$ uncertainty of the derived molecular gas cloud lifetime almost doubles from $t_{\rm g}^{\rm CO}=11.8_{-2.2}^{+2.7}$\,Myr to $12.8_{-3.6}^{+5.1}$\,Myr. \citet{chevance20} have shown that masking a very bright molecular cloud (referred to as the `headlight' cloud in \citealp{herrera20}) makes small differences in derived time-scales when obtained averaging over the entire galaxy. However, the difference becomes significant when focusing on smaller regions within the galaxy, which makes the headlight cloud to dominate more than 10\% of the total CO flux. Then, the measured molecular gas cloud lifetime changes from $t_{\rm g}^{\rm CO}=16_{-2.8}^{+4.3}$\,Myr with mask to $>25$\,Myr without masking the `headlight' cloud, when the analysis is performed on a radial bin with a width of 1\,kpc \citep{chevance20}.

In Table~\ref{tab:mask}, we show how the masking of bright peaks changes our time-scale measurements and the fraction of flux enclosed in the mask relative to the total emission in the galaxy disk with centers excluded. We show comparisons for all the galaxies that required additional masking of peaks compared to the masks adopted in \citet{kim22}. We find that our measurements with and without bright peaks masked agree within 1$\sigma$ uncertainties. The masking also does not seem to affect uncertainties, most likely because the bright peaks do not dominate the total flux as much as 30 Doradus (20\% in H$\alpha$ flux) and our values are obtained averaging over the whole galaxy.

\begin{deluxetable*}{lcccccccccc}
\tablewidth{0pt} 
\tablecaption{Comparison of our measurements of mid-IR emitting time-scale ($t_{\rm g}^{\rm 7.7\mu m}$), feedback time-scale ($t_{\rm fb}^{\rm 7.7\mu m}$), and region separation length ($\lambda^{\rm 7.7\mu m}$), with and without masking exceedingly bright emission peaks. $F_{\rm mask}$ shows the fraction of masked flux (in each 7.7\,$\mu$m and H$\alpha$ observation) relative to the total flux in the galaxy disk, excluding the central region.\label{tab:mask}}
\tablehead{&\multicolumn{3}{c}{With exceedingly bright peaks masked}&& \multicolumn{3}{c}{Without masking}&&\multicolumn{2}{c}{$F_{\rm mask}$} \\
\cline{2-4}  \cline{6-8} \cline{10-11}\\
\colhead{Galaxy} & \colhead{$t_{\rm g}^{7.7\mu m}$} & \colhead{$t_{\rm fb}^{7.7\mu m}$} & \colhead{$\lambda^{7.7\mu m}$}&  &\colhead{$t_{\rm g}^{7.7\mu m}$} & \colhead{$t_{\rm fb}^{7.7\mu m}$} & \colhead{$\lambda^{7.7\mu m}$}& &\colhead{7.7\,$\mu$m} & H$\alpha$\\
\colhead{} & \colhead{[Myr]} & \colhead{[Myr]} & \colhead{[pc]}& &\colhead{[Myr]} & \colhead{[Myr]} & \colhead{[pc]} & &\colhead{[\%]}&\colhead{[\%]}\\
}
\startdata 
NGC1087&$21.5_{-3.4}^{+2.3}$&$6.6_{-1.1}^{+1.1}$&$125_{-17}^{+23}$&&$20.5_{-4.0}^{+3.1}$&$6.4_{-1.1}^{+1.2}$&$122_{-23}^{+86}$&&2.6&5.3\\
NGC1300&$19.7_{-1.7}^{+1.8}$&$<7.4$&$<150$&&$18.1_{-1.7}^{+2.2}$&$<7.4$&$<167$&&2.9&5.5\\
NGC1365&$18.2_{-2.7}^{+4.1}$&$<6.9$&$<237$&&$17.5_{-2.8}^{+3.5}$&$<6.8$&$<249$&&1.7&6.2\\
NGC1512&$11.7_{-1.2}^{+1.8}$&$<5.5$&$<238$&&$13.8_{-1.7}^{+1.5}$&$<5.7$&$<215$&&7.9&16.4\\
NGC1672&$19.0_{-3.3}^{+2.2}$&$7.0_{-0.9}^{+1.1}$&$194_{-30}^{+104}$&&$23.6_{-5.8}^{+3.0}$&$8.7_{-1.9}^{+0.2}$&$182_{-36}^{+100}$&&5.3&11.3\\
NGC2835&$15.7_{-2.1}^{+2.1}$&$<4.9$&$<148$&&$16.8_{-2.5}^{+2.5}$&$<5.3$&$<156$&&4.1&8.4\\
\enddata
\end{deluxetable*}

\bibliographystyle{aasjournal}
\bibliography{mybib}{}
\suppressAffiliationsfalse
\allauthors


\end{document}